# Phase fluctuations in a conventional s-wave superconductor: Role of dimensionality and disorder

*A Thesis*

*Submitted to the*

*Tata Institute of Fundamental Research, Mumbai*
*for the degree of Doctor of philosophy in Physics*

*by*

Mintu Mondal

Department of Condensed Matter Physics and Materials Science
Tata Institute of Fundamental Research
Mumbai

March, 2013

*To*

*my parents and teachers…*

# Contents

# Chapter 2: Our model system: NbN thin films …...……...……… ……………..………....87

# DECLARATION

This thesis is a presentation of my original research work. Wherever contributions of others are involved, every effort is made to indicate this clearly, with due reference to the literature, and acknowledgement of collaborative research and discussions.

The work was done under the guidance of Prof. Pratap Raychaudhuri, at the Tata Institute of Fundamental Research, Mumbai.

**Mintu Mondal**

In my capacity as supervisor of the candidate's thesis, I certify that the above statements are true to the best of my knowledge.

**Prof. Pratap Raychaudhuri**

Date:

# Preface

The work presented here, was carried out for the partial fulfillment of the requirements for the degree, doctor of philosophy in physics from Tata Institute of Fundamental Research, Mumbai, India.

The superconductivity in a clean conventional superconductor is well described by Bardeen-Cooper-Schrieffer (BCS) theory where fluctuations are unimportant except very close to $T_c$. However in case of reduced dimensionality or very high disorder, the scenario becomes considerably different and phase fluctuations play an important role in determining the superconducting properties. In this work, I have investigated the effect of phase fluctuations in two dimensional and strongly disordered three dimensional thin films of conventional s-wave superconductor, NbN by measuring electrodynamics response using low frequency mutual inductance technique and high frequency broadband microwave Corbino spectrometer under supervision of Prof. Pratap Raychaudhuri.

My thesis is organized in the following way:

In Chapter 1, I will introduce the motivation behind this work and electrodynamics of superconductors. In chapter 2, I will give over view of basic properties of NbN thin films which were used as a model system to study the fundamental properties related to superconductivity.

Chapter 3 contains the experimental details. In section 1, I will give brief introduction about the sample preparation. Section 2 deals with the development of low frequency mutual inductance technique to study the electrodynamics response of superconducting thin films in kHz frequency range. In section 3, I will provide detailed overview of broadband microwave Corbino spectrometer which was developed in our lab to study the high frequency electrodynamics response of superconductors in the frequency range 10 MHz to 20 GHz.

In chapter 4, I will elucidate the nature of phase disordering transition induced by reduced dimensionality in ultrathin superconducting NbN films, belonging to Berezinskii-Kosterlitz-Thouless (BKT) universality class.

Chapter 5 deals with the effect of phase fluctuations on superconducting properties of 3D strongly disordered epitaxial NbN films through measurements of finite frequency electrodynamics response. The thickness (~50 nm) of NbN films used for this study are almost 10 times thicker than the coherence length (~5 nm), therefore all films are effectively in 3D limit.

Then I will summarize my findings of my investigations carried out in last four and half years in chapter 6.

In the end, I will describe one interesting work carried out on single crystal of noncentrosymmetric superconductor, BiPd using **Andreev reflection point contact spectroscopy**. This work doesn't have direct correlation with the rest of my PhD thesis; therefore I will put it in the appendix.

# Statement of joint work

The work presented here is carried out in the Department of Condensed Matter Physics and Materials Science, at the Tata Institute of Fundamental Research, Mumbai in close collaboration with my other colleagues under supervision of Prof. Pratap Raychaudhuri. Results of major proportion of the work have already been published in peer reviewed journals.

Although most of the experiments and analysis presented here have been carried out by me, for the sake of completeness I have included some of the works done by my other group members. Details about the collaborative work done with others as follows:

The NbN thin films depositions and its structural characterizations were done in close collaboration with Dr Sanjeev Kumar, John Jesudasan and Vivas C. Bagwe. Transport, magnetotransport and Hall effect measurements were carried out in collaboration with Madhavi Chand. Scanning Tunneling Spectroscopy was performed by Anand Kamlapure and Garima Saraswat. Measurement of electrodynamics response was carried out by me. All theoretical analysis was done in close collaboration with Prof. Vikram Tripathi, Department of Theoretical Physics, TIFR, Dr. Lara Benfatto, University of Rome, Rome, Italy and Prof. G. Seibold, Institut Für Physik, BTU Cottbus, Germany.

Point contact Andreev reflection spectroscopy was carried out on very good quality single crystal of noncentrosymmetric superconductor, BiPd grown by Bhanu Joshi in collaboration with Prof. Srinivasan Ramakrishnan and Dr. Arumugam Thamizhavel.

The whole investigation was performed under supervision of Prof. Pratap Raychaudhuri.

# Acknowledgements

*Throughout my academic carrier, numerous persons in many occasions provided much needed support and help. Without their help this thesis would not have been possible. I would like to take this opportunity to express my gratitude towards them and thank them all.*

*First and foremost, I would like to thank my thesis supervisor, **Prof. Pratap Raychaudhuri** for his invaluable support, encouragement and guidance throughout my PhD and giving me the opportunity to work with him. I am thankful to him for his valuable lesson in research. He made me realize the importance of hard work and dedication and always pushed me to do better with my full effort. Most importantly, I am very grateful to him for his constant help wherever I faced trouble even in my personal issues.*

*I am extremely thankful to my lab mates **Madhavi Chand, Anand Kamlapure, Garima Saraswat, Vivas C. Bagwe, John Jesudasan, Dr. Sanjeev Kumar** and **Dr. Parasharam Shirage** for their constant help and support.*

*I would like to thank **S. P. Pai** (Excel Instruments, Mumbai) and **Atul Raut** (DCMP&MS workshop, TIFR) for technical help in instrument fabrications and I sincerely acknowledge the team responsible for **Low Temperature Facility**, TIFR for a steady supply of liquid $^4$He throughout my PhD.*

*I am thankful to **Prof. Vikram Tripathi** and **Dr. Lara Benfatto** for their constant theoretical support and useful discussion. I am also thankful to **Prof. G. Seibold** and **Prof. Sudhansu Mandal** for their theoretical support.*

*I would like to thank **Bhanu Joshi**, **Prof. Srinivasan Ramakrishnan** and **Dr. Arumugam Thamizhavel** for providing me a very good quality single crystal of BiPd.*

*During my master's study at IIT Kanpur, my senior, **Dibyendu Hazra** not only helped me a lot in my difficult time but also inspired me to carry out my study. I am very much thankful to him.*

*I owe a lot to the **Monks** of **Ramakrishna Mission Calcutta Students' Home, Belgharia, Kolkata** (http://www.rkmstudentshome.org), for constant inspiration and support throughout my academic carrier. They cared and loved me like my parents for 3 years during my BSc. Without* *their help and inspiration, probably I would have to discontinue my study. My deepest gratitude goes to revered **Swami Bhadratmananda (Suprakash Maharaj)** who took special care of me. I am also thankful to **Swami Annapurnanda, Swami Nilkhantananda (Bilas Maharaj), Swami Nityasattyananda (Bishyarup Maharaj)** and **Sanjeev Maharaj** for their teaching and spiritual enlightenment which gave me strength in this long journey.*

*I am extremely thankful to my teacher, **Ganesh Sarkar** and my senior from school, **Bijan Sarkar**. After my father's death, having a very poor economic background, I was having very hard time of my life.  At that much needed moment, beloved teacher, **Ganesh Sarkar** welcomed me with love and care and **Bijan Da** joined him to guide me and help me continue my education.*

*I am thankful to my **family**, **friends** and **relatives** for their constant support and help. I am especially thankful to my **Mother** and **Parul Didi** for their great care.*

*In the end, I want to thank my **Father**. It is unfortunate that I have lost him prematurely in my childhood. His words always inspire and guide me in this journey.*

# List of publications

## *Publications in refereed journals:*

1) **<u>Mintu Mondal</u>,** Anand Kamlapure, Somesh Chandra Ganguli, John Jesudasan, Vivas Bagwe, Lara Benfatto and Pratap Raychaudhuri, "Finite high-frequency superfluid stiffness in the pseudogap regime in strongly disordered NbN thin films",

   **Scientific Reports 3, 1357 (2013).**

2) **<u>Mintu Mondal</u>,** Bhanu Joshi, Sanjeev Kumar, Anand Kamlapure, Somesh Chandra Ganguli, Arumugam Thamizhavel, Sudhansu S. Mandal, Srinivasan Ramakrishnan, and Pratap Raychaudhuri, "Andreev bound state and multiple energy gaps in the noncentrosymmetric superconductor BiPd"

   **Phys. Rev. B 86, 094520 (2012).**

3) Madhavi Chand, Garima Saraswat, Anand Kamlapure, **<u>Mintu Mondal</u>,** Sanjeev Kumar, John Jesudasan, Vivas Bagwe, Lara Benfatto, Vikram Tripathi, and Pratap Raychaudhuri, "Phase diagram of a strongly disordered s-wave superconductor, NbN, close to the metal-insulator transition"

   **Phys. Rev. B 85, 014508 (2012).**

4) **<u>Mintu Mondal</u>,** Sanjeev Kumar, Madhavi Chand, Anand Kamlapure, Garima Saraswat, G. Seibold, L. Benfatto and Pratap Raychaudhuri, "Role of the vortex-core energy on the Beresinkii-Kosterlitz-Thouless transition in thin films of NbN"

   **Phys. Rev. Lett. 107, 217003 (2011).**

5) **<u>Mintu Mondal</u>,** Anand Kamlapure, Madhavi Chand, Garima Saraswat, Sanjeev Kumar, John Jesudasan, L. Benfatto, Vikram Tripathi, Pratap Raychaudhuri, "Phase fluctuations in a strongly disordered s-wave NbN superconductor close to the metal-insulator transition"

   **Phys. Rev. Lett. 106, 047001 (2011).**

6) **<u>Mintu Mondal</u>,** Madhavi Chand, Anand Kamlapure, John Jesudasan, Vivas C. Bagwe, Sanjeev Kumar, Garima Saraswat, Vikram Tripathi and Pratap Raychaudhuri, "Phase diagram and upper critical field of homogeneously disordered epitaxial 3-dimensional NbN films"

   **J. Supercond Nov Magn 24, 341 (2011).**

7) Anand Kamlapure, **<u>Mintu Mondal</u>,** Madhavi Chand, Archana Mishra, John Jesudasan, Vivas Bagwe, Vikram Tripathi and Pratap Raychaudhuri, "Penetration depth and tunneling studies in very thin epitaxial NbN films"

   **Appl. Phys. Lett. 96, 072509 (2010).**

8) Madhavi Chand, Archana Mishra, Y. M. Xiong, Anand Kamlapure, S. P. Chockalingam, John Jesudasan, Vivas Bagwe, **<u>Mintu Mondal</u>,** P. W. Adams, Vikram Tripathi, and Pratap Raychaudhuri, "Temperature dependence of resistivity and Hall coefficient in strongly disordered NbN thin films"

   **Phys. Rev. B 80, 134514 (2009).**

## *Conference Proceedings:*

1) **<u>Mintu Mondal</u>**, Sanjeev Kumar, Madhavi Chand, Anand Kamlapure, Garima Saraswat, Vivas C Bagwe, John Jesudasan, Lara Benfatto, Pratap Raychaudhuri,

   **Journal of Physics: Conference Series 400 (2), 022078 (2012).**

2) Anand Kamlapure, Garima Saraswat, Madhavi Chand, **<u>Mintu Mondal</u>**, Sanjeev Kumar, John Jesudasan, Vivas Bagwe, Lara Benfatto, Vikram Tripathi, Pratap Raychaudhuri,

   **Journal of Physics: Conference Series 400 (2), 022044 (2012).**

3) Madhavi Chand, **<u>Mintu Mondal</u>,** Anand Kamlapure, Garima Saraswat, Archana Mishra, John Jesudasan, Vivas C. Bagwe, Sanjeev Kumar, Vikram Tripathi, Lara Benfatto, and Pratap Raychaudhuri,

   **Journal of Physics: Conference Series 273, 012071 (2011).**

# Symbols and Abbreviations

## Symbols:

| | |
|---|---|
| $a$ | lattice constant or characteristic length scale of phase fluctuations |
| $d$ | dimension |
| $e$ | electronic charge |
| $E_F$ | Fermi energy |
| $\hbar = h/2\pi$ | $h$ is Planck's constant |
| $H_{c2}$ | upper critical field |
| $H_{peak}$ | position of MR peak |
| $J$ | superfluid stiffness |
| $k_B$ | Boltzmann constant |
| $k_F$ | Fermi wave-number |
| $k_F l$ | Ioffe Regel parameter |
| $l$ | Electronic mean free path |
| $m_e$ | mass of electron |
| $n$ | number density/ electronic carrier density |
| $n_H$ | electronic carrier density measured using Hall effect |
| $n_n$ | number of electrons that remain normal |
| $n_s$ | superfluid density |
| $N(0)$ | density of states at Fermi level |
| $R$ | resistance |
| $R_H$ | Hall coefficient |
| $t$ | sample thickness |
| $T$ | temperature |
| $T_{BCS}$ | Superconducting transition temperature expected within BCS theory |
| $T_{BKT}$ | Berezinskii-Kosterlitz-Thouless (BKT) transition temperature |
| $T_c$ | superconducting transition temperature |
| $v_F$ | Fermi velocity |
| $V_H$ | Hall voltage |
| $\xi$ | correlation length |
| $\xi_0$ | Pippard Coherence length |
| $\xi_{BCS}$ | BCS Coherence length |
| $\xi_{GL}$ | Ginzburg Landau coherence length |
| $\rho$ | resistivity |
| $\rho_m$ | maximum/peak resistivity |
| $\rho_n$ | normal state resistivity |
| $\rho_{peak}$ | peak resistivity |
| $\tau$ | electron scattering time |
| $\lambda$ | penetration depth |
| $\lambda\omega$ | Complex screening length |
| $\Phi_0$ | flux quantum |
| $\sigma$ | conductivity |
| $\sigma_0$ | minimum conductivity |
| $\Delta$ | superconducting energy gap |

# Abbreviations:

| | |
|---|---|
| 2D | two dimensions |
| 3D | three dimensions |
| AA | Altshuler and Aronov |
| AL | Aslamazov and Larkin |
| BCS | Bardeen, Cooper and Schreiffer |
| BKT | Berezinskii-Kosterlitz-Thouless |
| BTK | Blonder-Tinkham-Klapwijk |
| DOS | density of states |
| GL | Ginzburg Landau |
| HTS | High temperature superconductors |
| MR | magnetoresistance |
| MT | Maki and Thompson |
| NCS | noncentrosymmetric superconductor |
| PG | pseudogap |
| SC | superconducting |
| STM | scanning tunneling microscope |
| STS | scanning tunneling spectroscopy |
| TEM | transmission electron microscopy |
| XRD | X-ray Diffraction |

# Synopsis

## 1. Introduction

In a superconductor, Cooper pairs form a macroscopic quantum phase coherent state described by complex order parameter, $\psi = |\Delta|e^{i\theta}$, where $|\Delta|$ is the measure of binding energy of the Cooper pairs which manifests as a gap in the electronic excitation spectrum, and $\theta$ is the phase of the macroscopic condensate. In a clean conventional superconductor which is well described by Bardeen-Cooper-Schrieffer (BCS) theory [1], the superconductivity is destroyed at a characteristic temperature, $T_c$, at which $|\Delta|$ goes to zero and fluctuations are unimportant except very close to $T_c$[2]. However, in principle, superconductivity can also be destroyed by thermal or quantum phase fluctuations even if $|\Delta|$ remains finite. The energy cost of twisting the phase is given by the superfluid stiffness ($J$), given by the relations [3],

$$J = \frac{\hbar^2 a n_s}{4m}; \quad n_s = \frac{m}{\mu_0 e^2 \lambda^2},$$ 
(1)

where $m$ is electronic mass and $a$ is the characteristic length scale for phase fluctuations, $\lambda$ is the magnetic penetration depth and $n_s$ is the superfluid density. When the superconductor has a thickness, $t$, smaller than the coherence length, $\xi_0$, (2D limit) $a \approx t$ ; for a 3D superconductor $a \approx \xi_0$. In clean bulk conventional superconductors, $J >> T_c$, and therefore phase fluctuations play negligible role, consistent with BCS theory. However, when $t$ is decreased or $n_s$ is reduced through strong disorder scattering, $J/k_B$ decreases and eventually becomes smaller than the mean field $T_c$ (defined by BCS theory) at some critical value of thickness or disorder. In such a situation, superconductivity can get destroyed through phase disordering, giving rise to novel electronic states with finite density of Cooper pairs but no global superconductivity [3,4].

In two dimensional (2D) or quasi 2D superconductors, the phase disordering transition was predicted to be belongs to Berezinskii-Kosterlitz-Thouless (BKT) universality class [5,6,7,8,9] where superconductivity is destroyed due to proliferation of vortices in the system. However in real superconductor this phase disordering transition appears to be nonuniversal in nature due to additional complicacies such as intrinsic inhomogeneity, which tends to smear the sharp signatures of BKT transition compared to the clean case and difference in vortex-core energy, μ, from the predicted value within the 2D XY model originally investigated by Kosterlitz and Thouless [5]. This can give rise to somehow different manifestation of vortex physics, even without the change of the order of transition [10]. Recently, the relevance of μ for the BKT transition has attracted a renewed interest in different contexts, ranging from the case of layered high-temperature superconductors [11,12,13] to the superconducting interfaces in artificial hetero structures [14,15,16,17] and liquid gated interface superconductivity[18].

In recent scanning tunneling spectroscopy (STS) measurements on strongly disordered *s*-wave superconductors [19,20,21,22] (TiN, InO$_x$ and NbN), reveal the appearance of pseudogap (PG) state characterized by a gap in electronic spectrum which persist at temperature well above $T_c$, which suggests the existence of Cooper pairs, but no global superconductivity.

All these phenomena have raised renewed wave of interest about the understanding of the nature of superconductivity in a 2D or quasi-2D superconductor and in very strongly disordered 3D superconductor. The electrodynamics response of superconductors provides an ideal tool to explore the role of phase fluctuations in superconductivity. In this thesis, I will present an investigation on the role of phase fluctuations, through the measurement of λ using low frequency mutual inductance technique and the microwave complex conductivity using broadband microwave Corbino spectrometer, in thin films of the conventional superconductor NbN both in 2D and 3D limit [21,22,23,24]. Our study elucidates interplay of quasiparticle excitations (QE) and phase fluctuations in strongly disordered and low dimensional superconductors.

## 2. Experimental details

## 2.1. Sample preparation and its characterization

Epitaxial NbN films used in our studies were grown by reactive DC magnetron sputtering of Nb target in an Ar/N$_2$ gas mixture on oriented MgO (100) substrate. The disorder, in the form of Nb vacancies in the crystalline NbN lattice, was controlled by changing the sputtering power or Ar/N$_2$ ratio in the gas mixture [25,26]. To study the effect of reduced dimensionality, the deposition conditions were optimized to obtain the highest possible $T_c$ (~16.5 K) for a 50 nm thick film. Then the thickness (*t*) of films was varied for a fixed disorder level by changing the deposition time and by keeping other deposition conditions fixed. To study the effect of disorder, we deposited another set of films where disorder level was tuned by changing the deposition conditions by keeping the thickness, $t \geq 50$ nm such that all our films are in 3D limit ($t >> \xi$).

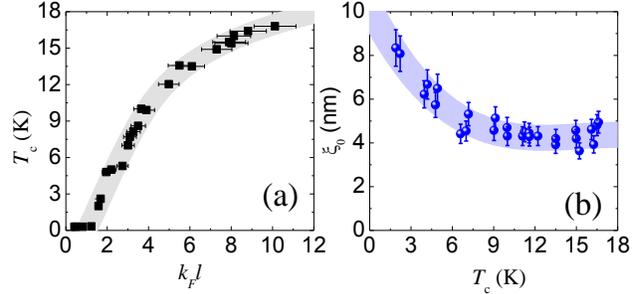

Figure 1. (a) Superconducting transition temperature, $T_c$ as a function of $k_F l$. (b) Coherence length, $\xi_0$ as a function of $T_c$.

All the films were characterized by transport measurements such as resistivity, magneto resistance and Hall carrier density. The resistivity ($\rho$) was measured using four probe technique. Hall carrier density ($n_H$) was calculated from the measured hall coefficient ($R_H = -1/n_H e$) by sweeping the magnetic field ($H$) from 12T to -12T at different temperatures. We define superconducting transition temperature ($T_c$) of our films where resistance drops below our measureable limit as $T$ decrease. To quantify disorder in our NbN samples, we have used Ioffe Regel parameter, $k_F l$ where $k_F$ is Fermi wave vector and $l$ is the mean free path. We have calculated the value of $k_F l$ from measured $\rho$ and $n_H$ using the relation, $k_F l = \left\{ \hbar \left( 3\pi^2 \right)^{2/3} \right\} / \left[ \left\{ n_H \left( 285K \right) \right\}^{1/3} \rho \left( 285K \right) e^2 \right]$ considering free electron model. In presence of electron-electron interaction which is very much present in our system [26], the relation $R_H = -1/n_H e$ is not truly valid. Therefore we calculate $k_F l$ using $R_H$ and $\rho$ at the highest temperature of our measurements i.e. at 285K, where the electron-electron interaction is expected to be small [27]. Most interesting part of NbN thin films is that we can tune disorder over a very large range by changing the deposition condition only and with increasing disorder the value of $k_F l$ varies from $k_F l$ ~10 for moderately clean sample to below Mott limit, $k_F l$ ~1 for very high disordered sample. Fig. 1. (a) shows the $T_c$ as a function of $k_F l$ for a set of 3D disordered NbN films. As we increase the disorder in the system, $T_c$ decreases and above a critical disorder level, sample becomes non-superconducting.

The upper critical field ($H_{c2}$) as a function of temperature ($T$) was measured for several samples from $R$-$T$ scans at different $H$. Since all our films are in the dirty limit, $l << \xi_0$, we have estimated $H_{c2}(0)$ and $\xi_0$ using dirty limit relation [28,29]:

$$H_{c2}(0) = 0.69 T_c \frac{dH_{c2}}{dT}\bigg|_{T=T_c} \quad ; \quad \xi_0 = \left[\frac{\phi_0}{2\pi H_{c2}(0)}\right]^{1/2} , \qquad (2)$$

Fig. 1.(b) shows the measured coherence length ($\xi_0$) which is characteristic length of fluctuations as function of $T_c$ for a set of disordered 3D NbN films.

Superconducting energy gap ($\Delta$) was measured for a set of NbN films with different level of disorder using planner tunnel junctions and home built scanning tunneling microscope (STM) down to 300 mK [21,22].

## 2.2. Low frequency electrodynamics response

To probe the low frequency electrodynamics, we have developed a two coil mutual inductance technique which is a very powerful technique for measuring magnetic penetration depth ($\lambda$) of superconducting thin films with thickness, $t \leq \lambda/2$ [30,31,32]. The main advantage of this technique is that the absolute value of $\lambda$ can be measured over the entire temperature range up to $T_c$ without any model dependent assumptions. In this technique, an 8 mm diameter superconducting film is sandwiched coaxially between a quadrupole primary coil and a dipole secondary coil (see Fig. 2). Then the mutual inductance ($M = M' + iM''$) between primary and secondary coil is measured as function of temperature by passing a small ac excitation current (0.5mA) through the primary coil and measuring the induced voltage at secondary coil using lock-in amplifier. $\lambda$ is determined by evaluating the mutual inductance for different values of $\lambda$ by numerically solving the London and Maxwell coupled equations and comparing with the experimentally measured value. For details about the measurement of $\lambda$ see the ref. 30, 31 and 32.

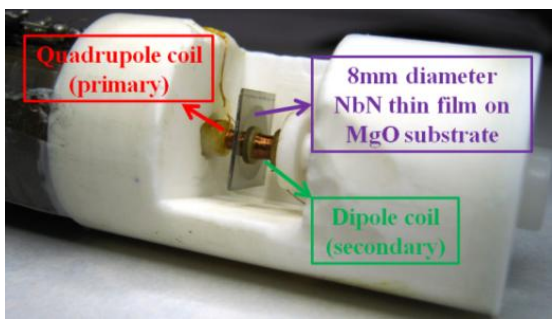

Figure 2. Coil assembly of our low frequency mutual inductance setup. The quadrupole (primary) coil has 28 turns with the half closer to the film wound in one direction and the farther half wound in the opposite direction. The dipole (secondary) coil has 120 turns wound in the same direction in 4 layers.

## 2.3. High frequency electrodynamics response

To probe the high frequency electrodynamics of superconducting thin films we have developed a broadband microwave Corbino spectrometer [33] (Fig. 3). The complex conductivity ($\sigma(\omega)=\sigma_1$-$i\sigma_2$) was measured using our spectrometer in the frequency range from 10 MHz to 20 GHz. This setup works based on broadband measurement of reflection coefficient, $S_{11}$ of microwave transmission line terminated by a Corbino shaped sample (Fig. 3.(c)). The sample impedance can be calculated from $S_{11}$ using standard formula, $Z_S = \frac{1+S_{11}}{1-S_{11}} Z_0$ where $Z_0$=50$\Omega$ is characteristic impedance of transmission line. For sample of thickness much smaller than the screen depth, the complex conductivity is given by $\sigma = \frac{\ln{(b/a)}}{2\pi\, d\, Z_S}$ where $a$ and $b$ are the inner and outer radii of the sample (Fig. 3.(c)). While the principle of this technique is simple, the real difficulty lies in determining the true $S_{11}$ from the measured reflection coefficient, $S_{11}^m$ which contains the contribution from both the sample and the long co-axial cable. The contribution from sample was extracted from $S_{11}^m$ through extensive calibration of our microwave probe using three known references: (i) a thick gold film as a short, (ii) a Teflon disk as an open and (iii) a 20nm NiCr film as a load [33].

The two-probe resistivity ($\rho$) of the sample was measured in-situ using the bias-tee of the network analyzer in the same run with microwave measurement.

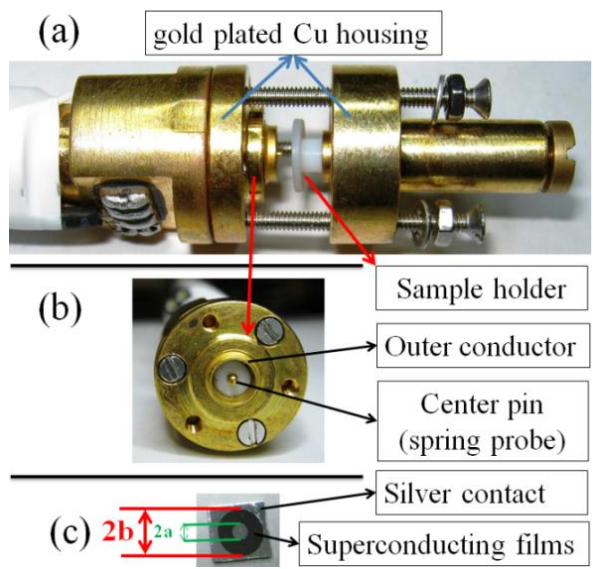

Figure 3. (a) Main head of our broadband microwave spectrometer; (b) Coaxial microwave probe: (c) Corbino shaped sample with silver contact pad.

## 3. Phase fluctuations in 2D superconductor

In strictly 2D superconductors, the superconducting transition has been proposed to belong to BKT universality class [34,35]. In this kind of phase transition, when the phase stiffness, $J$, becomes comparable to $k_B T$, thermally excited vortex-antivortex pair unbinds, thereby destroying the superconducting state due to vortex proliferation. The temperature at which this phase transition occurs is given by,

$$T_{BKT} \equiv \frac{\pi}{2} J(T_{BKT}^-) , \qquad (3)$$

above this temperature proliferation of vortices drives $n_s$ abruptly to zero therefore $J$ also approaches to zero. Although superfluid He films follows this behaviour quite precisely [9], the BKT transition in 2D superconductors has remained controversial [36]. For instance, the jump in $n_s$ is often observed at a temperature lower than the expected $T_{BKT}$ and at a $J(\propto n_s)$ larger than expected from eqn. (3). By a systematic study of $\lambda^{-2}(T)$ and $\rho(T)$, we will show that these discrepancies result from the effect of quasiparticle excitations which modifies the vortex core energy ($\mu$) from the value expected from 2D XY model and intrinsic disorder in the system.

To demonstrate the BKT transition in NbN thin films we have plotted $\lambda^{-2}(T) \propto n_s(T)$ as a function of temperature in figure Fig. 4(a-b). Since in our films the electronic mean free path,

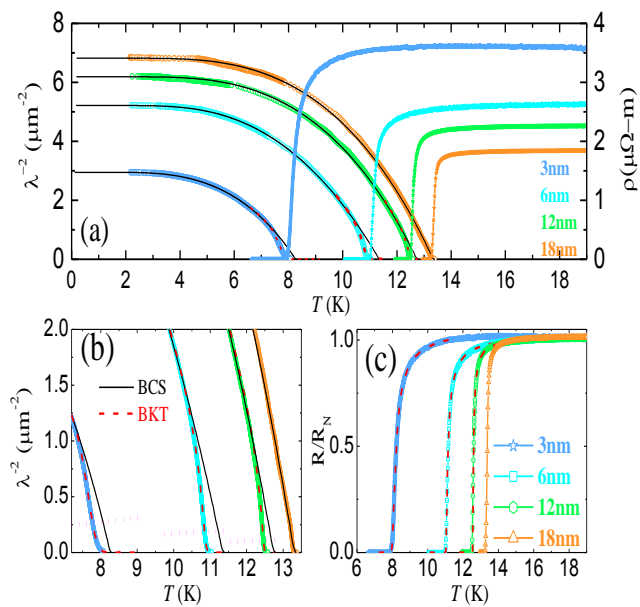

Figure 4. (a) Temperature dependence of $\lambda^{-2}(T)$ and $\rho(T)$ for four NbN films with different thickness. The (black) solid lines and (red) dashed lines correspond to the BCS and BKT fits of the $\lambda^{-2}(T)-T$ data respectively. (b) Shows an expanded view of $\lambda^{-2}(T)$ close to $T_{BKT}$; the intersection with (magenta) dotted line where the universal BKT transition is expected. (c) Temperature variation of $R/R_N$. The (red) dashed lines show the theoretical fits to the data, as described in the text.

$l<<\xi$, we fit the temperature variation of $\lambda^{-2}(T)$ with the dirty limit BCS expression [37],

$$\frac{\lambda^{-2}(T)}{\lambda^{-2}(0)} = \frac{\Delta(T)}{\Delta(0)} \tanh\left(\frac{\Delta(T)}{2k_bT}\right), \tag{4}$$

using $\Delta(0)$ as a fitting parameter. We observe that for thinner films $\lambda^{-2}(T)$ starts to deviate downwards from expected BCS behavior close to $T_c$. The jump in $\lambda^{-2}(T)$ close to $T_c$ which signifies the BKT transition becomes more and more prominent as we decrease the film's thickness. However, the jump in $n_s$ is observed at a temperature lower than the expected $T_{BKT}$ and therefore at a larger $J$ than expected from eqn. (3). The above discrepancy can be reconciled by taking into account that $J(T)$ is not only affected by the presence of quasiparticles excitation but also by the presence of thermally excited vortex-antivortex pairs in the system. When $\mu$ is large, the latter effect is negligible for $T<T_{BKT}$. However in a superconductor, the presence of quasiparticle excitations reduces the vortex core energy from the value expected from the 2D XY model. Therefore $J(T)$ gets renormalized due to increase of thermally excited vortex-antivortex pairs even below $T_{BKT}$. To take into account the effect described above, we have numerically solved the renormalization group equations of the original BKT formalism [38] using only one free parameter: $\mu/J$, where $J$ is obtained from BCS fit to the experimental data (Fig. 4) as $T\rightarrow 0$.

Table 1 Magnetic penetration depth $(\lambda(T\rightarrow0))$, $T_{BTK}$, $T_{BCS}$ along with the best fit parameters obtained from BKT fits of the $\lambda^{-2}(T)$ and $R(T)$ data for NbN thin films of different thickness. $T_{BCS}$ corresponds to the mean field transition temperature obtained by extrapolation of the BCS fit of $\lambda^{-2}(T)$ at $T<T_{BKT}$.

| $d$ | $\lambda(0)$ | $T_{BKT}$ | $T_{BCS}$ | From best fit of $\lambda^{-2}(T)$ | | | From best fit of $\rho(T)$ | |
|---|---|---|---|---|---|---|---|---|
| (nm) | (nm) | (K) | (K) | $\mu/J$ | $\delta/J$ | $b_{theo}$ | $A$ | $b$ |
| 3 | 582 | 7.77 | 8.3 | 1.19±.06 | 0.020±0.002 | 0.108 | 1.35±0.14 | 0.108±0.006 |
| 6 | 438 | 10.85 | 11.4 | 0.61±.05 | 0.005±0.0007 | 0.048 | 1.30±0.13 | 0.067±0.008 |
| 12 | 403 | 12.46 | 12.8 | 0.46±.05 | .0015±0.0003 | 0.027 | 1.21±0.12 | 0.039±0.006 |
| 18 | 383 | ---- | 13.4 | ---- | ---- | ---- | ----- | ----- |

To take into account the effect of inhomogeneity [39] in $J$, we average over the distribution of $J$, assuming a Gaussian distribution around $J_{BCS}$ with relative width $\delta$ for simplicity. The best fit values are listed in Table 1. Fig. 4(b) shows that the above procedure leads to excellent fits although the ratio $\mu/J$ (Table 1) is small compared to the value, $\mu_{XY}/J = \pi^2/2 = 4.9$ expected from 2D XY model. The fact that in a superconductor $\mu$ is small explains why the downturn is observed at higher superfluid density though the BKT transition happens at the point predicted by eqn. (3). To further establish our findings, we have analyzed our resistivity data by considering BKT fluctuations and G-L fluctuations. In 2D, the contribution of SC fluctuations to conductivity can be encoded in the temperature dependence of SC correlation length, $\delta\sigma \propto \xi^2(T)$. Due to proximity effect between the $T_{BKT}$ and $T_{BCS}$, it is expected that most of the fluctuations regime will be accounted for by G-L fluctuations while KT fluctuations will be relevant only between $T_{BKT}$ and $T_{BCS}$. We interpolate between these two regimes using the Halperin-Nelson [34,35] interpolation formula for the correlation length,

$$\frac{\xi}{\xi_0} = \frac{2}{A} \sinh\left(\frac{b}{\sqrt{t_r}}\right),$$  (5)

where $t_r = (T\text{-}T_{BKT})/T$ and A is a constant of order unity. $b$ is the most relevant parameter and related to vortex core energy by $b \approx (4\mu\sqrt{t_c} / \pi^2 J)$ [39], where $t_c = (T_{BCS}\text{-}T_{BKT})/T_{BKT}$ (Calculated $b$ using the best fit value of the superfluid density data is defined as $b_{\text{theo}}$ shown in Table 1.). The resistivity corresponding to the SC correlation length is given by

$$\frac{\rho}{\rho_N} = \frac{1}{1 + (\Delta\sigma/\sigma_0)} \equiv \frac{1}{1 + (\xi/\xi_0)^2},$$  (6)

To take into account the sample inhomogeneity we correlate the distribution of local superfluid stiffness used to analyze the superfluid density data below $T_{KTB}$ with distribution of local normalized resistivity values $\rho_i = R_i/R_N$ according to eqn. (6), where local $T^i_{BCS}$ is attributed to a patch having local superfluid stiffness $J_i$. Then the overall normalized resistivity $\rho = R/R_N$ can be calculated using effective medium theory (EMT) [40] by modeling our system as random resistor network. We apply the above procedure to analyze our resistivity data using the values of $T_{BKT}$ and $T_{BCS}$ determined from the analyzed superfluid density data where A and $b$ are taken as free parameters. The resulting fits are in excellent agreement with the experimental data shown in Fig. 4. Considering that the interpolation formula is an approximation, the value of $b$ is in very good agreement with theoretically estimated value, $b_{theo}$ using best fit values of superfluid density data listed in table 1.

Once the robustness of our estimate of $\mu$ is established, we now discuss the values reported in Table I and their thickness dependence. We first notice that the values of $\mu$ obtained by our fit are of the order of magnitude of the standard expectation for a BCS superconductor. In this case, one usually estimates $\mu$ as the loss of condensation energy within a vortex core of the size of the order of the coherence length $\xi_0$ [24],

$$\mu = \pi \xi_0^2 \varepsilon_{cond} , \tag{7}$$

where $\varepsilon_{cond}$ is the condensation-energy density for the superconductor. In the clean superconductor, $\mu$ can be expressed in terms of $J$ by means of the BCS relations for $\varepsilon_{cond}$ and $\xi_0$. Since $\varepsilon_{cond} = N(0)\Delta^2/2$, where $N(0)$ is the density of states at the Fermi level and $\Delta$ is the BCS gap; $\xi_0 = \xi_{BCS} = \hbar v_F/\pi\Delta$, where $v_F$ is the Fermi velocity; and assuming that $n_s \sim n$ at $T = 0$, where $n = 2N(0)v_F^2 m/3$, one has $\mu_{BCS,}$

$$\mu_{BCS} = \frac{\pi \hbar^2 n}{4m} \frac{3}{\pi^2} = \pi J \frac{3}{\pi^2} \approx 0.95 J , \tag{8}$$

so that it is quite smaller than $4.9J$ expected from XY-model. While the exact determination of $\mu$ depends on small numerical factors that can slightly affect the above estimate, the main ingredient that we should still account for the effect of disorder that can alter the relation between $\varepsilon_{cond}$, $\Delta$ and $J$ and explain the variations observed experimentally.

To properly account for it, we computed explicitly both $\mu$ and $J$ within the attractive two dimensional Hubbard model with local disorder [24]. The resulting value of $\mu/J$ at $T = 0$ is plotted in Fig. 5. (a). It is of the order of BCS estimate and it shows a steady increase as disorder increases, in agreement with the experimental results, shown in Fig. 5 (b), where we take the normal-state sheet resistance $R_s$ as a measure of disorder as the film thickness decreases. This behavior can be understood as a consequence of the increasing separation with disorder between the energy scales associated, respectively, to the $\Delta$ which controls $\varepsilon_{cond}$ and $J$, as it is shown by the ratio $\Delta/J$ that we report in the two panels of Fig. 5 for comparison. Notice that, the values of $\Delta/J$ are much larger than experimental ones because the calculation was done at strong coupling strength as compared to our NbN samples due to constraint in numerical analysis. Nonetheless, our approach already captures the experimental trend of $\mu/J$.

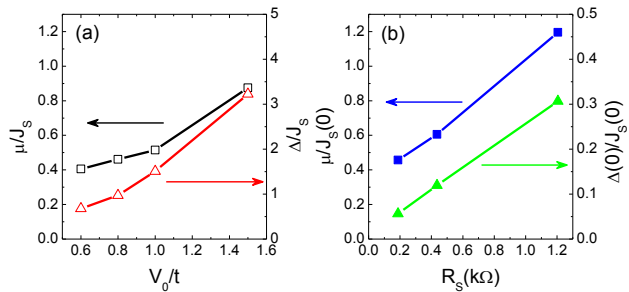

Figure 5. (a) Numerical results for the disorder dependence of $\mu/J$ and $\Delta/J$ as a function of disorder for the attractive Hubbard model. (b) Experimental values for the same ratios in our NbN films, plotted as a function of the normal state sheet resistance $R_s$.

In summary, we have shown that the phase transition in 2D or quasi-2D superconductor can be reconciled with the standard BKT physics when the small vortex core energy is taken into consideration. Our work finally provides a complete paradigm description of the BKT transition in real superconductors.

## 4. Phase fluctuations in strongly disordered 3D NbN films

According to the BCS theory the superconducting transition temperature is determined by superconducting energy gap, $\Delta$. The superconductor to normal transition occurs at a characteristic temperature, $T_c$ where $\Delta$ goes to zero. However, in scanning tunneling spectroscopy (STS) measurements, it was observed that for low disorder sample $\Delta$ goes to zero as $T$ goes to $T_c$ as expected from BCS theory but as we increase the disorder, $T_c$ is suppressed although $\Delta$ remains finite well above $T_c$ and give rise to pseudogap (PG) [19,20,21,22] like state contrary to BCS prediction[1].

A natural explanation of this observation can be from phase-fluctuations where the superconducting state is destroyed from phase disordering at $T_c$ before $|\Delta|$ goes to zero at temperature $T=T^*$ called PG temperature. In 3D disordered superconductor, there are two types of phase fluctuations about the BCS ground state which can destroy superconductivity: (i) the classical (thermal) phase fluctuations (CPF) and (ii) the quantum phase fluctuations (QPF) associated with number phase uncertainty. QPF results from the fact that, there will be Coulomb energy cost associated with number fluctuations when phase coherence is established between neighboring regions. Therefore if the electronic screening is poor, such as in a strongly disordered system it becomes energetically favorable to relax the phase in order to decrease number fluctuations. Here I will explore the effect of phase fluctuations induced by disorder by a thorough study of low and high frequency electrodynamics responses of disordered 3D NbN thin films.

## 4.1. Low frequency electrodynamics response

To study the effect of phase fluctuations, $\lambda(T)$ was measured using low frequency two coil mutual inductance technique for a series of 3D NbN films with progressively increasing disorder with $T_c$ varying from 16 K to 2.27 K. Fig. 6(a) and (b) show the temperature variation of $\lambda^{-2}(T)$ for a set of NbN films with different $T_c$. For the films with low disorder, the temperature variation of $\lambda^{-2}(T)$ follow the dirty-limit BCS behavior (black solid line) but as we increase the disorder $\lambda^{-2}(T)$ starts to deviate from the expected BCS temperature variation and shows a gradual evolution towards a linear-$T$ variation which saturates at low temperatures for samples with $T_c \leq 6$ K. This trend is clearly visible for the strongly disordered sample with $T_c \sim 2.27$ K (Fig. 4(d)). Now we concentrate on the value of $\lambda^{-2}(T)$ as $T \rightarrow 0$. In absence of phase

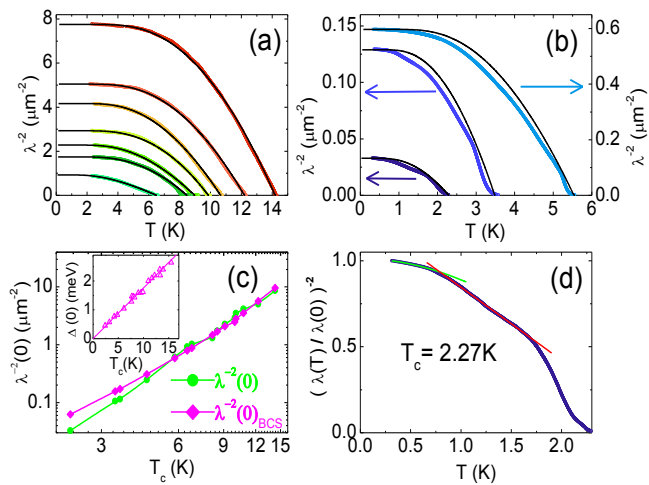

**Figure 6.** (a)-(b) $\lambda^{-2}(T)$ vs $T$ for a set of disordered NbN films; the solid black lines are the expected temperature variations from dirty limit BCS theory. (c) $\lambda^{-2}(0)$ and $\lambda^{-2}(0)_{BCS}$ as function of $T_c$; the *inset* shows the $\Delta(0)$ as function of $T_c$ . (d) Temperature variation of $(\lambda(T)/\lambda(0))^{-2}$ for film with $T_c = 2.27$ K; the solid lines (green) are fits to the $T^2$ dependence of $\lambda^{-2}(T)/\lambda^{-2}(0)$ at low temperature $(T \leq 0.65K)$ and the $T$ dependence (red) at higher temperature.

fluctuations, the disorder scattering reduces $\lambda^{-2}(0)$ according to the BCS relation [41],

$$\lambda^{-2}(0)_{BCS} = \frac{\pi\mu_0\Delta(0)}{\hbar\rho_0},\qquad(9)$$

where $\rho_0$ is the resistivity just above $T_c$. For NbN, we find $\Delta(0) \approx 2.05 k_B T_c$ from tunneling measurements performed at low temperatures ($T < 0.2T_c$) on planar tunnel junctions fabricated on a number of samples with different levels of disorder [42] (see the inset of Fig. 6(c)). Fig. 6(c) shows the $\lambda^{-2}(0) \approx \lambda^{-2}(0)_{BCS}$ within experimental error for samples with $T_c > 6$K. However, as we approach the critical disorder $\lambda^{-2}(0)$ becomes gradually smaller than $\lambda^{-2}(0)_{BCS}$, reaching a value which is 50% of $\lambda^{-2}(0)_{BCS}$ for the sample with $T_c \sim 2.27$K.

Since the suppression of $\lambda(0)^{-2}$ from its BCS value and linear-$T$ dependence of $\lambda^{-2}(T)$ are characteristic features associated with QPF and CPF [43] respectively, we now try to quantitatively analyze our data. The importance of QPF and CPF is determined by two energy scales: The Coulomb energy $E_c$, and the superfluid stiffness, $J (\propto n_s)$ [3,21]. The suppression of $\lambda^{-2}(0)$ due to QPF was estimated using the self consistent harmonic approximation [21,44] which gives (in 3-D) $n_s(T=0)/n_{s0}(T=0) \approx 0.02$. While this value is likely to have some inaccuracy due to the exponential amplification of any error in our estimation of $E_c$ or $J$, the important point is that this suppression is much larger than our experimental estimation, $\lambda^{-2}(0)/\lambda_{BCS}^{-2}(0) \approx 0.5$. On the other hand the crossover temperature from QPF to CPF is estimated to be about 75 K which implies that CPF cannot be responsible for the observed linear temperature dependence of $\lambda^{-2}(T)$ in this sample.

These two apparent contradictions can be resolved by considering the role of dissipation. In $d$-wave superconductors, the presence of low energy dissipation has been theoretically predicted [45] and experimentally observed from high frequency conductivity [46,47] measurements. In recent microwave experiment [48] on amorphous InO$_x$ films reveals that low energy dissipation can also be present in strongly disordered $s$-wave superconductors. While the origin of this dissipation is not clear at present, the presence of dissipation has several effects on phase fluctuations: (i) QPF are less effective in suppressing $n_s$; (ii) QPF contribute to a $T^2$ temperature dependence of $n_s$ of the form $n_s/n_{s0} = 1 - BT^2$ at low temperature where B is directly proportional to the dissipation and (iii) the crossover to the usual linear temperature dependence of $n_s$ due to CPF, $n_s / n_{s0} = 1 - (T / 6J)$, occurs above a characteristic temperature that is much smaller than predicted temperature. In the sample with $T_c \sim 2.27$ K, the $T^2$ variation of $\lambda(T)^{-2} / \lambda(0)^{-2}$ can be clearly resolved below 650 mK. In the same sample, the slope of the linear-$T$ region is 3 times larger than the slope estimated from the value of $J$ calculated for $T = 0$. This discrepancy is however minor considering the approximations involved. In addition, at finite temperatures $n_{s0}$ gets renormalized due to QE. With decrease in disorder, QE eventually dominates over the phase fluctuations, thereby recovering the usual BCS temperature dependence at low disorder. Since CPF eventually lead to the destruction of the superconducting state at temperature less than the mean field transition temperature, the increased role of phase fluctuations could naturally explain the observation of a PG state in strongly disordered NbN films. We would also like to note that in all the disordered samples $\lambda^{-2}$(T) shows a downturn close to $T_c$, reminiscent of the BKT transition, in ultrathin superconducting films. However, our samples are in the 3 D limit where a BKT transition is not expected. At present we do not know the origin of this behavior.

Further confirmation of the appearance of PG state due to phase fluctuations comes from the comparison of two energy scales: superfluid stiffness, $J$ and superconducting energy gap, $\Delta$. Using relation (1) we have estimated the values of $J$ (Fig. 7) using experimentally measured $\xi_0$ (ref. 28) and $\lambda^{-2}$(0). As expected, in the low disorder regime, $J$ is very large and thus the effect of phase fluctuations is negligible. However, as the disorder increases, $J$ rapidly reduces and becomes comparable to $\Delta$ for sample with $T_c \leq 6$K and as a result the phase fluctuations are expected to play a significant role in superconductivity which is consistent with the observed PG state in strongly disordered sample with $T_c \leq 6$ K [21,22].

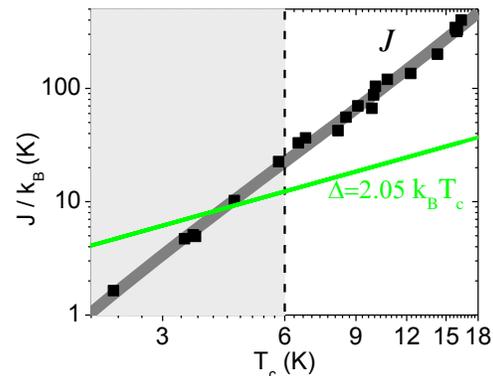

Figure 7. Superfluid stiffness ($J/k_B$) for NbN films with different $T_c$. The solid line corresponds to $\Delta = 2.05$ $k_B T_c$.

In summary, we have observed a progressive increase in phase fluctuations and the formation of a PG state in strongly disordered NbN thin films. The above observations lead us to conclude that the superconducting state in strongly disordered superconductors is destroyed by phase fluctuations. In scanning tunneling spectroscopy measurements [21,22], it was observed that at strong disorder the superconductor spontaneously segregates into domains separated by regions where the superconducting order parameter is suppressed. One would expect that the phase fluctuations between these domains result in destruction of the global superconducting state whereas Cooper pairs continue to survive in localized islands. In this scenario in a very strongly disordered system the superfluid stiffness, $J$ is expected to be strongly frequency dependent above $T_c$ and it will be zero over a large length scale but will remain finite in shorter length scale in the PG regime.

## 4.2. High frequency electrodynamics response

To confirm our phase fluctuations scenario we have studied the high frequency electrodynamics responses of disordered superconductor through measurement of ac complex conductivity, $\sigma(\omega) = \sigma_1(\omega) - i\sigma_2(\omega)$ using our broadband microwave Corbino (BMC) spectrometer [33] in the frequency range 0.4-20 GHz. Samples used in this study consist of a set of epitaxial NbN thin films with different levels of disorder having $T_c$ varying in the range $T_c \approx 15.7$ - 3.14 K. The advantage of this technique is that it is sensitive to the length scale set by the probing frequency, given by the relation, $L(\omega) = [D/(\omega/2\pi)]^{1/2}$. Here, $D$ is the electronic diffusion constant given by, $D \approx v_F l/d$, where $v_F$ is the Fermi velocity, $l$ is the electronic mean free path and $d$ is the dimension of the films.

Fig. 8(a) and (b) show the representative data for $\sigma_1(\omega)$ and $\sigma_2(\omega)$ as functions of frequencies at different temperatures for the sample with $T_c \sim 3.14$ K. At low temperatures $\sigma_1(\omega)$ shows a sharp peak at $\omega \rightarrow 0$ whereas $\sigma_2(\omega)$ varies as $1/\omega$ (dashed line), consistent with the expected behavior in the superconducting state. Well above $T_c$, $\sigma_1(\omega)$ is flat and featureless and $\sigma_2(\omega)$ is within the noise level of our measurement, consistent with the behavior in a normal metal.

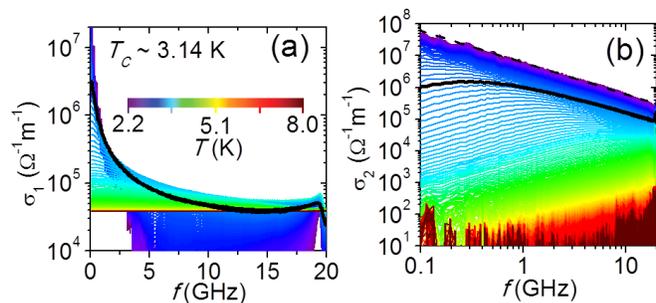

Figure 8. Frequency dependence of real and imaginary part of conductivity for a disordered NbN film with $T_c$=3.14 K. The solid black lines are the conductivity at $T_c$. In panel (b) dashed black line is $1/\omega$ fit to $\sigma_2$ below $T_c$. The residual features in conductivity about 19GHz is due to the imperfection calibration of our spectrometer.

In the superconducting state where phase coherence is established at all length and time scales, the superfluid density ($n_s$) and $J$ can be determined from $\sigma_2(\omega)$ using the relation,

$$\sigma_2(\omega) = \frac{n_s e^2}{m\omega} \text{ and } J = \frac{\hbar^2 n_s a}{4m},\qquad (10)$$

where $e$ and $m$ are the electronic charge and mass respectively, and $a$ is the characteristic length scale associated with phase fluctuations which is of the order of the dirty limit coherence length, $\xi_0$. Fig. 9(a)-(d) shows $\sigma_1(\omega)$-$T$, $\sigma_2(\omega)$-$T$ and $J$-$T$ at different frequencies for four samples with different $T_c$. One should notice that, all samples show a dissipative peak in $\sigma_1(\omega)$ close to $T_c$ and the peak becomes more and more prominent as we increase the disorder in our samples. In low disorder samples for all frequencies, $\sigma_2(\omega)$ dropped close to zero at $T_c$. On the other hand samples with higher disorder show an extended fluctuation region where $\sigma_2(\omega)$ remains finite up to a temperature well above $T_c$. We convert $\sigma_2(\omega)$ into $J$ (from eqn. 10) using the experimental values of $\xi_0$ [28]. For $T < T_c$, $J$ is frequency independent, showing that the phase is rigid at all length and time scales. However, for the samples with higher disorder (Fig. 9(c) and 9(d)), $J$ becomes strongly frequency dependent above $T_c$. While at 0.4 GHz $J$ falls to zero very close to $T_c$, with increase in frequency, it acquires a long tail and remains finite well above $T_c$.

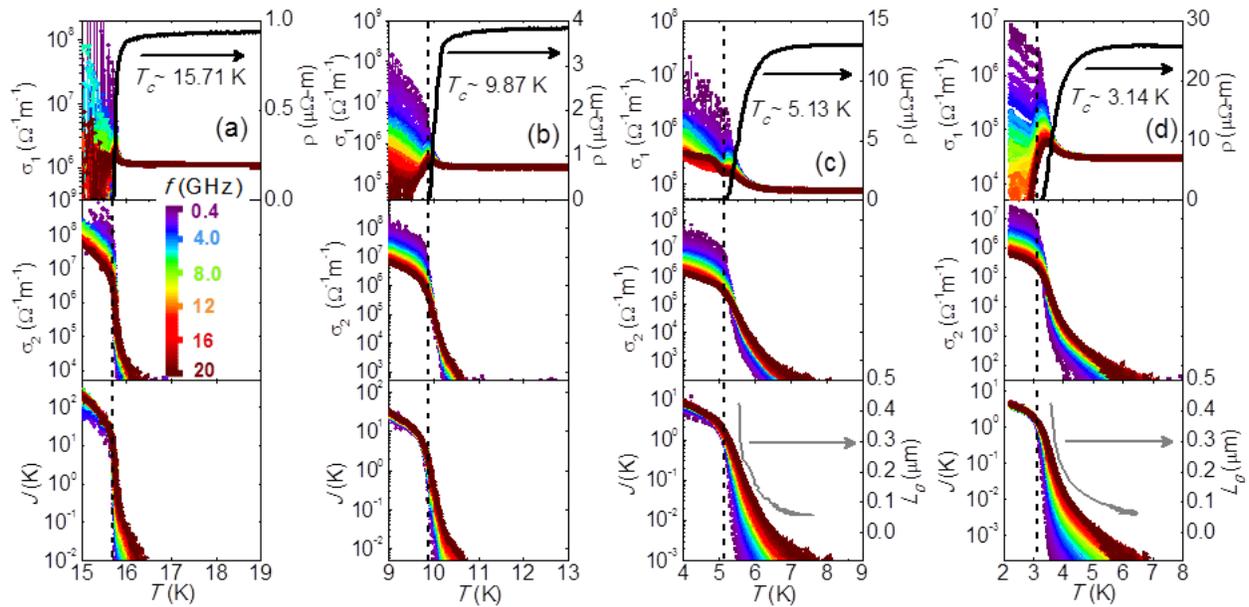

**Figure 9.** Temperature dependence $\sigma_1$ (upper panel), $\sigma_2$ (middle panel) and $J$ (lower panel) at different frequencies for four samples with (a) $T_c \sim 15.7$ K (b) $T_c \sim 9.87$ K (c) $T_c \sim 5.13$ K and (d) $T_c \sim 3.14$ K. The color scale representing different frequencies is displayed in panel (a). The solid (black) lines in the top panels show the temperature variation of resistivity. Vertical dashed lines correspond to $T_c$. The solid (gray) lines in the bottom panels of (c) and (d) show the variation of $L_0$ above $T_c$.

It has been shown from STS measurements that disordered NbN films [21,22] with $T_c \lesssim$ 6K show a pronounced PG state above $T_c$. To understand the relation between these observations and the PG state observed in STS measurements, we compare $T^*$ with the temperature, $T_m^*$, at which $J$ goes below our measurable limit at 20 GHz. In Fig. 10, we plot $T_m^*$ and $T_c$ for several samples obtained from microwave measurement along with the variation of $T^*$ and $T_c$ obtained from STS measurements, as a function of $k_F l$. Within the error limits of determining these temperatures, $T^* \approx T_m^*$, showing that the onset of the PG state in STS measurements and onset of the finite $J$ at 20 GHz take place at the same temperature. Furthermore, only the samples in the disorder range where a PG state appears, show a difference between $T_c$ and $T_m^*$. We therefore attribute the frequency dependence of $J$ to a fundamental property related to the PG state.

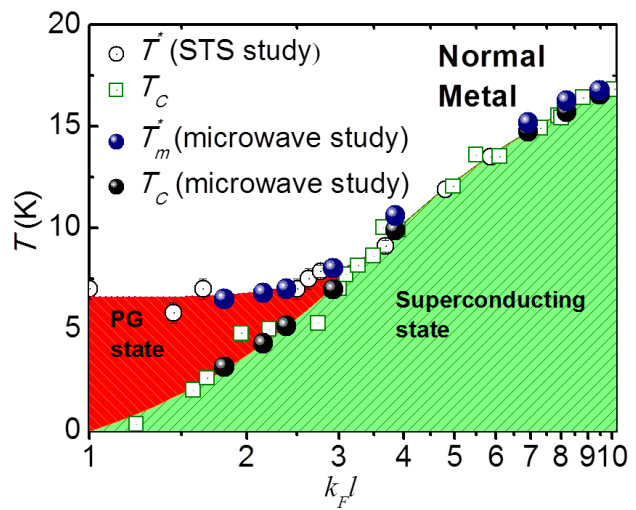

**Figure 10.** Phase diagram showing $T_c$ and $T^*$ obtained from STS measurements along with $T_c$ and $T_m^*$ obtained from microwave measurements.

Having established the relation between the PG state and the finite high frequency phase stiffness, we now concentrate on the fluctuation region above $T_c$. A superconductor above $T_c$ shows excess conductivity due to presence of unstable superconducting pairs induced by fluctuations. The first successful theoretical understanding of this excess conductivity in a dirty superconductor is provided by Aslamazov and Larkin (AL) [49]. They have attributed this excess conductivity (AL-term) to acceleration of superconducting pairs induced by fluctuations, as follows:

$$^{2D\ AL}\sigma_{fl}^{dc} = \frac{1}{16}\frac{e^2}{\hbar t}\varepsilon^{-1},$$ (11)

$$^{3D\ AL}\sigma_{fl}^{dc} = \frac{1}{32}\frac{e^2}{\hbar \xi_0}\varepsilon^{-1/2},$$ (12)

where $\varepsilon = \ln(T/T_c)$, $t$ is the thickness of the sample and $\xi_0$ is the BCS coherence length. In the excess conductivity, the AL contribution comes from direct acceleration of the superconducting pairs induced by fluctuations. These accelerated superconducting pairs have finite life time and in their way, they decay into quasiparticles of nearly opposite momentum. However due to the time reversal symmetry, they remain in the state of small total momentum and in spite of impurity scattering the resultant quasiparticles continue to be accelerated like their parent pairs. Quasiparticles also have a finite life time and ultimately they decay back into superconducting pairs. In the case of a dirty superconductor, contribution from quasiparticles acceleration is negligible but in clean superconductor it gives a finite second order correction to the fluctuation conductivity which is predicted by Maki-Thomson (MT)[50], as follows:

$$^{3D\ MT}\sigma_{fl}^{dc} = \frac{1}{8}\frac{e^2}{\hbar t}\frac{1}{\varepsilon-\delta}\ln\left(\frac{\varepsilon}{\delta}\right),$$ (13)

$$^{3D\ MT}\sigma_{fl}^{dc} = \frac{1}{8}\frac{e^2}{\hbar \xi_0}\varepsilon^{-1/2},$$ (14)

where $\delta$ is the Maki-Thompson pair breaking parameter. The AL and MT contributions are additive and they jointly explain the large magnitude of excess conductivity above $T_c$ in clean superconductor such as Al, In etc.

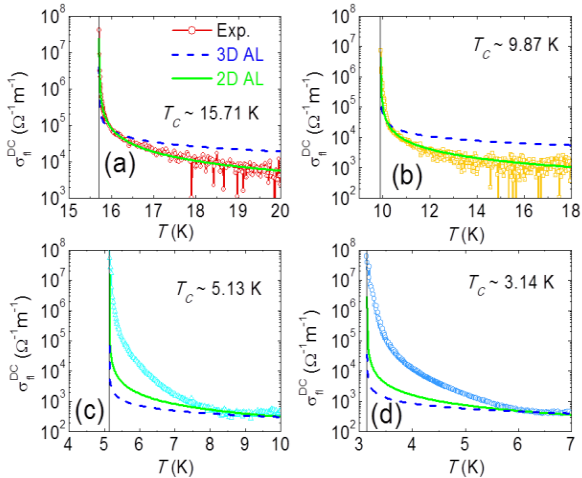

Figure 11. Temperature dependence of DC fluctuation conductivity {$\sigma_{fl}^{dc}=\sigma^{dc}(T)-\sigma_N$} for four samples with (a) $T_c \sim 15.7$ K (b) $T_c \sim 9.87$ K (c) $T_c \sim 5.13$ K and (d) $T_c \sim 3.14$ K. The scattered color plots are experimental data. The solid green lines are theoretical fit according to 2D AL prediction and dashed blue lines are for 3D Al prediction of fluctuation conductivity.

Fig. 11 (a)-(d) show the DC fluctuation conductivity ($\sigma_{fl}^{dc}$) for four samples on which we have done microwave measurements. Since all of our films are in the dirty limit, the MT contribution is negligible. In the case less disordered sample, the $\sigma_{fl}^{dc}$(T) follow the 2D AL prediction very well [Fig. 11 (a) and (b)] instead of the 3D AL prediction as the correlation length above $T_c$ becomes very large and effectively the sample behaves as 2D. However when we increase the disorder, the temperature dependence of $\sigma_{fl}^{dc}$(T) starts to deviate from AL predictions and in very strong disorder sample, the $\sigma_{fl}^{dc}$(T) decreases with temperature in much slower rate than the expected temperature variation from AL predictions. This anomalous behavior of $\sigma_{fl}^{dc}$(T) with respect to temperature, leads us to believe that in a strongly disordered superconductor, the amplitude fluctuations alone can't explain the fluctuation region above $T_c$.

For further understanding about the fluctuation region, we now concentrate on the frequency dependence of fluctuation conductivity. Using the time dependent Ginzburg-Landau equation, Schmidt [51] has calculated the frequency dependent AL term of the fluctuation conductivity in 2D and 3D limit as follows:

$$\sigma_{fl}^{2D\,AL}(\omega) = {}^{2D\,AL}\sigma_{fl}^{DC}\,S^{2D\,AL}\left(\frac{\omega}{\omega_0}\right) \;;\; \omega_0 = \frac{16k_B T_c}{\pi\hbar}\varepsilon$$

$$S^{2D\,AL}(x) = \left\{\frac{2}{x}\tan^{-1}x - \frac{1}{x^2}\ln(1+x^2)\right\} + i\left\{\frac{2}{x^2}(\tan^{-1}x - x) + \frac{1}{x}\ln(1+x^2)\right\} , \qquad (15)$$

$$\sigma_{fl}^{3D\,AL}(\omega) = {}^{3D\,AL}\sigma_{fl}^{DC}\,S^{3DAL}\left(\frac{\pi\hbar\omega}{16k_BT_c\varepsilon}\right)$$

$$S^{3D\,AL}(x) = \left\{\frac{8}{3x^2}\left(1-(1+x^2)^{3/4}\cos(\frac{3}{2}\tan^{-1}x)\right)\right\} + i\left\{\frac{8}{3x^2}\left(-\frac{3}{2}x+(1+x^2)^{3/4}\sin(\frac{3}{2}\tan^{-1}x)\right)\right\}, \quad (16)$$

For a clean superconductor the frequency dependence of the MT-term was calculated by Aslamazov and Varlamov [52]. They have shown that in 2D and 3D limits the frequency dependence of MT term is additive to the AL-term, as follows:

$$\sigma_{fl}^{2D\,AL+MT}(\omega) = {}^{2D\,AL+MT}\sigma_{fl}^{DC}\,S^{2D\,AL+MT}\left(\frac{\pi\hbar\omega}{16k_BT_c\varepsilon}\right);$$

$$S^{2D\,AL+MT}(x) = \left\{\operatorname{Re}S^{2D\,AL}(x)+\frac{2\pi x-2\ln(2x)}{1+4x^2}\right\} + i\left\{\operatorname{Im}S^{2D\,AL}(x)+\frac{\pi+4x\ln(2x)}{1+4x^2}\right\}, \quad (17)$$

$$\sigma_{fl}^{3D\,AL+MT}(\omega) = {}^{3D\,AL+MT}\sigma_{fl}^{DC}\,S^{3D\,AL+MT}\left(\frac{\pi\hbar\omega}{16k_BT_c\varepsilon}\right);$$

$$S^{3D\,AL+MT}(x) = \left\{\operatorname{Re}S^{3D\,AL}(x)+\frac{4-4x^{1/2}+8x^{3/2}}{1+4x^{1/2}}\right\} + i\left\{\operatorname{Im}S^{3D\,AL}(x)+\frac{4x^{1/2}-8x+8x^{3/2}}{1+4x^{1/2}}\right\}, \quad (18)$$

Fig. 12. (a)-(b) show the frequency dependence of scaling function, $S(x)$ expected from AL and AL+MT predictions in 2D and 3D limit. All predictions for fluctuations conductivity described above are for Gaussian fluctuations only. However after the discovery of high-$T_c$ superconductors, the fluctuation phenomena become much more important where not only Gaussian fluctuations, phase fluctuations also play an important role in dynamical properties of superconductor. In high-$T_c$ superconductor, the low superfluid density, shorter coherence length and quasi-two dimensionality enhance the fluctuation region above $T_c$ where the theory of Gaussian fluctuations only does not hold valid. To study the critical fluctuation region, Fisher, Fisher and Huse [53] have proposed a dynamical scaling theory where

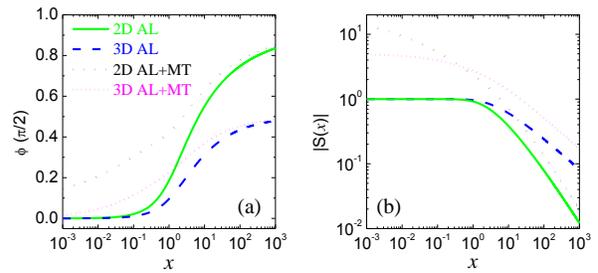

Figure 12. Phase, $\phi(x) = \tan^{-1}(S_{im}/S_{real})$ and amplitude, $|S(x)|$ of theoretical scaling functions for the AL term and the AL + MT term.

the fluctuation conductivity, $\sigma_{fl}(\omega)$ is predicted to scale as,

$$\sigma_{fl}(\omega)\big/\sigma_{fl}(0) = S(\omega/\omega_0), \qquad (19)$$

where $\omega_0$ is the characteristic fluctuation frequency, $\sigma_{fl}(0)$ is the zero frequency fluctuation conductivity at that temperature and $S$ is the universal scaling function. This scaling theory is the general one which contains Gaussian fluctuations and also any other means of fluctuations.

We experimentally obtain $\sigma_{fl}(\omega)$ from measured $\sigma(\omega)$ by subtracting the normal state dc conductivity, $\sigma_N$ at temperature above $T_m^*$. Since the phase angle $\phi(\omega) = \tan^{-1}(\sigma_{fl}^2/\sigma_{fl}^1)$ is the same as the phase angle of $S$, $\phi(\omega)$ expected to collapse into single curve by scaling $\omega$ differently at each temperature. For the amplitude the data would similarly scale when normalized by $\sigma_{fl}(0)$. Such a scaling works for all the samples as shown in Fig. 13(a) and (b) for the samples with $T_c \sim 15.7$ K and 3.14 K respectively. Fig. 13(c) and (d) show the variation of $\omega_0$ and $\sigma_{fl}(0)$ obtained for the best scaling of the data. In both plots $\omega_0 \rightarrow 0$ as $T_c$ is approached from above showing the critical slowing of fluctuations as the superconducting transition is approached. We observe a perfect consistency between the temperature variation of $\sigma_{fl}(0)$ and dc fluctuation conductivity, $\sigma_{fl}^{dc}$ obtained from $\rho$ - $T$ measured in the same run. Comparing the scaled phase and amplitude with various theoretical models of amplitude fluctuations, i.e. Ashlamazov-Larkin (AL) and Maki-Thompson (MT) in various dimensions, we observe that the sample with $T_c \sim 15.7$ K matches very well with the AL prediction in 2D in agreement with earlier measurements on low-disorder NbN films [54]. On the other hand, the corresponding curve for the sample with $T_c \sim 3.14$ K does not match with any of these models showing that amplitude fluctuations alone cannot explain the fluctuation conductivity in the region where samples are showing a PG state.

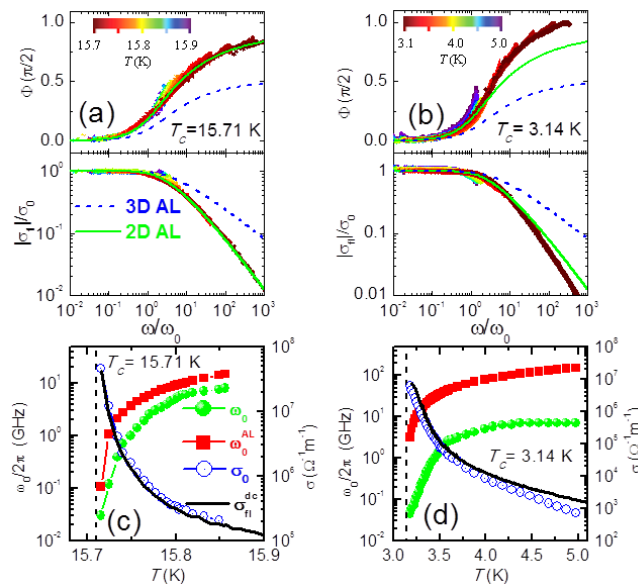

**Figure 13.** (a)-(b) Rescaled phase (upper panel) and amplitude (lower panel) of $\sigma_{fl}(\omega)$ using the dynamic scaling analysis on two films with $T_c \sim 15.7$ and 3.14 K respectively. The solid lines show the predictions from 2D and 3D AL theory. The color coded temperature scale for the scaled curves is shown in each panel. (c)-(d) Show the variation of $\omega_0$, $\omega_0^{AL}$ $\sigma_0$ and $\sigma_{dc}$ as function of temperature. The dashed vertical lines correspond to $T_c$.

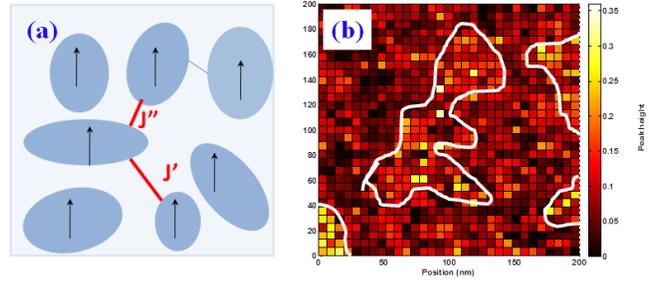

Figure 14. (a) Cartoon diagram of our system similar to Josephson junctions network but with large distribution in puddles size and coupling strength, $J$ (b) The spatial variation of coherence peak height measured using STM as the measure of local order parameters at 500 mK, is shown for sample with $T_c \sim 2.9$ K over a 200 nm × 200 nm area.

We can now put these observations in perspective. It has been shown from STS measurements that in the presence of strong disorder the spatial landscape of the superconductor become highly inhomogeneous [21,22] thereby forming domain like structures, tens of nm in size, where the superconducting OP is finite and regions where the OP is completely suppressed see [Fig. 14. (b)]. Therefore, one can visualize the superconducting state in strongly disorder films, as a network of Josephson junctions with a large distribution in coupling strength [see Fig. 14.(a)], where the superconducting transition is determined by phase disordering. In this scenario, $T_c$ corresponds to the temperature at which the weakest couplings are broken. Therefore, just above $T_c$ the sample consist of large phase coherent domains (consisting of several smaller domains) fluctuating with respect to each other. As the temperature is increased further, the large domains will progressively fragment giving rise to smaller domains till they completely disappear at $T = T^*$. In such a scenario $J$ will depend on the length scale at which it is probed. When probed on a length scale much larger than the phase coherent domains, $J \rightarrow 0$. On the other hand, when probed at length scale of the order of the domain size $J$ would be finite, however $J(\omega)$ would vanish at a temperature where the phase coherent domain becomes much smaller than $L(\omega)$. Using this criterion we plot the upper bound of the domain size ($L_0$) as a function of temperature in the lower panel of Fig 9 (c) and 9(d). Taking $d=3$ (since $l \ll t$), the limiting value of $L_0$ at $T^*$ is between 50-60 nm which is in agreement with the domains observed in STS measurements [see Fig. 14 (b)] on NbN films with similar $T_c$.

In summary, we have shown that in strongly disordered NbN thin films which display a PG state, $J$ becomes dependent on the temporal and spatial length scale in the temperature range, $T_c < T < T^*$. The remarkable agreement between $T^*$ determined from STS and $T^*_m$ microwave measurements is consistent with the notion that the superconducting transition in these systems is driven by phase disordering. In this context, we would like to note that the conventional phase disordering transition in 2D systems, namely the BKT transition cannot explain the large temperature range over which the high frequency $J$ is finite. In 2D NbN thin films [24], we have shown that the BKT fluctuation regime is restricted to a narrow temperature range above $T_c$.

## 5. Summary

We have shown that the superconductivity can be destroyed by phase fluctuations induced by reduced dimensionality or disorder although $|\Delta|$ remains finite well above $T_c$ contrary to BCS prediction. In 2D, we have shown that this phase disordering transition belonging to BKT universality class when the low vortex core energy of the superconductor is taken into account. In 3D disordered superconductor, it gives rise to a pseudogap state with frequency dependent superfluid stiffness above $T_c$ but no global superconductivity. Finally, I would like to note that many of these observations are similar to under doped high-temperature superconducting cuprates where the mechanism of superconductivity is still hotly debated. It would therefore be interesting to explore, through similar measurements, whether the superconducting transition is driven by phase disordering even in those materials.

---

# CHAPTER 1

## Introduction

The discovery of superconductivity by H. Kamerlingh Onnes [1] in 1911 opened a great era of excitement and challenges in condensed matter physics. Superconductivity is a quantum phenomena which manifests in macroscopic scale and characterized by zero electrical resistance observed in certain materials when cooled below a critical temperature, $T_c$ known as superconducting transition temperature (usually extremely low). This above single sentence explains the excitement and challenges for both experimentalists and theoreticians for the last 100 years following the discovery of superconductivity. It remains till date an area of active research.

The superconducting (SC) state is characterized by two main properties: (i) Pairs of electrons which form bound states i.e. Cooper pairs and (ii) the condensation of Cooper pairs into phase coherent macroscopic quantum state. The former manifests as the gap in electronic density of states (DOS), known as the superconducting energy gap, $\Delta$, and the latter gives rise to finite superfluid phase stiffness, $J$, which is the energy cost of twisting the phase of condensate. In conventional superconductors, the superconductivity is well described by Bardeen- Cooper-Schrieffer (BCS) theory [2] where the superconducting transition is governed solely by $\Delta$ and phase fluctuations are unimportant except very close to $T_c$. Consequently, it has been justified to expect the superconductor to normal transition at a temperature where $\Delta$ goes to zero. However after discovery of copper oxide based high temperature superconductors (HTS), it was observed that a soft gap called as pseudogap (PG), opens up in the electronic density of states (DOS) at a temperature well above $T_c$ [3,4,5,6]. The origin of this gap in electronic excitation spectra without zero resistance is a major outstanding issue in superconductivity. Having complicated quasi 2D layered structure and magnetic ordering coexisting with superconductivity, understanding the nature of superconductivity in HTS becomes very challenging. Since the PG state is viewed as the key to understand the nature of superconductivity in HTS [7], origin and physical interpretation of this PG state have become a major focus in superconductivity. For long time, this novel PG state was thought to be exclusive hallmark of HTS. However recent scanning tunneling spectroscopy (STS) measurements revealed the appearance of similar type PG state in more conventional strongly disordered s-wave superconductors such as TiN, InOx and NbN [8,9,10,11].

All these issues have renewed wave of interest in 2D or quasi-2D superconductors and in very strongly disordered 3D superconductors. A possible way of understanding the pseudogap state in these materials is the phase fluctuation scenario where superconductivity is destroyed by strong phase fluctuations although superconductors retain some of the Cooper pairs which manifest as the PG in electronic DOS. The electrodynamics response of superconductors provides an ideal tool to explore the role of phase fluctuations in superconductivity. In this thesis, I will present an investigation on the role of phase fluctuations, through measurements of magnetic penetration depth, $\lambda$, using low frequency mutual inductance technique and the complex conductivity using broadband microwave Corbino spectrometer, in thin films of conventional superconductor NbN both in 2D and 3D limit [10,11,12,13]. Our study elucidates interplay of quasi particle excitations (QE) and phase fluctuations in low dimensional and strongly disordered superconductors.

In this chapter, I will introduce the phenomenology of superconductivity and basic theoretical understanding. Then I will give brief overview of electrodynamics of superconductors and effect of SC fluctuations on various SC properties.

## 1.1.    Fundamentals of superconductivity

### 1.1.1. Zero resistance

The most important property of a superconductor is the zero electrical resistance in the SC state [see Fig. 1.1] first discovered by H. Kamerlingh Onnes in mercury [1]. Subsequently it was observed that many materials including metals, alloys and compounds undergo SC transition at characteristic transition temperature, $T_c$.

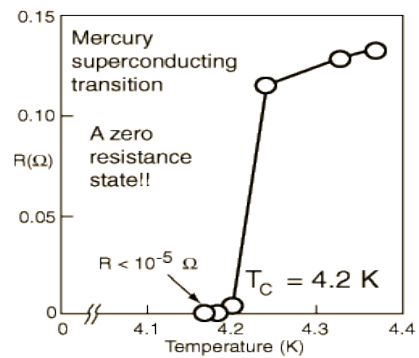

Figure 1.1. *R-T* curve of mercury showing SC transition at 4.2 K (after H. Kamerlingh Onnes (1911)).

## 1.1.2. Meissner effect

Another very fundamental property associated with the superconductivity is the perfect diamagnetism in Meissner state discovered by W. Meissner and R. Ochsenfeld [14] in 1933(see Fig. 1.2.). When a normal metal undergo SC transition in presence of weak magnetic field, it expels the magnetic field from inside the superconductor by introducing surface currents. Meissner effect is the unique property of superconductor, which cannot be explained by considering only a perfect conductor with zero resistivity.

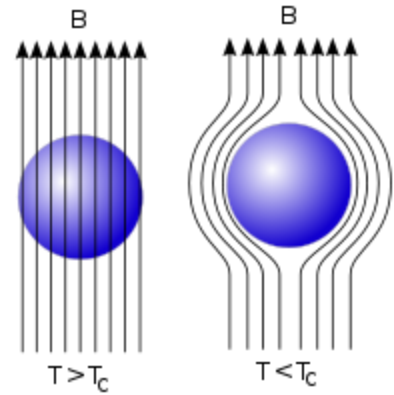

Figure 1.2. Showing the exclusion of magnetic field in the SC state under weak magnetic field (figure is adapted from Wikipedia).

### 1.1.2.1. Critical field

Although superconductors show perfect diamagnetism in weak magnetic field, strong enough magnetic field can destroy superconductivity. In most elemental superconductors the perfect diamagnetic state is destroyed abruptly at a magnetic field known as the critical field, $B_c$ and above $B_c$ these superconductors behave like normal metals. The applied magnetic field increases the free energy of the superconductor as,

$$\varepsilon_{cond} = f_s(B_c) - f_s(0) = \frac{B_c^2}{8\pi} \tag{1.1}$$

Since the free energy of the normal state is nearly independent of magnetic field, it follows that the free energy density of the superconductor is lower by an amount $B_c^2/8\pi$, which is known as the condensation energy, $\varepsilon_{cond}$.

## 1.1.3. London equations

To describe the observed zero resistance property and Meissner effect two brothers Fritz and Heinz London in 1935 [15,16] proposed a pair of equations,

$$\frac{\partial \vec{j}_s}{\partial t} = \frac{n_s e^2}{m} \vec{E} \quad \text{and} \quad \nabla \times \vec{j}_s = -\frac{n_s e^2}{m} \vec{B} \tag{1.2}$$

The first equation describes the acceleration of electrons rather than sustaining their velocity against resistance in presence of electric field, therefore represents the perfect conductivity of superconductors [16]. The second equation can be written by combining with Maxwell equation, $\nabla \times \vec{B} = \mu_0 \vec{J}$ as,

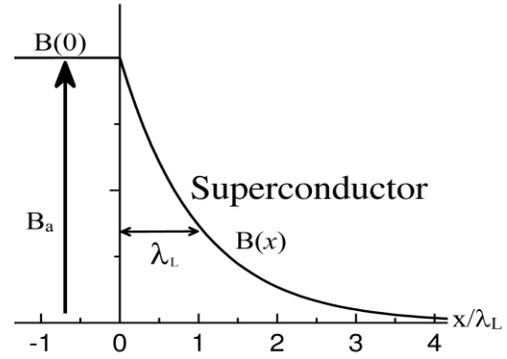

Figure 1.3. Variation of magnetic field at the boundary of a superconductor showing the exponential decay of magnetic field with characteristic length, $\lambda_L$ inside the superconductor.

$$\nabla^2 \vec{B} = \frac{\vec{B}}{\lambda_L^2} \qquad (1.3(a))$$

$$\lambda_L = \left( m / \mu_0 n_s e^2 \right)^{1/2} \qquad (1.3(b))$$

The above equation implies that there is a characteristic length, $\lambda_L$ known as London magnetic penetration depth, over which magnetic field decays inside a superconductor, thus represents the Meissner effect. The Meissner effect can be understood clearly assuming a superconductor placed in a applied magnetic field, $\boldsymbol{B_a} = B(0) \, \boldsymbol{\hat{z}}$ parallel to the surface of the superconductor. Then the eqn. 1.3 can be written as [17],

$$\frac{\partial^2 B(x)}{\partial x^2} = \frac{B(x)}{\lambda_L^2} \qquad (1.4)$$

Here $B(x)$ is the magnetic induction at a distance $x$ inside the superconductor from the surface. Solution of the above equation gives,

$$B(x) = B(0) \, e^{-x/\lambda_L} \qquad (1.5)$$

This implies that inside a superconductor the magnetic field exponentially decays (see Fig. 1.3.) with characterising length, $\lambda_L$. For uniform magnetic field in $z$ direction, the Maxwell equation, $\nabla \times \vec{B} = \mu_0 \vec{J}$ is reduced to $-\partial B / \partial x = \mu_0 J_y$. Thus using eqn. (1.5) we get the current density as a function of $x$,

$$J_y = \left( \frac{B(0)}{\mu_0 \lambda} \right) e^{-x/\lambda_L} \qquad (1.6)$$

Therefore the current flows at the surface of superconductor and exponentially decay inside the superconductor. Although the London equations are not derived from fundamental properties, they successfully explain the zero resistance property and Meissner effect in a superconductor.

### 1.1.4. Pippard's Coherence length

Coherence length, $\xi_0$, is the shortest length scale over which SC electron density can change. The concept of coherence length, $\xi_0$, was first introduced by A. B. Pippard when he proposed the non local generalization of London equations [18]. He argued that the response of super electrons in presence of field at point $\vec{r}$, depends on the surrounding SC wave functions within a volume of radius $\xi_0$ about that point. Using uncertainty principle he estimated $\xi_0$ as,

$$\xi_0 = a_0 \frac{\hbar v_F}{k_B T_c} \tag{1.7}$$

Here $a_0$ is a numerical constant of the order of unity. In presence of disorder, when the electronic mean free path, $l$, is smaller than $\xi_0$, the effective coherence length, $\xi_{eff}$ is modified according to relation [16],

$$\frac{1}{\xi_{eff}} = \frac{1}{\xi_0} + \frac{1}{l} \tag{1.8}$$

where $\xi_0$ is the coherence length of pure metal.

#### 1.1.4.1    Type I and Type II superconductor

The magnetic response of a superconductor depends on whether $\lambda < \xi$ or $\lambda > \xi$. Most of the elemental superconductors (except Nb) show perfect diamagnetism (Meissner effect) up to a critical magnetic field, $B_c$ and above which it becomes normal. These are the superconductors whose magnetic penetration depth, $\lambda < \xi$ and categorized as type I superconductors. However most of the superconducting alloys and compounds have $\lambda > \xi$ and are in type II category. In a Type II superconductor [see Fig. 1.4.(b)] , the response in magnetic field is much more complex. It shows perfect diamagnetism up to a very small characteristic field $B_{c1}$ known as lower critical field. When applied magnetic field exceeds $B_{c1}$, magnetic flux starts to penetrate inside the superconductor and give rise to vortex state where SC regions coexist with normal regions and the magnetic flux lines pass through the normal region. Further increase in magnetic field beyond a characteristic critical magnetic field $B_{c2}$, destroys the superconductivity completely and gives rise to normal state. The second characteristic critical magnetic field, $B_{c2}$ is known as upper critical.

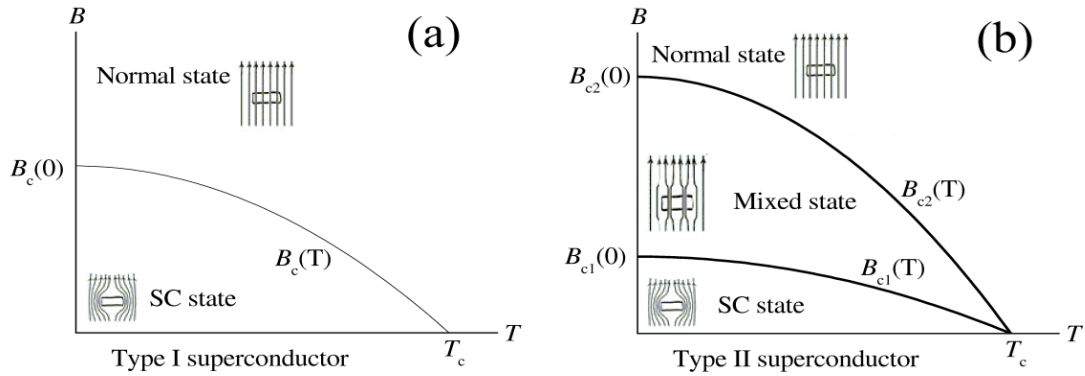

## 1.1.5. Isotope effect

In 1950, two groups independently showed that the $T_c$ of Hg depend on its isotopic mass [16,19] and follow the relation,

$$M^{\alpha}T_c=\text{Constant} \qquad (1.9)$$

where $\alpha = \frac{1}{2}$ for Hg. Subsequently, similar behavior has been observed in large number of superconductors. The isotope effect gives the indication about the role played by electron-phonon interaction, in occurrence of superconductivity in conventional superconductors.

## 1.1.6. Energy gap of single particle excitation at the Fermi level

At low temperature, electronic specific heat shows the exponential temperature dependence [20] which provides the evidence of energy gap at the DOS at the Fermi level. Existence of energy gap at the Fermi level also confirmed by electromagnetic absorption study in the frequency range, $\hbar\omega \sim k_B T_c$ carried out by Biondi et al [21]. The combined discovery of isotope effect and SC energy gap, leads to the discovery of famous BCS theory of superconductivity described in the following section.

## 1.2. BCS theory

To describe the novel phase of materials, superconductivity, Bardeen, Cooper and Schrieffer (BCS) proposed their most celebrated theory in condensed matter physics in 1957 [2,23]. According to their theory, even a weak net attractive interaction between two electrons caused by second order interaction mediated by a phonon gives rise to bound pair known as Cooper pair. Thus electrons in vicinity of the Fermi energy form bound Cooper pairs and subsequently condense into a phase coherent macroscopic quantum state and give rise to superconductivity. In this section, I will give brief overview of this elegant theory and its novelty to describe different properties of superconductors.

### 1.2.1. Cooper pairs

In 1950 Frohlich [22] showed that there can be a resultant attractive interaction between two electrons mediated by phonon in contrary to common belief of the repulsive interaction. The attractive interaction can be understood physically in the following way: when an electron passes through the lattice, it creates a positive charge imbalance by attracting positive ions towards it and this excess positive charge attracts another electron. If the attraction force is higher than the Coulomb repulsive force, it gives a net attractive interaction. In 1956 Cooper [23] showed that in presence of net attractive interactions, no matter how small it, two electrons with opposite momenta and spin can form a bound pair known as Cooper pair. Using uncertainty principle it can be shown that the size of a Cooper pair is of the order of BCS coherence length, $\xi_{BCS} = \hbar v_F / \pi \Delta$, much larger than the inter particles distance. Thus pairs are highly overlapping with each other and create a collective state.

### 1.2.2. BCS ground state at $T$=0

It was shown by L. N. Cooper in the same work that the Fermi sea is unstable against the formation of bound pairs induced by net attractive interaction [23]. In this situation the Cooper pairs are expected to condense until an equilibrium point is reached when the binding energy for additional pair goes to zero and gives rise to a macroscopic quantum state.

.      To describe this state, in 1957 Bardeen, Cooper and Schrieffer proposed a ground state wave function of the form [2,16],

$$\left|\psi_{BCS}\right\rangle = \prod_{\vec{k}}\left(\left|u_k\right| + \left|v_k\right|e^{i\varphi}c_{\vec{k}\uparrow}^{*}c_{-\vec{k}\downarrow}^{*}\right)\left|\Phi_0\right\rangle \tag{1.10}$$

where $c_{\vec{k}\uparrow}^{*}$ ($c_{-\vec{k}\downarrow}^{*}$) is creation operator which creates one electron (hole) of momentum $\vec{k}$ ($-\vec{k}$) and spin up (down). The coefficients $u_k$ and $v_k$ can be determined by minimizing the ground state energy $E = \left\langle \psi_{BCS}\left|H\right|\psi_{BCS}\right\rangle$ where $H$ is the BCS pairing Hamiltonian given by,

$$H = \sum_{\vec{k}\sigma}n_{\sigma\vec{k}}\varepsilon_{\vec{k}} + \sum_{\vec{k}\ \vec{l}}V_{\vec{k}\ \vec{l}}\ c_{\vec{k}\uparrow}^{*}c_{-\vec{k}\downarrow}^{*}c_{\vec{l}\uparrow}c_{-\vec{l}\downarrow} \tag{1.11}$$

The first term corresponds to the kinetic energy of noninteracting electron gas and second term is the pairing interaction via second order phonon scattering. Here, $V_{\vec{k}\ \vec{l}}$ is the electron-phonon scattering matrix element which is approximately constant and can be replaced by –V where V is a positive quantity.  Now applying variational method [16], the ground state energy can be determined as follows,

$$-\varepsilon_{cond} = f_s(0) - f_n(0) = -\frac{1}{2}N(0)\Delta^2(0) \tag{1.12}$$

where $f$ is the free energy per unit volume , $N(0)$ is the normal state electronic DOS at the Fermi level and $\Delta(0)$ is the SC energy gap equivalent to the pairing energy  given by,

$$\Delta(0) = 2\hbar\omega_D e^{-1/N(0)V} \tag{1.13}$$

where $\omega_D$ is the Debye frequency. Here $\varepsilon_{cond}$ is the condensation energy density of the superconductor at zero temperature, which is equivalent to the obtained condensation energy density, $\varepsilon_{cond} = B_{c2}^2(0)/8\pi$ in eqn. (1.1), from critical field.

### 1.2.3. Elementary excitations and loss of superconductivity at finite *T*

The elementary excitations in superconductors are electron and hole like quasi particles originating from breaking of the Cooper pairs. Although the variational method is workable in dealing with excited state, the most elegant way to approach the excited state problem is Bogoliubov-Valatin [24,25] self consistent approach [see ref. 16 and 26]. Using this method energy of these quasi particles can be determined as,

$$E_{\vec{k}} = (\xi_{\vec{k}}^2 + \Delta^2)^{1/2} \tag{1.14}$$

Here $\xi_{\vec{k}} = \varepsilon_{\vec{k}} - E_F$ is the single particle energy with respect to Fermi energy, $E_F$. Therefore even at the Fermi surface the minimum energy required for a quasi particle excitation is $\Delta$. Since quasi particles are Fermionic excitations and behave like electrons, there is a one to one correspondence between quasi particles in SC state and the electrons in normal state. Thus we can obtain the DOS of superconductors, $N_s(E)$ by equating, $N_s(E) \, dE = N_n(\xi) \, d\xi$. Since we are interested in the range of energy, $\xi$ within few meV about the Fermi energy, we can take $N_n(\xi) \approx N(0)$ and from eqn. 1.14 we obtain,

$$\frac{N_s(E)}{N(0)} = \frac{d\xi}{dE} = \begin{cases} \dfrac{E}{(E^2 - \Delta^2)^{1/2}} & (E > \Delta) \\ 0 & (E < \Delta) \end{cases} \tag{1.15}$$

The Fig. 1.5.(a) shows the theoretical BCS density of states compared to normal state. All states whose energy fall in the gap are shifted beyond $\pm\Delta$ about the Fermi energy as a result of long range coherence and away from the Fermi energy, the DOS decreases rapidly and asymptotically goes to normal state value. For temperature dependence of SC energy gap, the solution of Bogoliubov-Valatin self consistent equation gives the following functional form [16],

$$\frac{1}{N(0)V} = \int_0^{\hbar\omega_D} \frac{\tanh\frac{1}{2}\beta(\xi^2 + \Delta^2)^{1/2}}{(\xi^2 + \Delta^2)^{1/2}} d\xi \tag{1.16}$$

At finite temperature, the $\Delta$ can be calculated by solving the above equation numerically. The Fig. 1.5. (c) shows the BCS temperature variation (solid red line) of $\Delta$ obtained numerically from the above equation. In BCS theory the SC transition temperature ($T_c$) is the temperature where $\Delta(T)$ i.e. binding energy of Cooper pairs goes to zero. In the limit $\Delta(T) \rightarrow 0$, therefore the above integral gives,

$$k_B T_c = 1.13\, \hbar \omega_D e^{-1/N(0)V} \qquad (1.17)$$

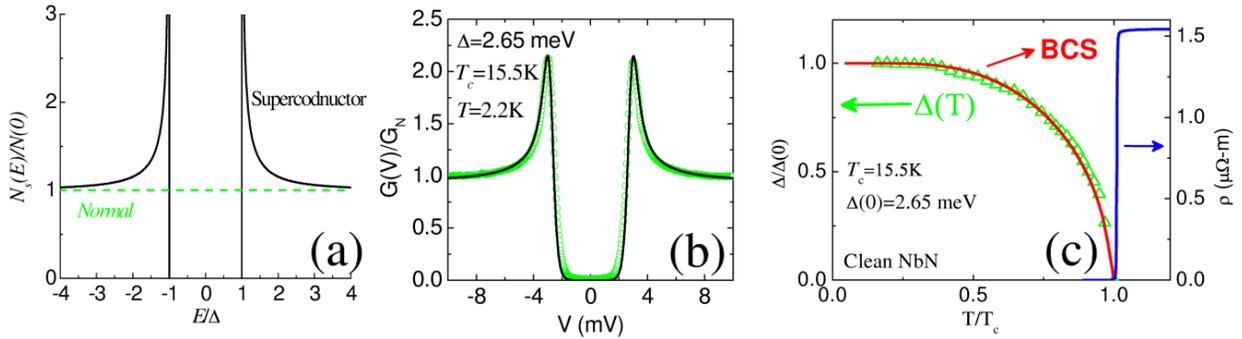

Therefore eqn. 1.13 and eqn. 1.17 lead to the famous BCS prediction,

$$\frac{\Delta(0)}{k_B T_c} = 1.764 \qquad (1.18)$$

Thus the ratio of $\Delta(0)$ to $T_c$ is a dimension less numerical constant which is the most significant prediction of BCS theory verified by many experiments on different superconductors.

### 1.2.4. Giaever tunneling and measurement of energy gap

According to BCS theory, an energy gap, Δ equivalent to pairing energy, opens up in electronic DOS at the transition from normal to superconductor. In 1960 Ivar Giaever directly measured the Δ of Al, Pb and some other superconductors [27] using tunnelling experiments in Superconductor-Insulator-Normal (SIN) tunnel junctions [for details see ref. 16]. The differential conductance of such type of SIN tunnel junctions, as function of bias voltage $V$ can be described using Fermi's golden rule as [16],

$$G_{ns} = \frac{dI_{ns}}{dV} = G_{nn} \int_{-\infty}^{\infty} \frac{N_s(E)}{N_n(0)} \left[ -\frac{\partial f(E+eV)}{\partial(eV)} \right] dE \qquad (1.19)$$

Here $G_{nn}$ is the differential conductance of the junction when superconductor becomes normal above $T_c$. $f(E+eV)$ is the Fermi-Dirac distribution function. $N_s$ and $N_n$ are the electronic DOS of SC and normal state respectively. Now at $T = 0$, $\partial f(E+eV)/\partial(eV)$ is a δ function at $E = eV$. Thus the differential conductance at $T = 0$,

$$G_{ns}\big|_{T=0} = G_{nn} \frac{N_s(e|V|)}{N_n(0)} = G_{nn} \frac{e|V|}{(e|V|^2 - \Delta^2)^{1/2}} \qquad (1.20)$$

Thus at low temperature, the differential conductance directly measure the BCS DOS. A representative of typical tunnelling conductance spectra of a SIN tunnel junction is shown in Fig. 1.5.(b) where NbN thin films with $T_c$~15.5 K was used as superconductor, oxide layer of Nb as insulator and Ag as normal metal [for details see ref. 28]. The temperature variation of extracted SC energy gap, Δ by fitting the conductance spectra using expression of BCS DOS is shown in Fig. 1.5.(c). Temperature dependence of Δ perfectly follows the BCS prediction and the superconductivity is destroyed at $T_c$, where Δ goes to zero, consistent with the BCS theory.

## 1.2.5. Pseudogap state

In BCS theory, the pairing energy of Cooper pairs, manifests as the energy gap in electronic DOS at the Fermi level, known as superconducting energy gap, Δ. Thus within BCS theory Δ is the signature of Cooper pairs which give rise to superconductivity. Superconductivity is destroyed when Δ goes to zero. However after the discovery of copper oxide based high temperature superconductors (HTS) by J. G. Bednorz and K. A. Müller in 1986, the above scenario has been challenged when a soft gap called as pseudogap (PG) was observed in electronic energy spectrum at temperature $T^*$ well above the transition temperature, $T_c$, where resistance goes to zero [29]. The Fig. 1.6.(a) shows the observed PG in tunnelling DOS and the

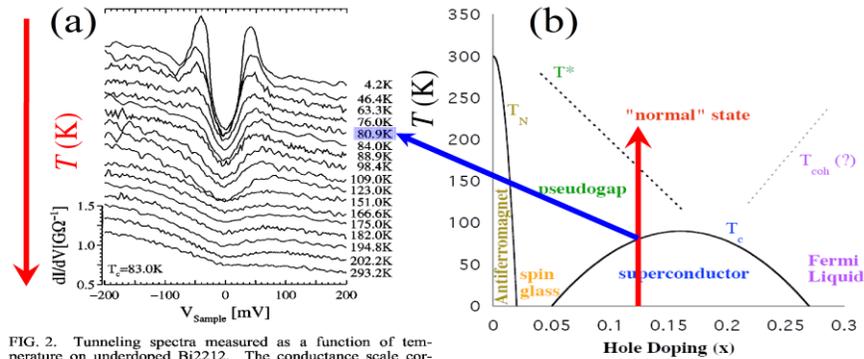

FIG. 2. Tunneling spectra measured as a function of temperature on underdoped Bi2212. The conductance scale corresponds to the 293 K spectrum, the other spectra are offset vertically for clarity.

Figure 1.6. (a) shows the PG in tunnelling DOS above $T_c$ in underdoped Bi2212. (b) The phase diagram showing the PG phase. (figure is adapted from ref. 6).

Fig. 1.6.(b) shows the schematic phase diagram of HTS with PG phase extending up to room temperatures. This PG phase was observed in various experiments such as tunnelling spectroscopy, specific heat measurements, ARPES (Angle Resolved Photoemission Spectroscopy) etc. [29] The existence of energy gap in conventional superconductors helps us to understand the mechanism of superconductivity through BCS theory, on the other hand the gap in electronic DOS above $T_c$ in HTS creates more confusion [30,31,32,33]. To explain the origin of PG in HTS, two main interpretations have been proposed by various groups:

(a) **The scenario of preformed pairs:** Superconductivity is destroyed due to phase fluctuations at temperature $T=T_c$ but Cooper pairs survive up to temperature $T^*$ called PG temperature well above $T_c$, which manifest as the energy gap in the electronic DOS although superconductivity has been destroyed [34,35].

(b) **Non-superconducting origin:** Formation of electronic stripes, anti-ferromagnetic ordering, exotic order parameter competing with superconductivity, give rise to PG state which has nothing to do with superconductivity [30,36].

The so called PG in DOS above $T_c$ was thought to be exclusive hallmark of HTS. However recently similar type of PG also observed in more conventional disordered superconductors such as TiN, InO$_x$ and NbN [8,9,10,11] contrary to BCS predictions. These observations are central to this thesis as will be discussed in later chapters.

## 1.3. Ginzburg-Landau theory

The microscopic BCS theory provides excellent explanations of various properties of superconductors such as nuclear relaxation, energy gap, elementary excitations etc where energy gap $\Delta$ is constant over space. However in inhomogeneous superconductors where $\Delta$ changes spatially and fluctuations are involved, BCS theory becomes very complicated. In such a situation, another exciting theoretical description proposed by Ginzburg and Landau (GL) in 1950 [16,37] much before the development of BCS theory, provides an elegant description of superconductivity close to $T_c$. GL theory concentrates entirely on super electrons rather than excitations and is generalized to deal with spatially varying and time dependent order parameter. In their phenomenological theory, Ginzburg and Landau introduced a complex order parameter, $\psi = |\psi| e^{i\theta}$ associated with SC state. The amplitude of $\psi$ represents the local SC electron density,

$$n_s = \left| \psi(r) \right|^2 \tag{1.21}$$

In presence of magnetic field, the free energy of SC ground state can be written as a series expansion of $\psi$ and it's gradient as follows [16]:

$$f = f_0 + \alpha |\psi|^2 + \frac{\beta}{2} |\psi|^4 + \frac{1}{2m^*} \left| \left( \frac{\hbar}{i} \vec{\nabla} - \frac{e^*}{c} \vec{A} \right) \psi \right|^2 + \frac{B^2}{8\pi} \tag{1.22}$$

where $\alpha$ and $\beta$ are the temperature dependent expansion coefficients. $\beta$ is positive throughout the transition but $\alpha$ changes sign to keep the free energy of the system minimum,

$$\alpha = \alpha_0 (T / T_c - 1) \qquad (\alpha_0 > 0) \tag{1.23}$$

The minimization of free energy, $f$ with respect to $\psi$, leads to the celebrated GL equation,

$$\alpha |\psi| + \beta |\psi|^2 \psi + \frac{1}{2m^*} \left( \frac{\hbar}{i} \vec{\nabla} - \frac{e^*}{c} \vec{A} \right)^2 \psi = 0 \tag{1.24}$$

In absence of magnetic field the above equation reduces to,

$$\alpha |\psi| + \beta |\psi|^2 \psi - \frac{\hbar^2}{2m^*} \nabla^2 \psi = 0 \tag{1.25}$$

Thus, it introduces a characteristic length scale for the spatial variations of $\psi$ (different from Pippard's coherence length) given by,

$$\xi_{GL} = \frac{\hbar^2}{2m^*|\alpha(T)|} \tag{1.26}$$

Physically $\xi_{GL}$ is the characteristic length scale over which the order parameter can vary without any cost of free energy. In the mixed state in Type II superconductors, $\psi$ is zero at the center of vortex core but gradually increases to the bulk value outside the vortex. Thus, $\xi_{GL}$ represents the size of vortex core. The upper critical filed ($H_{c2}$) is the field at which the vortex density increases to a critical value such that vortices start to overlap with each other and destroy the superconductivity. Using the above simple argument, $\xi_{GL}$ can be correlated to $H_{c2}$ through the following relation [16],

$$\xi_{GL}(T) = \left| \frac{\Phi_0}{2\pi H_{c2}(T)} \right|^{1/2} \tag{1.27}$$

It can be shown that near $T_c$, $\xi_{GL}$ behave differently in pure and dirty limit as [16],

$$\xi_{GL}(T) = 0.74 \frac{\xi_0}{(1-t)^{1/2}} \qquad \text{(pure, } l{>}\xi_0) \tag{1.28}$$

$$\xi_{GL}(T) = 0.855 \frac{(\xi_0 l)^{1/2}}{(1-t)^{1/2}} \qquad \text{(dirty, } l{<}\xi_0) \tag{1.29}$$

In clean superconductors well below $T_c$, $\xi_{GL}$ is of the order of Pippard's coherence length, $\xi_0$. Although GL theory was a phonological theory, it represents the nature of the macroscopic quantum mechanical properties of SC state in very simple way. Later in 1959, L. P. Gor'kov [38] showed that GL theory is the limiting form of microscopic BCS theory where $\psi(r)$ is proportional to $\Delta(r)$, valid at temperature very close to $T_c$, which established GL theory as a universally accepted profound theory of superconductivity.

The $\xi_{GL}$ can be obtained from experimentally measured upper critical field, $H_{c2}$ through eqn. (1.27). The zero temperature upper critical field can be determined from measured $H_{c2}(T)$ at temperature close to $T_c$ using Werthamer-Helfland-Honenberg (WHH) [39] relation in dirty limit,

$$H_{c2} = 0.693 \, T_c \left. \frac{dH_{c2}}{dT} \right|_{T=T_c} \qquad \text{(dirty, } l{<}\xi_0) \tag{1.30}$$

## 1.4.    Electrodynamics of superconductors

London equations provide fairly good description of Meissner effect but it is a phenomenological theory which treats electrons as classical object. Superconductivity is a quantum mechanical phenomenon and exhibit long range order which needs non local treatments of electrodynamics response of electrons which is absent in London equations. Apart from this, superconductors have other complicacy such as at finite $T$, the superconductors always can have excited quasi particles which behave like normal electrons and give dissipation in time ac electric filed. When the frequency of ac field becomes comparable to SC energy gap, the cooper pair can absorb microwave photon and give pair breaking effect, which creates a pair of electron like and hole like quasi particles. In this section following the ref. 16, I will first introduce two-fluid model which is a very simple way to understand the response of superconductors at low frequency and then I will discuss about the electrodynamics response of superconductors within BCS theory.

### 1.4.1. Two fluid model and complex conductivity

In this simplified model, the total carrier density, $n$, is assumed to be the sum of super electrons, $n_s$, responsible for superconductivity and normal electrons, $n_n$, given by [see the ref. 16 and 17],

$$n = n_s + n_n \tag{1.31}$$

The super and normal electrons have different relaxation times, $\tau_s$ and $\tau_n$ respectively. Here, $n_s$ is modeled as super electrons assuming $\tau_s \rightarrow \infty$. This model is valid only at much lower frequency than the energy gap and for $\omega\tau_n \ll 1$. The response of superconductors in time varying filed can be describe through complex conductivity,

$$\sigma(\omega) = \sigma_1(\omega) - i\sigma_2(\omega) \tag{1.32(a)}$$

$$\sigma_1 = (\pi n_s e^2 / 2m)\delta(\omega) + n_n e^2 \tau_n / m \tag{1.32(b)}$$

$$\sigma_2 = n_s e^2 / m\omega \tag{1.32(c)}$$

Although it is an oversimplified model, it is very useful to describe qualitatively the dissipation in superconductors at microwave frequency at finite temperature.

## 1.4.2. Electrodynamics of superconductors within BCS theory

More accurate description of electrodynamics response is given by BCS theory. The electrodynamics response can be studied by considering a perturbation term in BCS Hamiltonian in weak magnetic field [16],

$$H_1 = \frac{ie\hbar}{2m} \sum_i (\vec{\nabla}_i \cdot \vec{A} + \vec{A} \cdot \vec{\nabla}_i) \tag{1.33}$$

Here, the field $B = \vec{\nabla} \times \vec{A}$ is the total magnetic field including the field due to screening current.

### 1.4.2.1. Low frequency electrodynamics

In very low frequency limit ($\hbar\omega << 2\Delta$), the electrodynamics response of superconductors reduced to dissipation less diamagnetic response (Meissner effect). Therefore the solution of the above perturbation term in eqn. (1.33) gives the following analytical form of the temperature dependence of magnetic penetration depth, $\lambda$ [16],

$$\lambda^{-2}(T) = \lambda^{-2}(0) \left[ 1 - 2 \int_\Delta^\infty \left( -\frac{\partial f}{\partial E} \right) \frac{E}{(E^2 - \Delta^2)^{1/2}} \, dE \right] \qquad (\hbar\omega << 2\Delta) \tag{1.34}$$

The above relation holds for pure metal. In disordered superconductors where electronic mean free path is less than coherence length i.e. $l < \xi_0$, the temperature dependence of $\lambda$ can be shown to be by the relation,

$$\frac{\lambda^{-2}(T)}{\lambda^{-2}(0)} = \frac{\Delta(T)}{\Delta(0)} \tanh\left( \frac{\Delta(T)}{2k_B T} \right) \qquad (\hbar\omega << 2\Delta) \tag{1.35(a)}$$

Here,
$$\lambda^{-2}(0) = \frac{\pi\mu_0 \Delta(0) \sigma_n}{\hbar} \tag{1.35(b)}$$

where $\sigma_n$ is the normal state conductivity above $T_c$.

### 1.4.2.2. High frequency electrodynamics

In the high frequency regime the ($\hbar\omega \sim 2\Delta$), dissipation and absorption of electromagnetic fields become important. At finite $T$ superconductors always have quasi particles which behave like electrons in time varying field and give rise to dissipation. If frequency is sufficiently high such as $\hbar\omega > 2\Delta$, Cooper pair breaks into pair of quasi particles by absorbing a microwave photon. The response of superconductors at $T$=0 in extremely dirty limit i.e. for $l<\xi_0$, is given by following relations obtained from solution of perturbation term in eqn. (1.33) [see ref. 16],

$$\left.\frac{\sigma_1}{\sigma_n}\right|_{T=0} = \left(1+\frac{2\Delta}{\hbar\omega}\right)E(k)-\frac{4\Delta}{\hbar\omega}K(k) \qquad (\hbar\omega \geq 2\Delta) \qquad (1.36(a))$$

$$\left.\frac{\sigma_2}{\sigma_n}\right|_{T=0} = \frac{1}{2}\left(1+\frac{2\Delta}{\hbar\omega}\right)E(k')-\frac{1}{2}\left(1-\frac{2\Delta}{\hbar\omega}\right)K(k') \qquad (1.36(b))$$

$$k = \left|\frac{\hbar\omega-2\Delta}{\hbar\omega+2\Delta}\right| \text{ and } k' = (1-k^2)^{1/2} \qquad (1.36(c))$$

where $K(k)$(first kind) and $E(k)$ (second kind) are complete elliptic integral. At finite $T$ the thermally excited quasi particles contribute to the absorption, in that case the complex conductivity can be obtained numerically. The typical frequency dependence of real and imaginary part of $\sigma$ at $T$=0, is shown in Fig. 1.7.

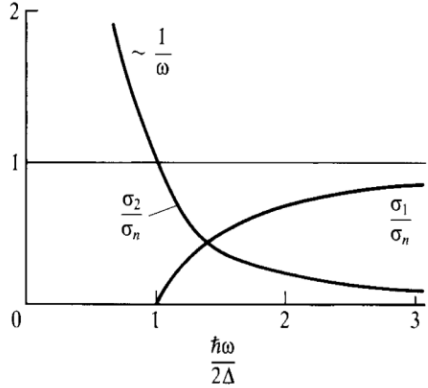

### 1.4.3. Sum rule

In Fig. 1.7 we can see that in SC state some area under the curve $\sigma_1/\sigma_n$ is missing respect to normal state value when an energy gap, $\Delta$ opens up in the SC state. The consequence of this missing area can be understood using oscillator strength sum rule, described below. The complex Drude conductivity of a material is given by,

$$\sigma(\omega) = \sigma_1(\omega)-i\sigma_2(\omega) \qquad (1.37(a))$$

$$\sigma_1(\omega) = \frac{\sigma_0}{(1+\omega^2\tau_n^2)} \qquad (1.37(b))$$

$$\sigma_2(\omega) = \frac{\sigma_0 \omega \tau_n}{(1 + \omega^2 \tau_n^2)} \tag{1.37(c)}$$

The schematic in Fig. 1.8.(a) shows the typical complex conductivity in normal state. According to oscillator strength sum rule, the total spectral weight under the curve $\sigma_1(\omega)$ vs $\omega$ should be constant given by [40],

$$\int_0^\infty \sigma_1(\omega) \, d\omega = \frac{\pi n e^2}{2m} \tag{1.38}$$

Here $n$ is the total carrier density. When a strongly correlated systems undergoes phase transition such as normal to superconductor, metal to insulator etc, the spectral weight gets redistributed, however the total oscillator strength always remain the same and determined by total carrier density. In superconductors, some of the spectral weight shifts to zero frequency which physically corresponds to the absorption of energy from dc electric field to increase the kinetic energy of super electrons. The schematic in Fig. 1.8.(b) represents the typical behaviors of complex conductivity in a superconductor at $T = 0$. The shaded region represents the missing area. Using Kramers-Kronig relations, Ferrel, Glover and Tinkham [41] showed that the missing area at finite frequency below the energy gap in SC state accumulate at $\omega = 0$ and appears as $\delta$ function. They associated this missing area to the superfluid density in SC state as,

$$\delta A = \int_{0+}^\infty \mathrm{Re}\left[\sigma_N(\omega) - \sigma_S(\omega)\right] d\omega \;\; = \;\; \frac{\pi n_s e^2}{2m} \tag{1.39}$$

where $\sigma_S(\omega)$ and $\sigma_S(\omega)$ are the complex conductivity at normal and SC state respectively. Using this sum rule we can estimate the approximate value of superfluid density in very simple way. In clean limit case when $2\Delta/\hbar \sim 1/\tau_n$, in SC state whole area under the curve $\sigma_1(\omega)$ shifts to zero frequency, therefore $n_s \approx n$. However in extremely dirty limit, $2\Delta/\hbar \ll 1/\tau_n$, fraction of total carrier density, $n$ transforms into super electrons equivalent to the missing area (see the shaded region in schematic Fig. 1.8.(b)),

$$\delta A = \sigma_0 \times \left(\frac{2\Delta}{\hbar}\right) \tag{1.40}$$

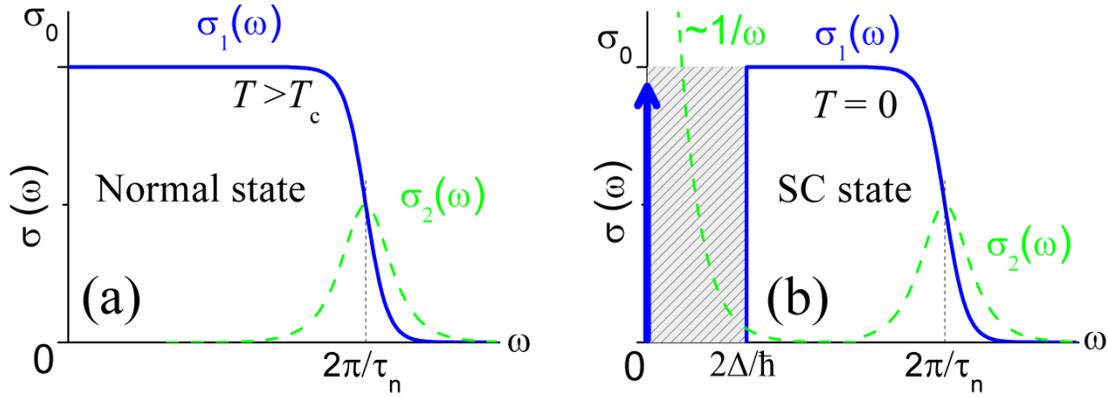

Figure 1.8. (a) Schematic diagram of complex conductivity at normal state ($T > T_c$). (b) The schematic of complex conductivity of a superconductor at $T = 0$ in dirty limit $\Delta/\hbar \ll 1/\tau_n$ (i.e. $l \ll \xi_0$).

Now comparing eqn. 1.39 and eqn. 1.40 we get the approximate value of superfluid density and magnetic penetration depth as follows,

$$n_s = \frac{4}{\pi} \frac{m\Delta\sigma_0}{\hbar e^2} \tag{1.41(a)}$$

$$\lambda^{-2}(T=0) = \frac{4}{\pi^2}\left(\frac{\pi\mu_0\Delta(0)\sigma_0}{\hbar}\right) \tag{1.41(b)}$$

Thus the sum rule gives the upper bound of the superfluid density. The disorder scattering decreases the $\sigma_0$, therefore $n_s$ is reduces with increasing disorder, which explains the low superfluid density in strongly disordered system.

We can see that the obtained expression of $\lambda$ by using sum rule which is similar to the expression obtained by complicated calculations using BCS theory (see eqn. 1.35), except the numerical constant ($4/\pi^2$) of the order of unity. Thus sum rule provides an intuitive way to calculate and understand different quantity associated with superconductivity.

In a superconductor, the redistribution of oscillator strength to zero frequency is the direct consequence of long range coherence in the SC state. Superconductor can be gapless in presence of magnetic impurities but it remain superconductor as far as there is still missing area under $\sigma_1(\omega)$ curve.

## 1.5. Phase stiffness

The phase stiffness is the measure of rigidity of phase of SC order parameter against any distortion in phase. The expression of phase stiffness can be obtained using analogy with the XY model described below.

Traditionally, the physics of phase fluctuations in superconductors is studied within framework of XY model consists of planer rotors placed in a square lattice. The rotors can be represented by complex parameter, $S = |S|e^{i\theta}$ where the direction of rotors is represented by angle $\theta$ (for details of 2D XY model see section 1.6.2.3). The Hamiltonian of this model is given by,

$$H = -J\sum_{<ij>}\hat{S}_i \cdot \hat{S}_j = -J\sum_{<ij>}\cos(\theta_i - \theta_j) \qquad (1.42)$$

Here $J$ represents the stiffness against any non uniform change in $\theta$. Assuming the direction of rotors smoothly varies from site to site, the Hamiltonian can be expressed in terms of Taylor expansion of cosine term as,

$$H \approx -J\sum_{<ij>}(1 - \frac{1}{2}\Delta\theta_{ij}^2) \approx H_0 + \frac{1}{2}J\sum_{<ij>}(\Delta\theta_{ij}^2) \qquad (1.43)$$

Thus to introduce the non uniform change in $\theta$, the energy cost which will be stored as elastic energy in the system is given by,

$$\Delta E = H - H_0 \approx \frac{1}{2}J\sum_{<ij>}(\Delta\theta_{ij}^2) \qquad (1.44)$$

In superconductors, there is also an energy cost to apply twist or gradient on the phase of order parameter, $\psi = |\psi|e^{i\theta}$. This energy is utilised to increase the kinetic energy of the super electrons given by,

$$H_s = n_s\int d^3r \left(\frac{1}{2}mv_s^2\right) \qquad (1.45(a))$$

$$v_s \approx \frac{\hbar}{2m}[\nabla\theta] \qquad (1.45(b))$$

where $n_s$ is the superfluid density, $m$ is the electronic mass and $v_s$ is the velocity of super electrons associated with the phase gradient. Since the phase $\theta$ of superconductors is a periodic variable with period of $2\pi$ and there is a short distance spatial cut off, *a,* over which phase can be twisted, the above integration can be regarded as the sum of finite difference in $\theta$ in lattice model with lattice constant, *a,*

$$H_s = \frac{1}{2}\frac{\hbar^2 n_s a}{4m}\sum_{<ij>}\Delta\theta_{ij}^2 \tag{1.46}$$

Therefore comparing eqn. 1.44 and 1.46, the superfluid phase stiffness, *J* equivalent to the increase in kinetic energy of super electrons [42] is given by,

$$J = \frac{\hbar^2 a n_s}{4m}; \quad n_s = \frac{m}{\mu_0 e^2 \lambda^2}, \tag{1.47}$$

where *a* is the minimum distance cutoff or lattice constant in lattice model, represents the characteristic length scale for phase fluctuations. $n_s$ is the superfluid density and $\lambda$ is the magnetic penetration depth. In 3D superconductors, the minimum distance cutoff given by the coherence length, $\xi_0$ however in 2D superconductors where thickness, $t < \xi_0$, the thickness play the role of characteristic length scale. Thus the minimum distance cut off, $a \approx \min(t, \xi_0)$.

## 1.6.    Superconducting fluctuations

Superconductors poses highly ordered state with long range correlation, therefore in clean conventional superconductors the thermodynamic fluctuations are unimportant except very close to $T_c$ [43]. However in a superconductor with low dimensionality and strong disorder the fluctuations are enhanced.  In this section I will give brief overview of different kind of fluctuations and their effect on various properties of superconductor.

### 1.6.1. Amplitude fluctuations

Fluctuations in order parameter influence various SC properties such as transport properties, diamagnetism, specific heat etc [43]. In this section, I will concentrate only on the effect of amplitude fluctuations on electrical conductivity above $T_c$ studied through transport measurements and high frequency complex conductivity measurements.

A superconductor above $T_c$ shows excess conductivity due to presence of unstable SC pairs induced by fluctuations. The first successful theoretical understanding of this excess conductivity in a dirty superconductor is provided by Aslamazov and Larkin (AL) [44]. This excess fluctuation conductivity, $\sigma_{fl}^{DC}(T) = \sigma(T) - \sigma_n$, is a result of direct acceleration of Cooper pairs, given by,

$$^{2D\,AL}\sigma_{fl}^{DC} = \frac{1}{16}\frac{e^2}{\hbar t}\varepsilon^{-1} \tag{1.48}$$

$$^{3D\,AL}\sigma_{fl}^{DC} = \frac{1}{32}\frac{e^2}{\hbar \xi_0}\varepsilon^{-1/2} \tag{1.49}$$

where $\varepsilon \equiv \ln(T/T_c)$, $t$ is the thickness of the sample and $\xi_0$ is the BCS coherence length. These accelerated SC pairs have finite life time and in their way, they decay into quasi particles of nearly opposite momentum. However due to the time reversal symmetry, they remain in the state of small total momentum. Therefore the resultant quasi particles continue to be accelerated like their parent pairs, though they get scattered by impurity. Quasi particles also have a finite life time and ultimately they decay back into SC pairs ( Skocpol and Tinkham (1975)) [43]. In the case of a dirty superconductors, contribution from quasi particles acceleration is negligible due to strong scattering but in clean superconductors, it gives a finite second order correction to the fluctuation conductivity which is predicted by Maki-Thomson (MT) [45] as follows,

$$^{2D\,MT}\sigma_{fl}^{DC} = \frac{1}{8}\frac{e^2}{\hbar t}\frac{1}{\varepsilon-\delta}\ln\left(\frac{\varepsilon}{\delta}\right) \tag{1.50}$$

$$^{3D\,MT}\sigma_{fl}^{DC} = \frac{1}{8}\frac{e^2}{\hbar \xi_0}\varepsilon^{-1/2} \;, \tag{1.51}$$

where $\delta$ is the Maki-Thompson pair breaking parameter. The AL and MT contributions are additive and they together explain the large amount of excess conductivity above $T_c$ in clean superconductors such as aluminum, tin, indium etc.

To understand the fluctuation phenomena further, we now concentrate on the frequency dependence of fluctuation conductivity. Using the time dependent Ginzburg-Landau equation, Schmidt [46] calculated the frequency dependent AL term of the fluctuation conductivity in 2D and 3D limit as follows [47],

$$\sigma_{fl}^{2D\,AL}(\omega) = \sigma_{DC,fl}^{2D\,AL} S^{2D\,AL}\left(\frac{\omega}{\omega_0}\right)\,;\; \omega_0 = \frac{16k_B T_c}{\pi\hbar}\varepsilon$$

$$S^{2D\,AL}(x) = \left\{\frac{2}{x}\tan^{-1}x - \frac{1}{x^2}\ln(1+x^2)\right\} + i\left\{\frac{2}{x^2}(\tan^{-1}x - x) + \frac{1}{x}\ln(1+x^2)\right\}$$

(1.52)

and

$$\sigma_{fl}^{3D\,AL}(\omega) = \sigma_{DC,fl}^{3D\,AL} S^{3D\,AL}\left(\frac{\pi\hbar\omega}{16k_B T_c\varepsilon}\right)$$

$$S^{3D\,AL}(x) = \left\{\frac{8}{3x^2}\left(1-(1+x^2)^{3/4}\cos(\frac{3}{2}\tan^{-1}x)\right)\right\} + i\left\{\frac{8}{3x^2}\left(-\frac{3}{2}x + (1+x^2)^{3/4}\sin(\frac{3}{2}\tan^{-1}x)\right)\right\}$$

(1.53)

The frequency dependence of the MT-term was calculated by Aslamazov and Varlamov [48]. They have shown that in 2D and 3D limits the frequency dependence of MT term is additive to the AL-term, as follows [47],

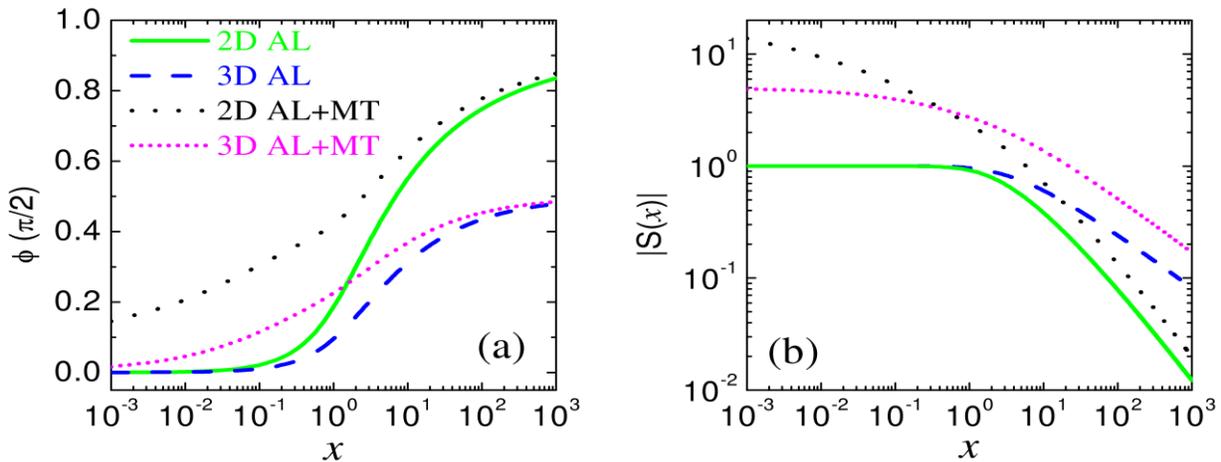

Figure 1.9. Shows the functional form of phase $\phi$ and amplitude $|S|$ of theoretically predicted functional form of $S$ from amplitude fluctuation theory, as a function of reduced frequency $x=\omega/\omega_0$ (for details see text in section 1.6.1).

$$\sigma_{fl}^{2D\,AL+MT}(\omega) = \sigma_{DC,fl}^{2D\,AL+MT}\,S^{2D\,AL+MT}\left(\frac{\pi\hbar\omega}{16k_BT_c\varepsilon}\right);$$

$$S^{2D\,AL+MT}(x) = \left\{\mathrm{Re}\,S^{2D\,AL}(x) + \frac{2\pi x - 2\ln(2x)}{1+4x^2}\right\} + i\left\{\mathrm{Im}\,S^{2D\,AL}(x) + \frac{\pi + 4x\ln(2x)}{1+4x^2}\right\}$$

$$(1.54)$$

and

$$\sigma_{fl}^{3D\,AL+MT}(\omega) = \sigma_{DC,fl}^{3D\,AL+MT}\,S^{3D\,AL+MT}\left(\frac{\pi\hbar\omega}{16k_BT_c\varepsilon}\right);$$

$$S^{3D\,AL+MT}(x) = \left\{\mathrm{Re}\,S^{3D\,AL}(x) + \frac{4-4x^{1/2}+8x^{3/2}}{1+4x^{1/2}}\right\} + i\left\{\mathrm{Im}\,S^{3D\,AL}(x) + \frac{4x^{1/2}-8x+8x^{3/2}}{1+4x^{1/2}}\right\}$$

$$(1.55)$$

Fig. 1.9(a) and (b) show the frequency dependence of the function, $S$(x) predicted from AL and AL+MT predictions in 2D and 3D limit.

### 1.6.2. Phase fluctuations

Phase stiffness, $J$ in clean conventional 3D superconductors is very high for example $J \sim 10^5$ K for Pb ($T_c \sim 7$ K) and $J \sim 10^4$ K for NbSe$_3$ ($T_c \sim 18$ K), therefore phase fluctuations play negligible role. However superconductors characterized by low phase stiffness, are susceptible to phase fluctuations.

#### 1.6.2.1. Longitudinal and transverse mode of phase fluctuations

The change in phase of SC order parameter, $\psi = |\psi|e^{i\theta}$ over a path, $C$ is given by [54],

$$\Delta\theta = \int_C d\vec{r} \cdot \vec{\nabla}\theta \tag{1.56}$$

For closed path, the total change in θ must by multiple of 2π. Therefore,

$$\oint_C d\vec{r} \cdot \vec{\nabla}\theta = 2\pi n_l \tag{1.57}$$

If the closed path C don't include any singularity, $n_l$=0, therefore applying Stokes law we get,

$$\vec{\nabla} \times \vec{\nabla}\theta = 0 \tag{1.58}$$

The integral in eqn. (1.57) will give finite value, if $C$ includes *singular points* which define the vortices. Now we can generalized the eqn.(1.58) to include vortices $n_i$ at position, $\boldsymbol{r}_i$ , as

$$\vec{\nabla} \times \vec{\nabla} \theta = \hat{\mathbf{z}} \ 2\pi \sum_i n_i \delta(\vec{r} - \vec{r}_i) \equiv 2\pi \ \vec{n}(\vec{r}) \tag{1.59}$$

Here **n(r)** is the vortex density. At low temperature, $\vec{\nabla} \theta$ gives the velocity of super electrons as $\vec{v} \equiv \hbar \vec{\nabla} \theta / 2m$. We can write $\vec{v}$ as the sum of **longitudinal** and **transverse** part as $\vec{v} \equiv \vec{v}_l + \vec{v}_t$ defined by,

$$\vec{\nabla} \times \vec{v}_l(\vec{r}) = 0 \qquad (\textbf{\textit{longitudinal}} \text{ phase fluctuations}) \tag{1.60}$$

$$\vec{\nabla} . \vec{v}_t(\vec{r}) = 0 \qquad (\textbf{\textit{transverse}} \text{ phase fluctuations}) \tag{1.61}$$

Here $\vec{\nabla} \theta_l \equiv 2m\vec{v}_l / \hbar$ represents the longitudinal mode and $\vec{\nabla} \theta_t \equiv 2m\vec{v}_t / \hbar$ transverse mode of phase fluctuations. From above equations, we can see that there is no contribution from vortex in longitudinal parts, whereas the transverse part entirely determined by vortices in the system and given by,

$$\vec{v}_t(\vec{r}) = \pi \int d^d r G_L(\vec{r} - \vec{r}\,') \vec{\nabla} \times \vec{n} \tag{1.62}$$

Here $G_L$**(r-r′)** is the Greens function for the Laplacian operator $-\nabla^2$ and *d* is the dimension.

### 1.6.2.2. Phase fluctuations in 3D disordered superconductors

In strongly disordered superconductors, disorder scattering suppress the superfluid density, therefore reduces the superfluid stiffness, *J,* through eqn.(1.47). The disorder also increases the electron-electron interaction by decreasing the Coulomb screening in strongly disordered system, therefore increases the energy cost of number fluctuations which are essential for phase coherence. In such a situation when the *J* becomes comparable to Δ or Coulomb screening is poor, phase fluctuations are expected to play significant role and may even destroy the superconductivity by phase disordering [34].

In 3D strongly disordered superconductors, the longitudinal phase fluctuations affect superconducting properties significantly, while transverse phase fluctuations are negligible. There are mainly two types of longitudinal phase fluctuations which influence the SC properties in disordered superconductor: quantum phase fluctuations (QPF) and thermal phase fluctuations or classical phase fluctuations (CPF).

*(i) Quantum phase fluctuations (QPF):*

The phase of SC state is the complex conjugate variable of number density. Hence it follows the number-phase uncertainty relation given by $\Delta\theta \cdot \Delta N \sim 1/2$. To have phase coherent quantum state, superconductors need to allow number fluctuations. In clean 3D conventional superconductors, Coulomb screening is very good; therefore the energy cost for charge pileup is negligible. However in strongly disordered or low dimensional system, energy cost for number fluctuations is enhanced due to low number density and poor Coulomb screening, as a result system relaxes its phase instead number fluctuations to minimize the energy cost. Therefore in systems with low superfluid stiffness and poor screening, quantum phase fluctuations play a significant role.

In disordered superconductors, QPF affect the SC properties by suppressing the zero temperature superfluid density from its bare value. The suppression in superfluid density can be estimated using the following relation predicted from self-consistent harmonic approximation [49,50]:

$$\frac{n_s(0)}{n_{s0}(0)} \approx e^{-\Delta\theta^2(0)/2d} \qquad (1.61(a))$$

$$\Delta\theta^2(0) = \frac{1}{2}\sqrt{E_c \Big/ J} \qquad (1.62(b))$$

where $n_s(0)$ is the bare superfluid density when there was no QPF, $d$ is the dimension of the system and $E_c$ is the Coulomb energy which can be estimated using following relation [50],

$$E_c = \frac{16\pi e^2}{\varepsilon_\infty a} \qquad (1.63)$$

Here $\varepsilon_\infty$ is the background dielectric constant and $a$ is the characteristic length scale of fluctuations.

*(ii) Classical phase fluctuations (CPF):*

At high temperature the CPF expected to take over the QPF and the crossover temperature from QPF to the CPF is given by Josephson plasma frequency, $\omega_p$, [50],

$$k_B T_{cross} \approx \hbar\omega_p = \sqrt{4\pi e^2 n_s / m^* \varepsilon_\infty} = \sqrt{E_c J} \qquad (1.64)$$

Above the crossover temperature, $T_{cross}$ CPF gives linear temperature dependence of superfluid density in the form [50],

$$\frac{n_s(T)}{n_s(0)} \approx 1 - (1/2dJ)T \qquad (1.65)$$

The schematic in Fig. 1.10 represents the typical temperature variation of superfluid density in presence of QPF and CPF. At low temperature, QPF is the dominant effect where superfluid density gets suppressed due to QPF. QPF does not dependent on temperature; therefore the superfluid density remains constant up to the characteristic crossover temperature, $T_{cross}$ determined by Josephson plasma frequency. Above the $T_{cross}$, CPF takes over the QPF and superfluid density follows the linear temperature dependence due to CPF.

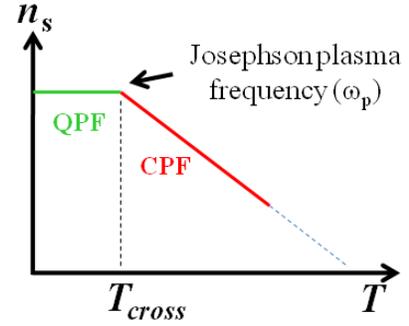

Figure 1.10. shows the schematic of typical temperature dependence of superfluid density in presence QPF and CPF (see text). Close to $T_c$, long wave length harmonic approximation breaks down, therefore it is shown by dashed line.

### 1.6.2.3. Phase fluctuations in 2D superconductors

The SC transition in strictly 2D superconductors is governed by transverse phase fluctuations (vortex) described within famous Berezeski, Kosterlitz and Thouless (BKT) theory [51,52]. However due to additional complicacy such as quasi particle excitations, low vortex core energy, intrinsic disorder in the system, the true nature of the BKT transition in real 2D superconductors appears somewhat different than the predicted behavior within 2D XY model [53].

In this section I will give detailed overview of BKT transition in 2D XY model and then I will discuss the nature of BKT transition in real 2D superconductors.

### 1.6.2.3.1. 2D XY model and BKT transition

According to Mermin-Wagner theorem [54], a system with continuous symmetry having dimension $d \leq 2$, can't have true long range order at finite $T$ due to low energy thermal excitations which destroy the long range order. A simplest example of 2D system with continuous symmetry is the 2D XY model consist of planer rotors [see Fig. 1.11.(a) where spin lies in xy plane, represented by order parameter, $\mathbf{S}=|\mathbf{S}|e^{i\theta}$ where angle $\theta$ is the variable which defines the symmetry of the system. In this kind of system, Berezeski, Kosterlitz and Thouless (BKT) proposed a universal phase transition at characteristic temperature, $T_{BKT}$, from quasi long range order at low $T$ to high temperature disordered phase [53].

The BKT transition in 2D XY model can be understood in a very simple way by considering the competition between vortex self energy and entropy of an isolated vortex. Self energy of an isolated vortex is given by [54],

$$E_{vor} = \pi J \ln\left(L/a\right) \tag{1.66}$$

where $L$ is system size and $a$ is the lattice spacing. The increase in entropy due to vortex excitation can be calculated by considering the number of way, $(L^2/a^2)$, the vortex can be placed in the square lattice of dimension $L^2$ as,

$$S = \ln\left(L^2/a^2\right) = 2\ln\left(L/a\right) \tag{1.67}$$

Hence the Helmholtz free energy of the system is given by,

$$F = E_{vor} - TS = (\pi J - 2T)\ \ln\left(L/a\ \right) \tag{1.68}$$

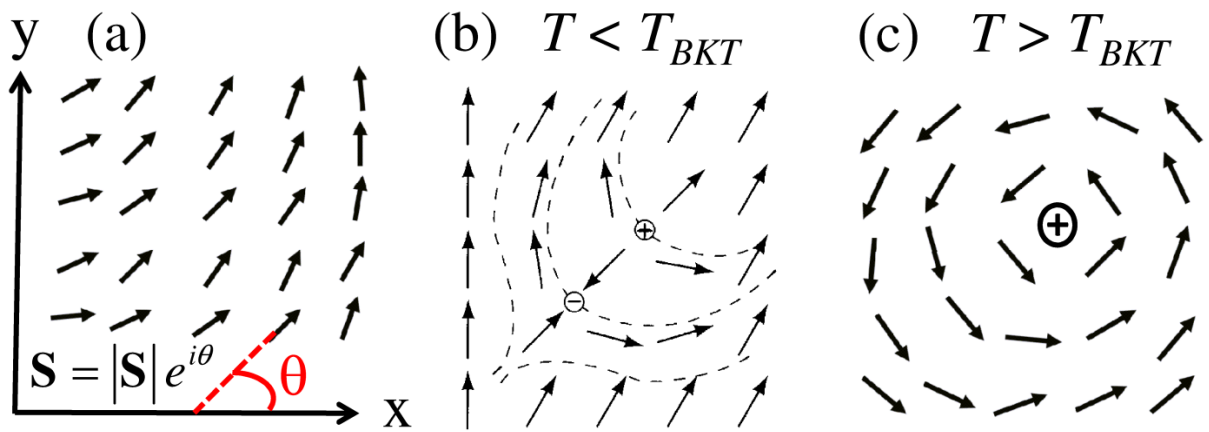

Figure 1.11. (a) 2D XY model, a simple example of a 2D system with continuous symmetry. (b) Thermally activated vortex-antivortex bound pair at $T<T_{BKT}$. (c) Shows the unbound vortex excitations which destroy the quasi long range order at $T>T_{BKT}$ (part of the figure is adapted from ref. 54).

Therefore the free energy depends on the system size logarithmically. For $T < \pi J/2$, $F \rightarrow \infty$ as $L \rightarrow \infty$, therefore free energy will be minimum when there is no vortex. For $T > \pi J/2$, $F \rightarrow -\infty$ as $L \rightarrow \infty$ which implies that system can lower its free energy by creating a vortex. However at temperature, $T > \pi J/2$, infinite number of vortices can proliferate in the system and destroy the quasi long range order. Therefore the logarithmic dependence of free energy on system size leads to the phase transition from quasi long range order to disorder phase at temperature,

$$T_{BKT} = \frac{\pi}{2} J(T_{BKT}^-) \qquad (1.69)$$

The $T_{BKT}$ changes depending on phase stiffness of the system, however the ratio, $J(T_{BKT})/T_{BKT}$ $=2/\pi$ is a constant which system independent.

### Renormalization group (RG) analysis:

Instead of single vortex-antivortex pair, in real system there can be large number of thermally activated vortex-antivortex pairs below $T_{BKT}$, however they don't destroy the quasi long range order although renormalize the phase stiffness, $J$ of the system. The detailed nature of BKT transition in real system can be realized from RG analysis using following famous renormalization group equations of reduced renormalized stiffness, $K = J/T$ and vortex fugacity, $y = e^{-\mu/T}$ [53,54],

$$\frac{dK^{-1}}{dl} = 4\pi^3 y^2 \quad \text{and} \quad \frac{dy}{dl} = \left(2 - \pi K\right) y \qquad (1.74)$$

Here $\mu$ is the vortex core energy. Within 2D XY model, $\mu$ is given by [55],

$$\mu_{XY} = \frac{\pi^2}{2} J \cong 4.9 J \qquad (1.73)$$

The above RG equations can be solved analytically only close to the critical point $K(l)=2/\pi$ and $y(l)=0$, however it can be solved numerically. The resultant RG flow diagram is shown in Fig. 1.12(a). The flow diagram shows a separatrix between the low temperature quasi ordered phase with no free vortices (shaded region in Fig. 1.12) and high temperature disordered phase with free vortices, passing through critical point $y(l) = 0$, $K(l)=2/\pi$. The dashed line represents the line of initial conditions at different temperatures. Here $y(l) = 0$ is the line of fixed points, which corresponds to no vortices. At low temperatures (points reside below the separatrix) $K^{-1}(l) < \pi/2$, all RG flows are towards $y(l)=0$ and finite $K^{-1}(l)$. That means vortex core energy flows to infinity, therefore it gives a state with finite superfluid stiffness with no unbound vortex in the system. However for a point just above the separatrix, whose $K^{-1}(l) > \pi/2$, RG flow is towards large value of $y$ and $K^{-1}$. That means the vortex core energy flows to zero which implies the proliferation vortices and break down the ordered phase. Therefore it gives a disordered state with zero phase stiffness. Hence there is a phase transition from quasi ordered phase with finite phase stiffness to a disordered phase with zero phase stiffness at a temperature, $T=T_{\text{BKT}}$ determined by the intersection of the line of initial conditions and separatrix. Therefore at the transition temperature, $T=T_{\text{BKT}}$ the reduced superfluid stiffness, $K = J/T$ shows a sharp jump from a finite value to zero given by,

$$\lim_{l\to\infty} K^{-1}(T_{BKT}^{-}) = \frac{2}{\pi},$$ (1.75(a))

$$\lim_{l\to\infty} K^{-1}(T_{BKT}^{+}) = 0,$$ (1.76(b))

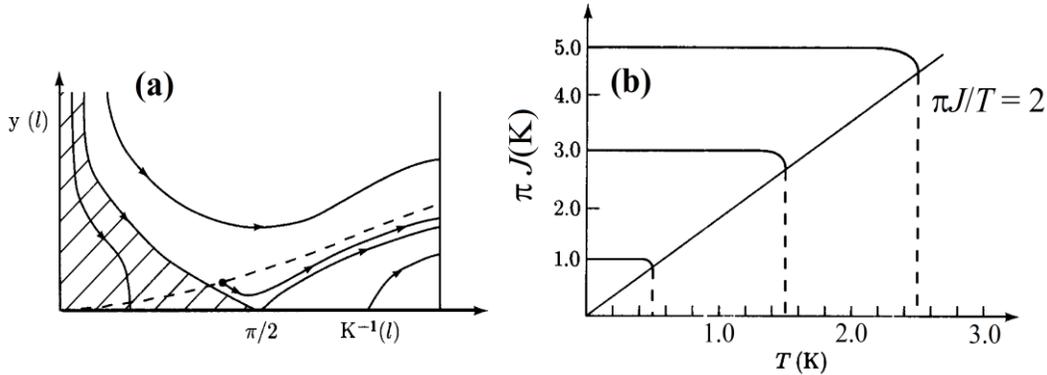

Figure 1.12.(a) The schematic RG flow diagram of the Kosterlitz-Thouless recursion relations. The dashed line is the line of initial conditions as *T* increases. The critical temperature is determined by the crossing of the dashed line and the separatrix (see text for details). (b) Schematic representation of superfluid stiffness as a function of temperature, *T*. All curves terminate at the universal line predicted within BKT model, showing the universality of BKT transition (figure is adapted from ref. 54).

One can notice that the jump in reduced superfluid stiffness is system independent, therefore represents the universal nature of BKT transition.

The Figure 1.12 (b) shows the typical temperature variation of $J$ in a system like 2D XY model which undergoes BKT transition. The $J$ remains constant as temperature increases but abruptly goes to zero when becomes equal to $2T/\pi$ due to proliferation of vortices predicted within BKT theory.

### 1.6.2.3.2. **BKT transition in 2D superconductors**

2D superconductors having continuous symmetry, are predicted to undergo BKT phase transition [51,52]. Unlike XY model, other than the vortex excitations, at finite temperature superconductors have also quasi particle excitations which decrease the superfluid density with increasing temperature predicted within BCS theory. Therefore the nature of BKT transition in 2D superconductor appears somewhat different from the predicted behavior within 2D XY model. The Fig. 1.13. shows the typical nature of BKT transition in 2D superconductors. The upper panel (green black line) shows the normalized superfluid density, $n_s$ ($\propto J$) as a function of temperature. The normalized superfluid density decreases with increasing temperature and suddenly goes to zero at $T=T_{BKT}$ due to proliferation of vortices as it touches the universal BKT line (red dashed line), instead of smoothly going to zero at $T = T_{BCS}$ (blue dashed line) as predicted within BCS theory. The lower panel represents the normalized resistivity data. The resistivity appears at $T=T_{BKT}$ when the superconductivity is destroyed due to vortex (transverse phase) fluctuations. Above $T_{BKT}$, the vortex fluctuations leave its signatures in various SC properties.

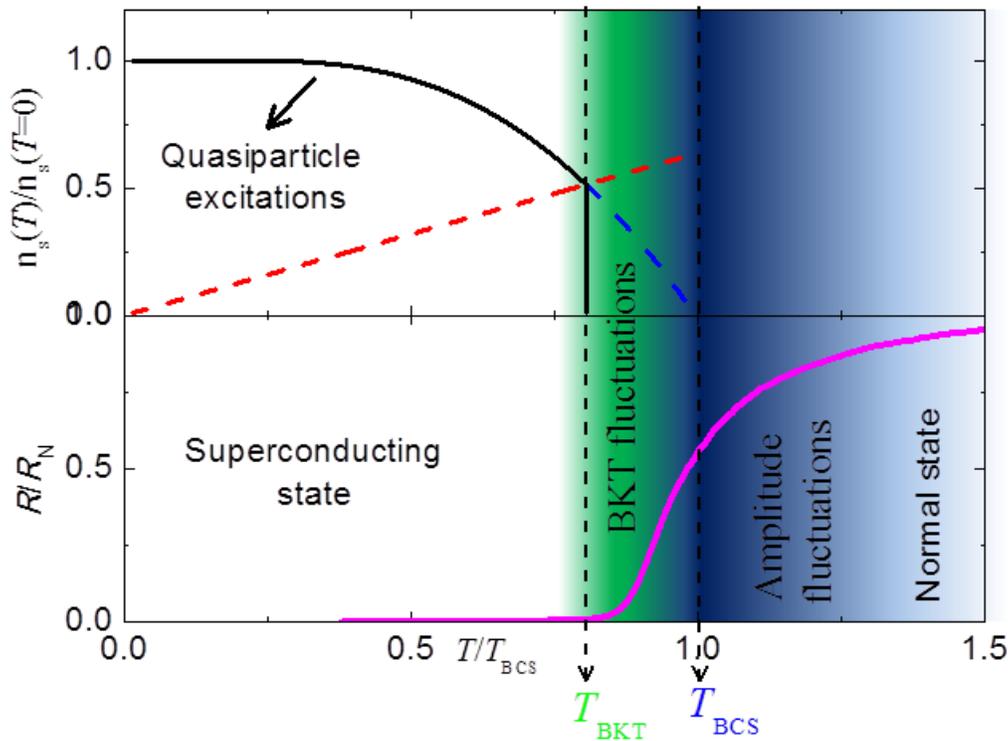

Figure 1.13. The schematic represents typical BKT transition in 2D superconductors. The upper panel shows that the normalized superfluid density is decreasing with increasing temperature and suddenly goes to zero at $T=T_{BKT}$ as it touches the universal BKT line (red dashed line) due to proliferation of vortices instead of smoothly going to zero at $T = T_{BCS}$ (blue dashed line) as predicted within BCS theory. The lower panel represents the normalized resistivity data. The resistance begins to appear at $T=T_{BKT}$ when vortex proliferation destroy the quasi long range order in the system (for details see text).

Therefore BKT transition in 2D superconductors manifests in various SC properties such as temperature variation of superfluid density when approaching towards $T_{BKT}$ from below and in resistivity, diamagnetism, Nernst effect etc when approaching towards $T_{BKT}$ from above. The SC fluctuations in between $T_{BKT}$ and $T_{BCS}$ (green region) are dominated by BKT fluctuations which give exponential temperature dependence of correlation function. Above $T_{BCS}$ (blue region), more conventional GL fluctuations which give power law temperature dependence of correlation function, take over. Above $T_{BKT}$ the effect of SC fluctuations on SC properties can be studied using interpolating formula of correlation length to take into account the crossover from BKT to GL fluctuations, proposed by Halperin and Nelson [52].

Although superfluid He films follow the BKT relation quite precisely [56] the BKT transition in 2D superconductors has remained controversial [57]. For instance, the jump in $n_s$ is often observed at a temperature lower than $T_{BKT}$ and at a $J(\propto n_s)$ larger than expected value within BKT theory[58,59]. In chapter 4, I will explore the true nature of BKT transition in real 2D superconductors using ultrathin NbN films as our model system.

# CHAPTER 2

## Our model system: NbN thin films

To study the effect of phase fluctuations on superconducting properties, we work on NbN thin films. NbN is a conventional s-wave superconductor with reasonably high $T_c \sim 16.5$ K. Magnetic penetration depth of bulk NbN is about 250 nm, electronic mean free path, $l \sim 4$ Å and coherence length, $\xi_0$ is about 4 nm. Thus it is an extremely dirty ($l << \xi_0$) type **II** ($\xi_0 < \lambda$) superconductor with fcc crystal structure like NaCl. Band structure calculation [1] shows that the only 4d band of Nb atom contributes to the carrier density which is about $2.33 \times 10^{29}$ /m$^3$ verified experimentally using Hall effect measurement [2].

Crystalline epitaxial NbN thin films are grown on (100) oriented MgO substrate using popular DC reactive magnetron sputtering technique. The sample growth using sputtering process provides a very good control over thickness and disorder in our films. NbN films are mechanically strong and chemically stable in ambient atmosphere and can be thermally recycled without any detectable degradation in film's properties, which makes it an ideal system to study fundamental physics related to superconductivity.

*Before discussing the electrodynamics of NbN thin films, for sake of completeness of my thesis, in this chapter, I will introduce basic normal and superconducting state properties of NbN thin films studied in collaboration with other group members.*

## 2.1. Tuning of disorder level

Properties of sputtered NbN thin films strongly depend on the sample growth conditions. It was observed by S. P. Chokalingam et al [2] that at the optimum deposition conditions (typically sputtering power $\approx$ 200 W and Ar:N$_2$ $\approx$ 84:16 at substrate temperature, $T = 600$ $^0$C), the deposited NbN is stoichiometric in nature and shows maximum $T_c$ ~ 16.5 K. When the sputtering power increases or $N_2$ ratio decreases in plasma, metallic $Nb_2N$ phase is formed. Since $Nb_2N$ is non-superconducting, $T_c$ drops and temperature dependence of resistivity shows strong metallic behavior. On the other hand, when sputtering powering decreases or $N_2$ ratio increases, Nb vacancies are created in epitaxial crystalline NbN films, which act as disorder and suppress $T_c$ by disorder scattering.

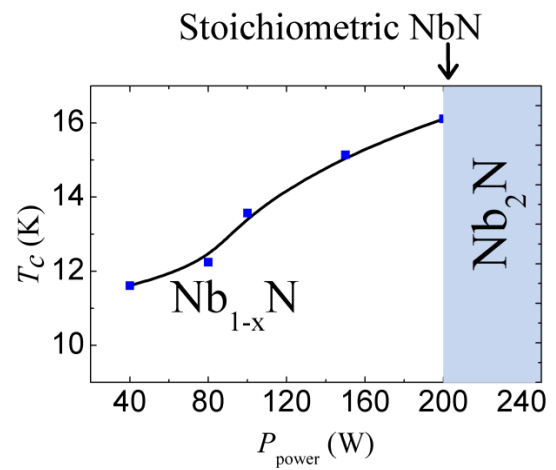

Figure 2.1. Superconducting transition temperature ($T_c$) as a function of sputtering power (figure is reproduced from ref. 2).

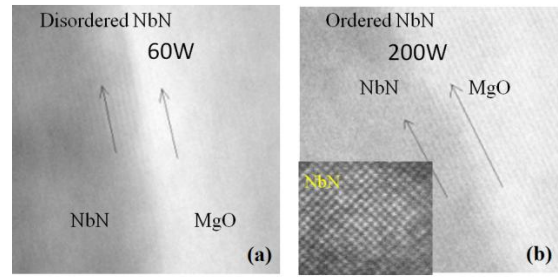

Figure 2.2. The TEM images showing the perfect lattice plane matching with MgO. (a) A disordered NbN film with $k_F l$ ~ 3.3. (b) A stoichiometric film with $k_F l$ ~ 9. The *inset* shows the perfect crystalline structure.(figure is adapted from ref. 4).

The Fig. 2.1. shows the variation of superconducting transition temperature as a function of sputtering power of a set of NbN films with thickness, $t \geq 50$ nm. Thus by tuning the deposition conditions, we can control the level of disorder in the form of Nb vacancies in epitaxial crystalline NbN films.

## 2.2. Structural property

Structural study using X-ray diffraction (XRD) confirmed the formation of crystalline NbN thin films on MgO substrates [2,3]. The φ-scan using a four circle goniometer showed the epitaxial nature of NbN films. The epitaxial nature was also confirmed by structural study using transmission electron microscope (TEM) [4]. Fig. 2.2. shows the cross sectional TEM images of a strongly disordered film and one less disordered film. It was observed that the epitaxial nature of the films remains intact even in very high disorder sample implying that the main sources of disorder in NbN films are not from structural granularity consistent with the X-ray data. For details about the structural characterization of NbN thin films, see the ref. 3 and 4.

## 2.3. Characterization

NbN thin films were characterized by transport measurements such as resistivity, magneto resistance (MR) and Hall carrier density by Madhavi Chand et al. [3,5].

The resistivity (ρ) as function of temperatures was measured using four probe technique. Hall carrier density ($n_H$) was calculated from the measured hall coefficient ($R_H = -1/n_H e$) by sweeping the magnetic field ($H$) from 12T to -12T at different temperatures.

The upper critical field ($H_{c2}$) as a function of temperature ($T$) was measured for several samples from R-T scans at different $H$. Since all our NbN films are in the dirty limit, $l << \xi_{GL}$, $H_{c2}(0)$ and $\xi_{GL}$ were estimated using dirty limit relation [6,7]:

$$H_{c2}(0) = 0.69 T_c \left. \frac{dH_{c2}}{dT} \right|_{T=T_c} \quad ; \quad \xi_{GL} = \left[ \frac{\phi_0}{2\pi H_{c2}(0)} \right]^{1/2} , \qquad (2.1)$$

## 2.3.1. Quantification of disorder

The disorder in NbN films is in the form of Nb vacancy in NbN crystalline films, introduced during the sample growth. To quantify disorder in NbN samples, the Ioffe Regel parameter, $k_F l$ was used, where $k_F$ is the Fermi wave vector and $l$ is the mean free path. Considering free electron model, the value of $k_F l$ was extracted from measured ρ and $n_H$ using the relation,

$$k_F l = \frac{\hbar \left( 3\pi^2 \right)^{2/3}}{\left\{ n_H \left( 285K \right) \right\}^{1/3} \rho \left( 285K \right) e^2} \qquad (2.2)$$

In presence of electron-electron interaction which is very much present in NbN films [4], the relation $R_H = -1/n_H e$ is not truly valid. Therefore $k_F l$ was calculated using $R_H$ and ρ at the maximum measurement temperature achievable in our experimental setup i.e. 285K, where the electron-electron interaction is expected to be small [8].

Most interesting part of NbN thin films is that the disorder can be varied over a very large range by changing the deposition conditions only and with increasing disorder the value of $k_F l$ varies from $k_F l$ ~10 for moderately clean sample to below Mott limit with $k_F l$ ~1 for a very high disordered sample.

## 2.4. Effect of disorder

In this section, I will review some of recent works carried out on effect of disorder on various SC properties of 3D NbN films. Since the coherence length, $\xi_0 \sim 5\text{nm}$ and the thickness of all the films was about 50 nm, all films are in 3D limit.

### 2.4.1. Basic parameters ($\rho$, $T_c$, $H_{c2}$, $\xi_{GL}$, $n_H$, $\Delta$, $\lambda$)

#### 2.4.1.1. Resistivity ($\rho$) and superconducting transition temperature ($T_c$)

The Fig. 2.3(a) shows the temperature dependence of $\rho$ for a set of NbN films with different level of disorder. The least disordered sample with $k_F l \sim 10.1$, follows the metallic behavior where resistivity increases linearly with temperature. However when the disorder is increases, $\rho(T)$-$T$ curves show negative temperature coefficient due to strong electron-electron interaction [4]. The negative temperature coefficient becomes progressively pronounced with increasing disorder. The inset of the Fig. 2.3(a) shows the expanded view of transition region at low temperatures. The figure Fig. 2.3(b) shows the $\sigma(T)$ as function of $T$ for three most disordered samples with $k_F l$ ~0.82, 0.49 and 0.42. The $\sigma(T)$ increases linearly with $T$ for samples with $k_F l$ <1 and extrapolated $\sigma(T)$ at $T \rightarrow 0$, give finite values. Although $k_F l$ ~1 is generally associated with Anderson metal-insulator transition, however all NbN films are in metallic regime, since $\sigma(T \rightarrow 0)$ are finite. This discrepancy may be due to the error in determining true values of $k_F l$. Since values of $k_F l$ are calculated using free electron model in a system characterized by strong electron-electron interaction, the error in determining true values of $k_F l$ probably reflected in bad metallic behavior for the sample with $k_F l$ < 1.

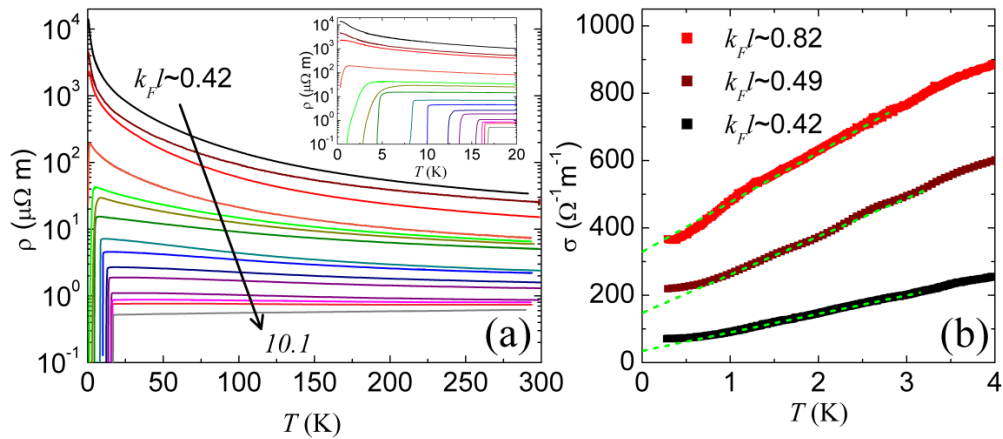

Figure 2.3. (a) shows the resistivity ($\rho$) vs temperatures ($T$) for a set of films with different level of disorder. The *inset* shows the expanded view of the superconducting transition region at low temperatures. (b) The conductivity, $\sigma(T)$ vs $T$ at low temperature for three most disordered samples. The (green) dashed lines show the extrapolated $\sigma$ at $T \rightarrow 0$ for samples with $k_F l \sim 0.82$, 0.49 and 0.42 (figure is reproduced from ref. 5).

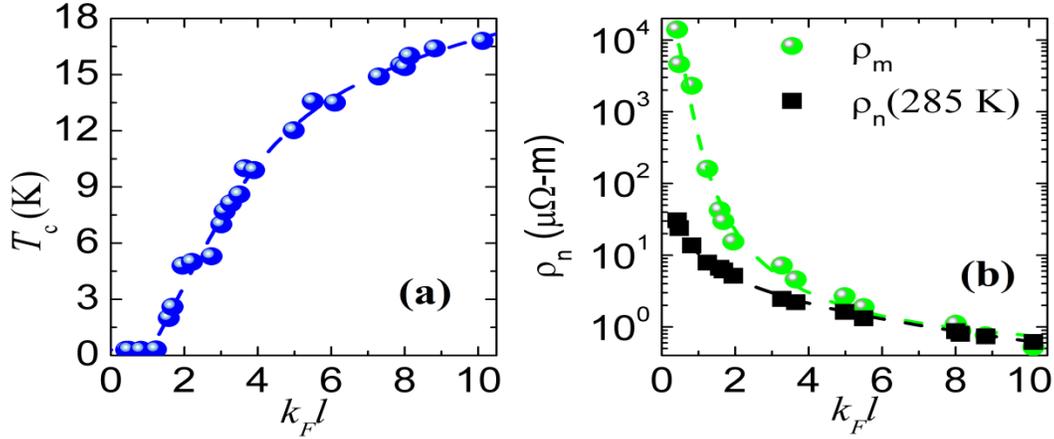

The superconducting transition temperature ($T_c$) is defined as the temperature at which resistance drops below the measurable limit of our measurement setup. The Fig. 2.4(a) shows the variation of $T_c$ with $k_F l$. For sample with $k_F l$ > 1, $T_c$ varies from 16.5 K to below 300 mK as $k_F l$ varies from 10.12 to $k_F l$ ~ 1. The samples with $k_F l$ < 1 remain non-superconducting down to 300 mK.

The normal state resistivity, $\rho_n$ at 285 K and maximum resistivity, $\rho_m$ as function of $k_F l$ are shown in Fig. 2.4(b). Here maximum resistivity, $\rho_m$ was taken as the peak value above $T_c$ for superconducting sample and for non superconducting sample with $k_F l$ < 1, the resistivity at 300 mK was considered as the maximum resistivity, $\rho_m$. The Fig. 2.4(b) shows that value of $\rho_m$ rapidly increases by 5 orders of magnitude with increasing disorder.

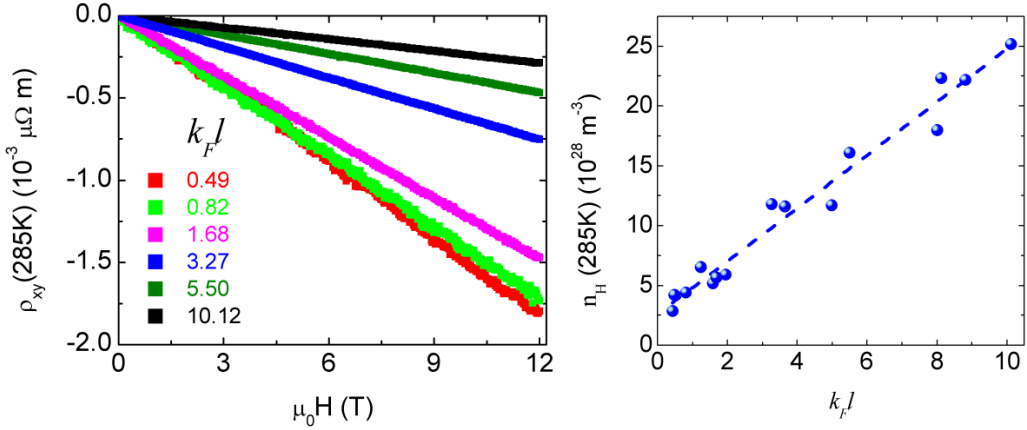

Figure 2.5. (a) shows the Hall resistivity $\rho_{xy}$ as function of magnetic field $H$ for a set of films with different level of disorder. (b) Hall carrier density, $n_H$ as function $k_F l$ for a set NbN films with thickness, $t \geq 50$ nm (figure is adapted from ref.3).

### 2.4.1.2. Hall carrier density ($n_H$)

The Fig. 2.5(a) shows $\rho_{xy}$ as function $H$ for a set of samples as $k_F l$ varies from 10.12 to 0.49. The extracted Hall carrier density, $n_H$ from measured Hall coefficient, $R_H$ as function of $k_F l$ is shown in Fig. 2.5(b). Since the relation, $R_H = -1/n_H e$ is not valid in presence of electron-electron (e-e) interaction, the $n_H$ was calculated at the highest temperature, 285 K possible in our measurement setup, considering the e-e interaction is small at high temperature [8]. The measured carrier density for stoichiometric NbN films with $T_c$ ~16.5 K, is in excellent agreement with band structure calculation [1]. The carrier density decreases by factor of 10 as the disorder increases from clean limit with $k_F l$ ~10.12 to strong disorder limit with $k_F l$ ~ 0.42. The band structure calculation shows that the only 4d orbital of Nb atom contributes to the conduction electrons which explains the decrease in carrier density with increasing disorder in the form of Nb vacancy in crystalline NbN films (see Fig. 2.5.(b)).

### 2.4.1.3. Upper critical field ($H_{c2}$) and GL coherence length ($\xi_{GL}$)

The upper critical filed, $H_{c2}(0)$ and extracted GL coherence length, $\xi_{GL}$ from $H_{c2}(0)$ are shown in Fig. 2.6 [3,6]. . The $H_{c2}(0)$ and $\xi_{GL}$ show nonmonotonic behavior with increasing disorder. This unusual behavior can be understood by taking into consideration the competition between the BCS coherence length, $\xi_{BCS}$ ($\approx \hbar v_F / \pi \Delta$) and mean free path, $l$. Since the $\xi_{BCS}$ does not change much in low disorder limit with $k_F l \geq 5.5$ but $l$ rapidly decreases with increasing disorder, therefore $\xi_{GL} \approx (\xi_{BCS} l)^{1/2}$ decreases and accordingly $H_{c2}(0)$ increases. In strong disorder limit with $k_F l \leq 5.5$, the $\xi_{BCS}$ increases rapidly which explains the increase in $\xi_{GL}$ and the decrease in $H_{c2}(0)$.

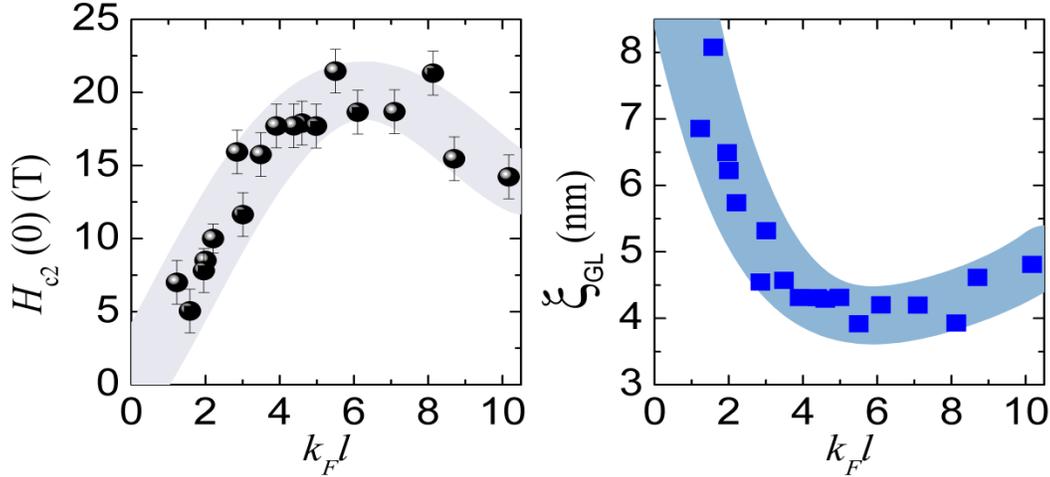

Figure 2.6. (a) Upper critical field $H_{c2}(0)$ as function of $k_F l$ for set of films with different level of disorder. (b) GL coherence length, $\xi_{GL}$ as function of $k_F l$ (figure is reproduced from ref. 3 and 6).

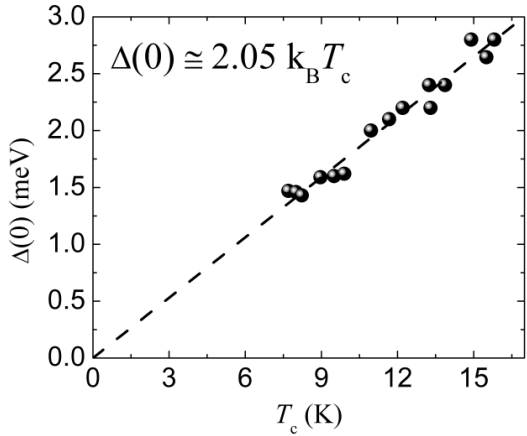

Figure 2.7. Superconducting energy gap, $\Delta(0)$ measured using planer tunnel junctions for a set of relatively less disordered NbN films with $T_c \geq 8.1$ K. The black dashed line is the linear fit to the data (figure is adapted from ref. 9).

### 2.4.1.4. Superconducting energy gap (Δ)

The Fig. 2.7 shows the superconducting energy gap, $\Delta(0)$ in the low disorder limit as function of $T_c$ measured by tunneling experiments at low temperature ($T < 0.2\ T_c$) on planar tunnel junctions fabricated using NbN thin films with different levels of disorder [3,9]. The disorder scattering increases the Coulomb interaction thus decreases the pairing energy $\Delta(0)$. It was observed that $\Delta(0)$ follows the relation, $\Delta(0) \cong 2.05\ k_B T_c$ for disordered NbN films with $T_c \geq 8$ K.

### 2.4.1.5. Magnetic penetration depth (λ)

Magnetic penetration depth of bulk NbN, $\lambda(0) \sim$ 250 nm is much greater than its coherence length $\xi_{GL} \sim 5$ nm, thus NbN is a type II superconductor. The Fig. 2.8 shows the experimentally measured magnetic penetration, $\lambda_{Exp.}(0)$ using two coil mutual inductance technique (see section 3.2) and the theoretically calculated magnetic penetration depth, $\lambda_{BCS}(0)$ using BCS relation (see the section 1.4.2.1)[10],

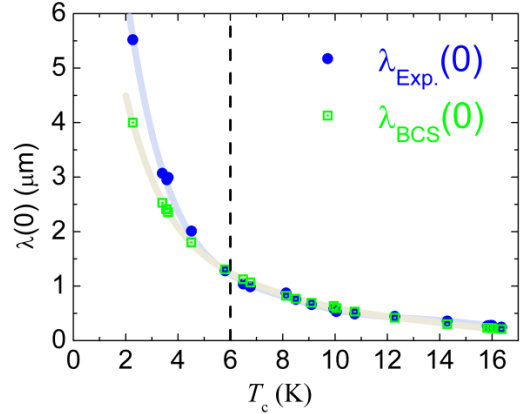

Figure 2.8. shows the magnetic penetration depth, $\lambda(0)$ for a set of films with $T_c$ varying from 2.27 K to 16.5 K.

$$\lambda_{BCS}^{-2}(0) = \frac{\pi\mu_0\Delta(0)}{\hbar\rho_m},$$ (2.3)

Here $\Delta(0)$ and $\rho_m$ experimentally measured superconducting energy gap at $T\rightarrow0$ and maximum resistivity just above $T_c$ respectively. The $\lambda_{Exp.}(0)$ matches quite well to $\lambda_{BCS}(0)$ (see Fig. 2.8.) for low disordered samples with $T_c > 6$ K, however as the disorder increases, $\lambda_{Exp.}(0)$ becomes progressively larger than $\lambda_{BCS}(0)$ in strongly disordered samples with $T_c \leq 6$ K. The deviation of $\lambda_{Exp.}(0)$ from $\lambda_{BCS}(0)$ can be understood by taking into account the effect of quantum phase fluctuations (see section 1.6.2) which will be explored in details in chapter 5.

### 2.4.2. Appearance of pseudogap state

In this section I will review our observation of PG state in strongly disordered NbN thin films [3,4,9], using scanning tunneling microscope (STM) and magneto-resistance (MR) measurements (MR) carried out in collaboration with other group members, namely, Anand Kamlapure (STM), Garima Saraswat (STM) and Madhavi Chand (MR).

### 2.4.2.1. Tunneling spectroscopy using STM

Fig. 2.9 shows the intensity plot of tunneling DOS (average over 32 equally spaced points along 150 nm line) as a function of bias voltage, $V$ for a set of films with $T_c$ varying from 11.9 K to below 300 mK (upper part of each panels). The corresponding resistance ($R$) vs temperature ($T$) plots in lower part of each panel. The color scale shows the conductance value normalized to the conductance at high bias voltage. The raw tunneling spectra for all samples, shows superconducting feature riding over a broader V-shaped background coming from Altshuler-Aronov(AA) type electron-electron interaction, extended up to high bias voltage [5,9]. In low disordered sample this V-shaped background is very small, however it becomes progressively pronounced as the disorder increases. The final tunneling conductance spectra (shown in Fig.

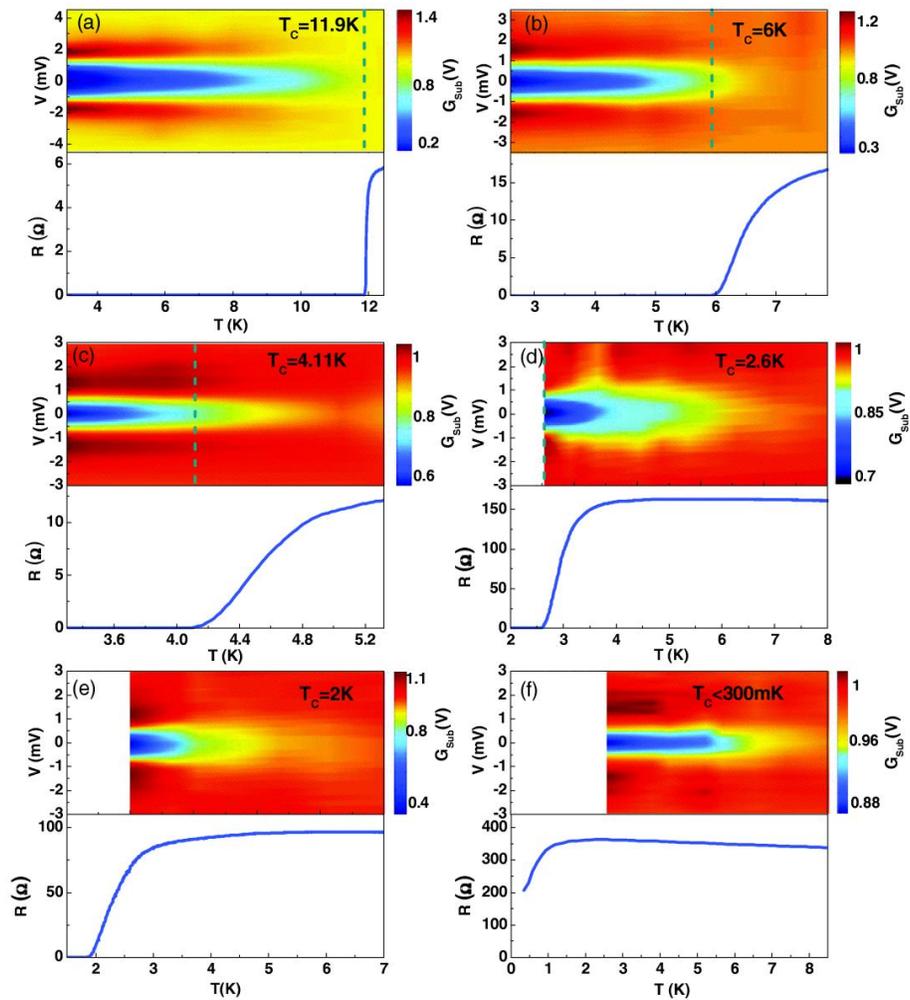

Figure 2.9. shows the color plots of normalized differential conductance, $G(V)$ as a function of bias voltage, $V$ for six samples with different level of disorder (upper part of each panel) and the resistance as a function of $T$ in the same temperature range for same set of samples (lower part of each panel). The vertical dashed lines in the upper parts correspond to $T_c$ (figure is adapted from ref. 5).

2.9.) was obtained after subtracting the nearly temperature independent V-shaped back ground measured at high temperature, $T > T^*$ from the raw conductance spectra.

Tunneling studies show that the energy gap in tunneling DOS, $\Delta$, manifestation of pairing energy in accordance with BCS theory, goes to zero as $T$ goes to $T_c$ in low disordered samples (for example see the tunneling DOS for sample with $T_c \sim 11.9$ K in Fig. 2.9.(a)). However when the disorder increases further, $\Delta$ remains finite up to temperature, $T^*$ well above $T_c$ in a strongly disordered system contrary to BCS prediction. For example, the most disordered sample with $T_c$ below 300 mK (Fig. 2.9.(f)), shows dip in tunneling DOS up to temperature, $T^* \sim 6$ K. Similar type of PG is also observed in some other strongly disordered superconductors such as TiN and $InO_x$ [11,12].

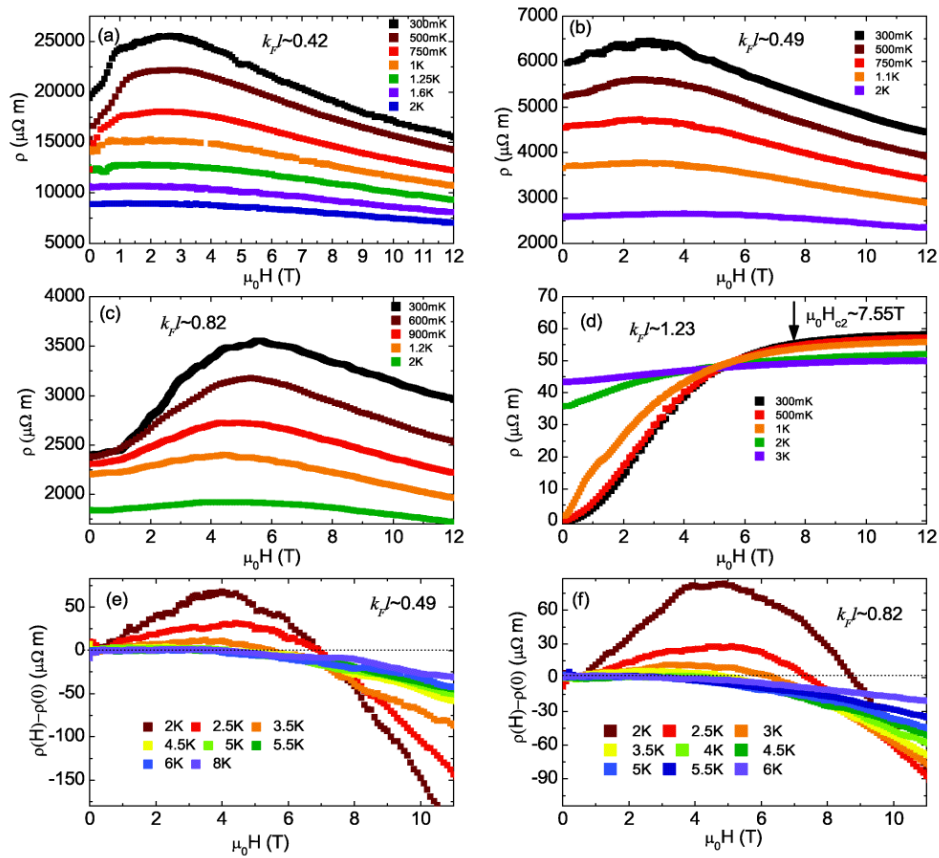

Figure 2.10. Resistivity ($\rho$) as a function of $H$ for four strongly disordered samples with (a) $k_F l \sim 0.42$, (b) $k_F l \sim 0.49$, (c) $k_F l \sim 0.82$ and (d) $k_F l \sim 1.23$. (e) and (f) show the expanded view of $\Delta\rho(H) = \rho(H) - \rho(0)$ as function of $H$ for samples with $k_F l \sim 0.49$ and $k_F l \sim 0.82$. MR data shows pronounced peak for all samples with $k_F l < 1$ (figure is adapted from ref. 5).

#### 2.4.2.2. Magneto resistance (MR) measurements

The Fig. 2.10 shows $\rho(H)$ vs $H$ for a set of disordered films with $k_Fl \sim 0.42$, 0.49, 0.82 and 1.23. All strongly disordered samples with $k_Fl < 1$, show a pronounced peak at characteristic magnetic field, $H_p$ and the peak gradually disappears with increasing temperature. The Fig. 2.10. (e) and (f) show the expanded view of temperature evolution of MR peak for samples with $k_Fl \sim 0.49$ and 0.82. This type of MR peak was also observed in other strongly disordered superconductors such as $InO_x$ [13,14], TiN [15] etc. The temperature evolution of MR peak shows that MR peak disappears at a temperature close to the PG temperature, $T*$ for the most disordered sample on which STS measurements were carried out. The MR peak is less pronounced in low disordered sample and disappears for sample with $k_Fl > 1$. The sample with $k_Fl \sim 1.23$, shows superconducting transition at 0.6 K and above $T_c$, a positive MR was observed due to superconducting fluctuations persist above $T_c$. Since the MR peak is not expected in disordered metal, therefore observed MR peak which vanishes at close to PG temperature, $T*$, suggest the presence of superconducting correlations in nonsuperconducting samples with $k_Fl < 1$.

## 2.4.3. Phase diagram of NbN

Based on transport, magneto transport and tunneling measurements, a phase diagram of disordered NbN thin films was proposed which is shown in Fig. 2.11 [5]. The phase diagram shows three distinct regimes of disorder.

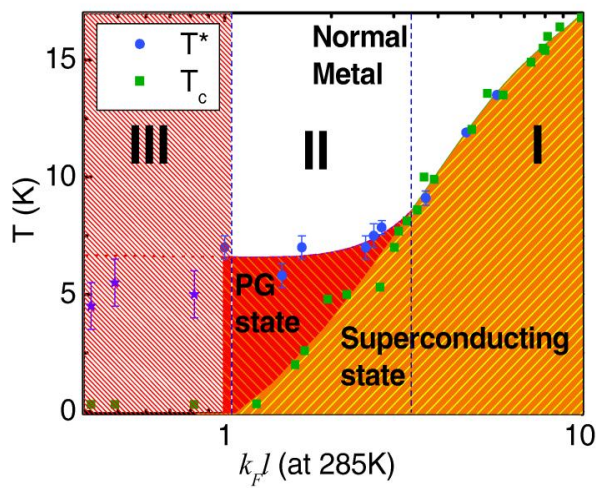

Figure 2.11. Phase diagram of disordered NbN films showing the appearance of pseudogap (PG) state in strongly disordered NbN films. The three regimes with increasing disorder are shown as I, II, and III. A PG state emerges between $T_c$ and $T*$ for samples with $T_c \leq 6$ K (Regime II). The temperature at which the peak in the MR vanishes for the strongly disordered samples (Regime III) is shown with purple stars (figure is adapted from ref. 5).

The intermediate disordered regime (marked as **I**), characterized by single energy scale $T_c$ where $T_c$ monotonically decreases with increasing disorder, however superconductors continue to follow conventional BCS behavior. The suppression of $T_c$ is mainly due to weakening of pairing interactions due to increase in repulsive e-e Coulomb interactions with increasing disorder. Apart from e-e interactions, localization of electronic states also affects the $T_c$ in disorder superconductor. The combining effect of e-e interactions and localizations suppress the $T_c$ in regime **I**, where superconducting energy gap, $\Delta$ in tunneling DOS disappears precisely at $T_c$.

With further increase in disorder, the superconductor enters in regime **II** where superconducting state characterized by two energy scale; $T^*$ corresponds to the temperature where the energy gap in tunneling DOS appears and $T_c$, where resistance goes to zero. The superconducting transition temperature, $T_c$ continues to decrease monotonically with increasing disorder, however the $T^*$ remains almost constant down to $k_F l \sim 1$ where superconducting ground state is destroyed. The phase diagram shows a pseudogap state (see Fig. 2.11.) characterized by a gap in electronic DOS in the temperature range, $T_c < T < T^*$, appears in strongly disordered samples with $T_c \leq 6K$.

As the disorder increases further, the superconductor enters in regime **III** where all samples are nonsuperconducting and characterized by a broad MR peak originated from superconducting correlations.

## 2.5. Effect of reduced dimensionality

The aim of this study was to observe the Berezeski-Kosterlitz-Thouless (BKT) transition due to unbinding of thermally activated vortex-antivortex pairs [17,18] in very low disordered ultrathin NbN films when thickness, $t$, becomes comparable to coherence length ~ 5 nm.

It has been observed that for films with thickness, $t < 50nm$, $T_c$ gets gradually suppressed [16] from its bulk value with decreasing thickness. With decreasing film thickness, the superconducting properties can be affected by various effects, such as phase disordering due to the unbinding of thermally activated vortex-antivortex pairs [17,18], formation of an inhomogeneous superconducting state [19], thermal phase fluctuations [20], eventually even resulting in Superconductor- Insulator transition [21,22].

In this study the effect of reduced thickness was investigated by reducing film thickness and keeping the structural and compositional disorder levels same. For that purpose, first the optimal deposition condition was achieved and then the deposition time was varied by keeping other deposition conditions fixed to get the sample of desired thickness.

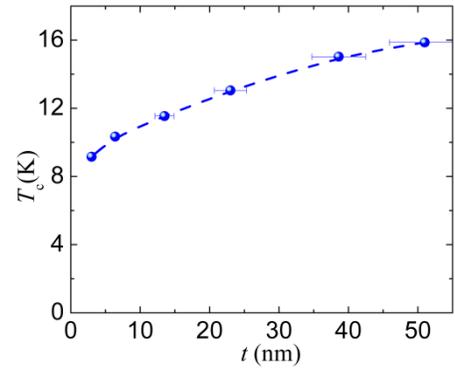

Figure 2.12. Variation of $T_c$ with film thickness for epitaxial NbN thin films.

The investigation was carried out through measurements of $\lambda(T)$ using two coil mutual inductance technique (see section 3.2) and $\Delta(T)$ using STM as a function of $T$ in epitaxial NbN thin films with thickness, $t$ varying between 3 to 51 nm [23]. Fig. 2.12 shows the variation of superconducting transition temperature, $T_c$ with film thickness. The $T_c$ varies from 15.87 K for a 50nm thick film to 9.16 K for 3 nm thick film.

## 2.5.1. Scanning tunneling spectroscopy

The Fig. 2.13 shows the STM measurements on the films with thickness, $t \sim 50$ nm and 5 nm. While the morphology of the thicker films ($t > 10$ nm) show a coarse surface (inset Fig. 2.13 (c)), for $t$<10nm step-like structures was observed (inset Fig. 2.13(d)), reflecting the step edges on the single crystalline MgO substrate. Figures 2.13(a-b) show the tunneling conductance spectra normalized to conductance measured at high bias voltage (~7 mV) at various temperatures. $\Delta$ is extracted by fitting these spectra using following tunneling equation,

$$G(V) = \frac{dI}{dV}\bigg|_V = G_N \frac{d}{dV}\left\{\int_{-\infty}^{\infty} N_s(E)\big\{f(E) - f(E - eV)\big\}dE\right\}, \qquad (2.4)$$

$$N_s(E) = \mathrm{Re}\left\{\big(|E| - i\Gamma\big)\Big/\big(\big(|E| - i\Gamma\big)^2 - \Delta^2\big)^{1/2}\right\}, \qquad (2.5)$$

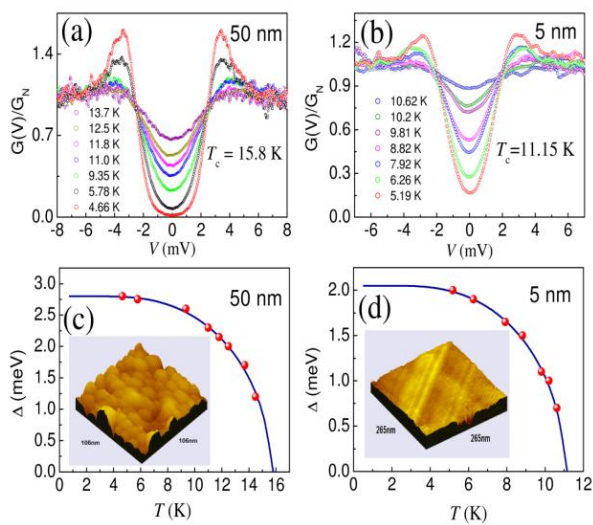

Figure 2.13.(a-b) show the temperature evolution of normalized tunneling conductance spectra for NbN films with thickness 50nm and 5nm respectively. The solid lines show the theoretical fits. (c-d) $\Delta(T)$ as function of $T$ (scattered plots) along with the expected BCS behavior (solid lines); the *insets* show the topographic image (106nm by106nm for (c) and 265nm by 265nm for (d)) at 5K of the surface (figure is adapted from ref. 23 ).

Here the broadening parameter $\Gamma(= \hbar/\tau)$ is formally incorporated [24] to take into account the lifetime ($\tau$) of the quasi particles which phenomenologically incorporates all sources of non-thermal broadening in the BCS DOS. Fig. 2.13(c-d) shows the temperature variation of $\Delta$ for these two films. For all films in this study we observe that the temperature variation of $\Delta$ closely follows the BCS behavior within experimental accuracy, even though NbN is a strong coupling superconductor with $2\Delta(0)/k_B T_c \sim 4.1$, there is no significant deviation in the temperature dependence of $\Delta(T)$ from weak coupling BCS behavior.

## 2.5.2. Magnetic penetration depth

Figure 2.14(a) shows the temperature variation of $\lambda^{-2}(T) \propto n_s(T)$ (where $n_s$ is the superfluid

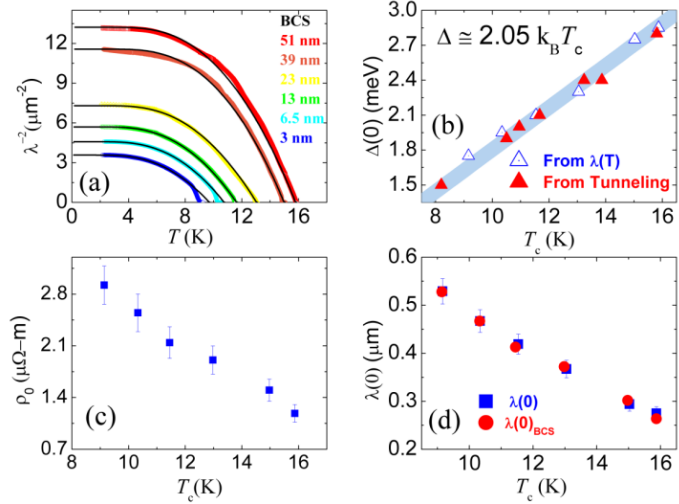

Figure 2.14. $\lambda^{-2}(T)$ vs $T$ for set of NbN films with different thickness; (b) The extracted $\Delta(0)$ from tunneling measurements (hollow triangle) and penetration depth (solid triangle) as a function of $T_c$; (c) $\lambda(0)_{exp}$ (solid square) and $\lambda(0)_{BCS}$ (solid circle) as a function of $T_c$; (d) normal state resistivity ($\rho_0$) just above $T_c$ as a function of $T_c$ (figure is adapted from ref. 23 ).

density). The λ increases with decreasing thickness from 275nm to 529nm in the thickness range 51nm to 3nm. The value of λ for the 51nm thick film is consistent with earlier measurements which vary between [25] λ~200–400nm for NbN films with similar thickness. Since in NbN films, the electronic mean free path, $l<<\xi$, temperature variation of $\lambda^{-2}(T)$ was fitted with the dirty limit BCS expression (see section 1.4.2.1) [26],

$$\frac{\lambda^{-2}(T)}{\lambda^{-2}(0)} = \frac{\Delta(T)}{\Delta(0)} \tanh\left(\frac{\Delta(T)}{2k_B T}\right), \tag{2.6}$$

where $\Delta(0)$ was used as a fitting parameter. The theoretical fits are excellent except for the two thinnest films where an abrupt drop in $n_s$ was observed close to $T_c$, associated with the BKT transition. The best fit values of $\Delta(0)$ plotted (Fig. 2.14(b)) along with the ones obtained from tunneling measurements (as a function of $T_c$) agree well with each other. A further consistency check was obtained by calculating $\lambda(0)_{BCS}$ using the dirty-limit BCS relation (eqn. 2.3.). The agreement between experimental values of $\lambda(0)$ and $\lambda(0)_{BCS}$ calculated [27] using $\rho_0$ (Fig. 2.14(d)) and $\Delta(0)$ (Fig. 2.14(b)) suggests that the evolution of the ground state properties of thin NbN films can be understood from weakening of the electron-phonon pairing interaction, possibly due to the increase in the Coulomb pseudopotential arising from loss of effective screening.

## 2.5.3. Observation of BKT transition

Now we concentrate on temperature variation of $\lambda^{-2}(T)$ close to $T_c$ more closely. While films with thickness, $t > 10$ nm show no significant departure from BCS theory, the two thinnest samples show a sharp drop in the $\lambda^{-2}(T)$ close to $T_c$ (Fig. 2.15). This abrupt drop in $n_s$ is the hallmark of Berezeski-Kosterlitz-Thouless (BKT) transition in superconducting films with thickness comparable or smaller than $\xi$ [17]. However, in these samples the transition happens at a superfluid density which is almost 4 times larger than the expected value from BKT prediction given by [17],

$$\lambda^{-2}(T) = \frac{8\pi\mu_0 k_B T}{d\phi_0^2},$$  (2.7)

where $\phi_0$ is the flux quantum. The same type of apparent discrepancy was also observed in other systems such as layered high-$T_c$ superconductors [28], InO$_x$ [29] etc. Since there are various additional effects in real superconductors such as intrinsic disorder, granularity, electron-electron interactions etc, which can modify the nature of BKT transition originally studied in 2D XY model [30], it demands further investigations to find out the true nature of BKT transition in real 2D superconductors.

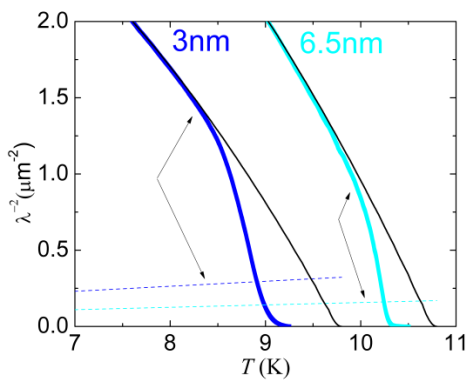

Figure 2.15. Temperature variation of $\lambda^{-2}(T)$ close to $T_c$ for 6.5nm and 3nm films. The solid lines show the expected temperature variation from BCS theory. Intersection with the dashed line is where the BKT transition is predicted (figure is reproduced from ref. 23 ).

## 2.6. Summary

*(i) Effect of disorder:*

In summary, with increasing disorder in 3D epitaxial NbN films evolves from conventional BCS superconductors in moderately clean limit to a situation where superconducting state is characterized by pseudogap state above $T_c$ in strongly disorder limit. The superconducting properties of moderately disordered films with $T_c \geq 6$ K (regime **I** in phase diagram), follow roughly BCS behavior. In the intermediate disorder limit (regime **II** in phase diagram), $T_c$ monotonically goes to zero with increase in disorder, however the pseudogap temperature, $T^*$ almost remains constant. In strongly disorder limit (regime **III** in phase diagram), the samples become nonsuperconducting characterized by a pronounced MR peak. The observation of pseudogap in more conventional strongly disordered superconductors raised the new questions about the mechanism which leads to the destruction of superconductivity in strongly disordered superconductors. In chapter 5, I will explore this issue through measurements of electrodynamics response of strongly disordered NbN thin films.

*(ii) Effect of reduced dimensionality:*

When the thickness reduces the low temperature values of $\Delta(0)$ and $\lambda(0)$ are governed primarily by weakening of the pairing interaction which in turn reduces the $T_c$. However, as the film thickness becomes comparable or smaller than $\xi_0$, the temperature variation of the superfluid density deviates from dirty limit BCS behaviour and shows an abrupt drop close to $T_c$, related to the phase disordering BKT transitions. However the jump in superfluid density appears to be at much higher value of superfluid density and lower temperature than the expected values within BKT formalism. This apparent nonuniversal nature of BKT transition in real 2D superconductors will be explored in details through measurement of $\lambda$ using low frequency mutual inductance technique in chapter 4.

# CHAPTER 3

## Experimental details

## 3.1. Sample preparation

To study the effect of phase fluctuations, epitaxial NbN thin films appears as an elegant system for its simple structure, durability and ease of fabrication. Epitaxial NbN thin films can be grown using popular thin film deposition techniques such as sputtering [1,2,3,4] and pulse laser deposition[5,6]. In this study NbN films were grown using reactive dc magnetron sputtering.

### 3.1.1. Sputtering

Sputtering is a very popular thin film deposition technique which relies on ejection of ions, atoms and clusters from target through bombardment by accelerated charge particles (usually inert gas Ar) and condensation on the substrate located in front of the target. It is a very fast and flexible deposition technique and provides very good control over deposition conditions and film thickness.

In this technique, a high negative voltage applied on the target (in our case Nb) produces plasma of positive ions of Ar gas introduced inside a vacuum chamber. These positive ions accelerate towards the target due to high negative voltage applied on the target and bombard with considerable momentum to knock out materials in the form of atoms and clusters from the target. These ejected materials travel across the plasma and condense on the substrate placed in front of the target in the form of thin film. During the bombardment process in addition to neutral materials, some charged ions, secondary electrons and photons are also emitted from the target. Some of the positive ions recombining with secondary electrons emit photons which make the plasma to glow with beautiful color depending on the gases inside the chamber.

To create plasma in the system, dc or rf voltage can be applied for electrically conducting target. However for an insulating target rf voltage is necessary. To guide the plasma, a static magnetic field is introduced by placing an array of small powerful magnets (in our case NdFeB magnets) behind the target. The magnetic field modifies the path of charged particles into closed loop to provide the repetitive bombardment resulting higher sputtering yield. In some cases, to produce thin film of compound of target materials, a reactive gas of choice (such as $O_2$, $N_2$ etc) is introduced in combination with inert gas in the chamber. The reactive gas reacts chemically with the target materials and form compound. The chemical reaction may take place at the target surface or during the flight from target to substrate or during the condensation on the substrate. Chemical compositions of the compound can be controlled by varying the relative partial pressure of reactive gas and inert gas. The reactive sputter deposition is the most popular technique for fabrication of oxide and nitride films.

The properties of thin films and sputtering yields strongly depend on the deposition conditions such as nature of substrate, sputtering power, sputtering pressure, different gases used, partial pressures of different gases, target to substrate distance, angle of incidence on the substrate, substrate temperature etc.

## 3.1.2. Fabrication of NbN thin films

NbN thin films were fabricated using reactive dc magnetron sputtering by sputtering Nb in $Ar/N_2$ gas mixture atmosphere using the sputtering system manufactured by Excel Instruments [7]. The sputtering chamber has four ports along the side for loading sputtering gun or viewing inside the chamber. It has one port at top of it which is used for substrate mounting and there are some additional smaller ports used for letting in the sputtering gas, venting, pressure gauge etc. In our system target to substrate holder distance was about 6.5 cm.

To fabricate NbN thin films, first a circular shaped target (2" diameter) of 99.999% pure Nb metal manufactured by Kurt-Lesker [8] was mounted on the sputtering gun. To get epitaxial NbN films, one side polished (100) oriented single crystal MgO substrate was used for having lattice constant, 4.212 Å closely matching with NbN.

To get the sample of desired size, the MgO substrates were cut using diamond cutter across the lattice plane. Then the substrate was cleaned using trichloroethylene (TCE) as cleaning solvent in an ultrasonic cleaner. The cleaned substrates were then loaded onto the sample holder using high temperature silver paste. Patterning of samples were done whenever necessary using shadow masking where a stainless steel mask of desired shape was placed on top of the substrate with the help of two stainless clips.

Once the sample was loaded, the chamber was evacuated using a Varian turbo pump followed by a rotary connected through a gate valve at the bottom of the chamber. After pumping for about 20 minutes, when vacuum reachs to roughly $\sim 5 \times 10^{-5}$ Torr, the substrate temperature was raised to 600 $^o$C using an embedded nichrome wire heater attached under the flat sample holder plate. For our all depositions, base pressure was achieved about $\sim 3.5 \times 10^{-6}$ Torr which takes typically 3 to 3.5 hours of pumping at full pumping speed of our turbo pump.

When the desired vacuum ($\sim 3.5 \times 10^{-6}$ Torr) was achieved, the rotation speed of the turbo pump was reduced to 70% of its maximum speed using standby mode of the turbo controller and then the mixture of Ar and $N_2$ gases was let in the chamber. The both gases are highly pure with purity level was 99.999%. The partial pressures of Ar and $N_2$ were controlled using two mass flow controllers. Once gases were introduced in the chamber, the total pressure was controlled by gate valve and was tuned to $5 \times 10^{-3}$ Torr in our all depositions.

When the system was stabilized after few minutes (typically 5 to 6 minutes) of introducing gases, a negative high voltage (typically about 0.35 kV) was applied using an Aplab[9] high voltage power supply to strike the plasma. The breakdown voltage of gas mixture depends on the nature of the gases, total pressure and also on the partial pressures of different gases used. The breakdown voltage for 84:16 (Ar:$N_2$ ) gas ratio at total pressure $5 \times 10^{-3}$ Torr is about 0.35 kV and it increases with decreasing partial pressure of Ar. Once the plasma was created, the sputtering power was controlled in the range from 15W to 250W by regulating the current passing through the plasma. Pre-sputtering was carried out for about 3 minutes, before the actual deposition to remove any impurity on the surface of the target. The substrate was then exposed for deposition for the desired time by opening the shutter used for blocking the materials reaching the substrate. After the deposition, the sample was cooled down in vacuum to room temperature before removing from the chamber.

Properties of NbN thin films and level of disorder in the films strongly depend on the deposition conditions. In this study, the sputtering power, Ar/$N_2$ gas ratio and time of depositions were varied to get the film of desired properties.

To study the effect of reduced dimensionality, the deposition conditions were optimized to obtain the highest possible $T_c$ ($\sim 16.5$ K) for a 50 nm thick film. Then the thickness ($t$) of films was varied for a fixed disorder level by changing the deposition time and keeping other deposition conditions fixed.

The disorder, in the form of Nb vacancies in the crystalline NbN lattice, was controlled by changing the sputtering power or $Ar/N_2$ ratio in the gas mixture. To study the effect of disorder, we deposited a set of films where disorder level was tuned by changing the deposition conditions by keeping the thickness, $t \geq 50$ nm such that all our films are in 3D limit ($t >> \xi$).

### 3.1.3. Thickness measurement

Thickness of all films was measured using Ambios XP2 Stylus profilometer. Thickness measurements were carried out at various positions of the sample and the mean value was taken as the film thickness. The Fig. 3.1. shows the thickness as function of deposition time for a set of films deposited under identical deposition conditions. The variation of 10% was observed in measured thickness values at different positions. Since the resolution limit of our profilometer ~10 nm and the thicknesses of films are close to the resolution limit, this variation was more likely from the measurement error and not from nonuniform thickness of our films. This is the major source of error in any measured quantity related to thickness such as resistivity, Hall effect etc. The thickness of thinner films was determined from the deposition time using linear fit passing through the origin for the measured thickness vs deposition time plot for films with thickness $(t) \geq 20$ nm.

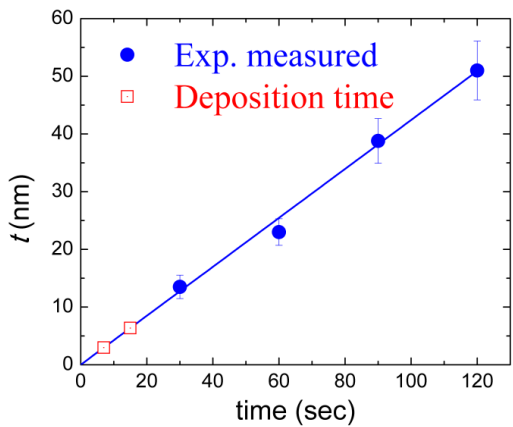

Figure 3.1. Thickness, $t$ as function of deposition time. The (blue) solid circular scattered plot is for measured thickness and (red) open square data points represent the estimated thickness from deposition time.

## 3.2. Low frequency electrodynamics response

Low frequency electrodynamics response provides important information about the superfluid density[10]. Historically low frequency electrodynamics response of superconductors has been investigated mainly through measurements of magnetic penetration depth, $\lambda$ using various techniques such as $\mu$SR measurements [11,12], self inductance method [13,14], mutual inductance technique [15,16], magnetic force microscopy and scanning SQUID microscopy [17,18] etc. In our lab, we have developed a two coil mutual inductance setup following S. J. Turneaure et al [19] to study the low frequency electrodynamics response through measurement of absolute value of $\lambda$.

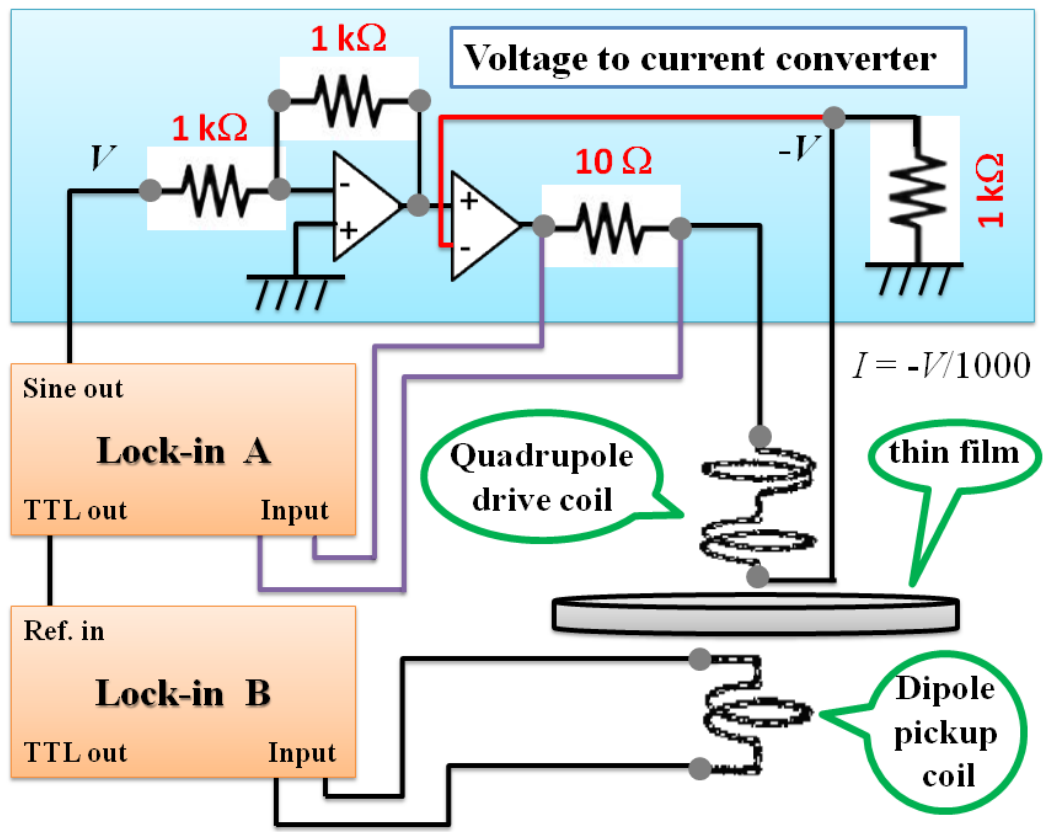

Figure 3.2. Schematic diagram of our two coil mutual inductance setup. The following Fig. 3.3. is the image of the two coil probe of our apparatus. For details about the coil configuration please see the text in section 3.2.1.2.

### 3.2.1. Two coil mutual inductance technique

The two coil mutual inductance technique is a very powerful nondestructive method for measuring λ of SC films [19,20,21,22]. The main advantage of this technique is that the absolute value of λ can be measured over the entire temperature range up to $T_c$ without any model dependent assumptions.

### 3.2.1.1. Experimental method

In our measurement setup, a SC film is sandwiched coaxially between a quadrupole primary coil and a dipole secondary coil (see Fig. 3.2. and Fig. 3.3.). The mutual inductance, $M = M_1 + iM_2$ between primary and secondary coil is measured as function of temperature by passing a small ac excitation current, $I_d$ of frequency, $f = \omega/2\pi$ through the primary coil and measuring the induced voltage, $V_p$ at secondary coil using lock-in amplifier,

$$M = M_1 + i\,M_2 = \frac{V_p}{\omega I_d} \tag{3.1}$$

Here, the first term corresponds to the inductive coupling and second term corresponds to the resistive coupling between two coils. Beside geometry of coil setup and SC film, the mutual inductance depends on the complex screening length, $\lambda_\omega$ of the film given by, $\lambda_\omega = (i\mu_0\omega\sigma)^{-1/2} = (\lambda^{-2} + i\,\delta^{-2})^{-1/2}$ where $\delta = (\mu_0\omega\sigma_1)^{-1/2}$. Except very close to $T_c$, the response of SC thin film is purely inductive, therefore $\lambda_\omega = (\mu_0\omega\sigma)^{-1/2} \approx (\mu_0\omega\sigma_2)^{-1/2} = \lambda$. Close to $T_c$, finite $\sigma_1$ contributes to the resistive coupling, therefore $M$ becomes complex.

In our experiments, the drive coil is supplied with a small ac current (~ 0.5mA) ensuring the excitation magnetic field is very low (~ 3.5 mOe). The drive ac current of frequency, $f = 60$

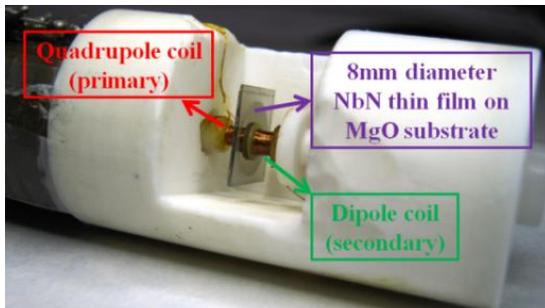

Figure 3.3. Coil assembly of our low frequency mutual inductance setup. The quadrupole (primary) coil has total 28 turns with the half closer to the film wound in one direction and the farther half wound in opposite direction. The dipole (secondary) coil has 120 turns wound in the same direction in 4 layers.

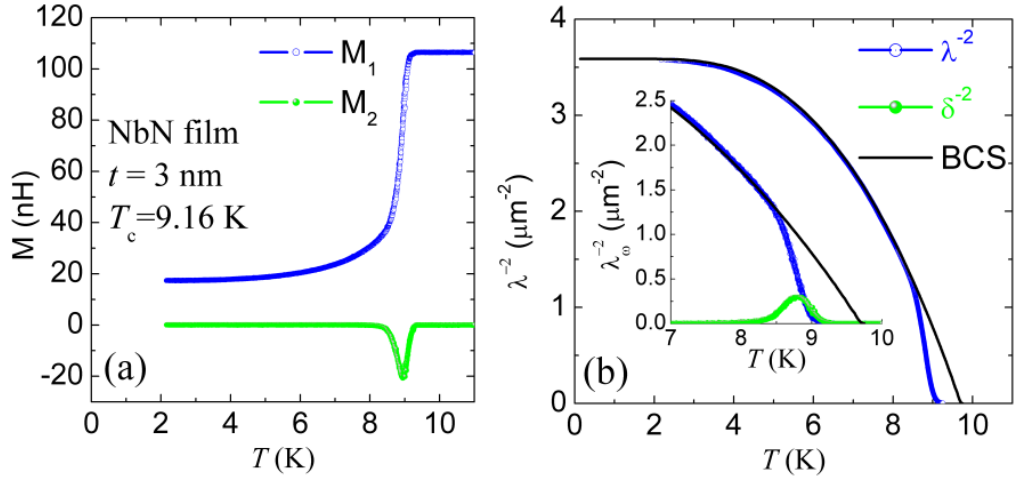

### 3.2.1.2. Coil description

The image of the probe head of our coil assembly is shown in Fig. 3.3. In our coil setup, the primary drive coil is attached with the cylindrical Macor* sample housing with the help of thread arrangement. The secondary coil is pressed from bottom using a spring loaded piston made of Macor. The spring arrangement help us to hold the SC film coaxially placed in between the drive and pickup coils and also takes care the thermal contraction of bobbins. The whole coil assembly is kept inside a heater can made of Cu in He gas environment which provides very good thermal equilibrium.

The primary drive coil (quadrupole) is coaxially centered above the film, has 28 turns in one layer. The secondary pickup coil (dipole) is coaxially centered below the film, has 120 turns in 4 layers. In the primary coil, the half of the coil nearer to the film is wound in one direction and the farther half is wound in the opposite direction. The radiuses of the drive and pickup coil are $r_d$ ~1.0 mm and $r_p$ ~1.0 mm respectively. The separation between the film and the nearest turn in the drive and pickup coils are given by $h_d \sim S_d + 0.4$ mm and $h_p \sim 0.4$ mm respectively where $S_d$ is the substrate thickness. The location of every turn of a coil can be found by adding multiples of the wire separation $dh_d \sim 71.4$ μm for the drive and $dh_p \sim 66.6$ μm for the pickup coil to the $h_d$ and $h_p$ respectively. Length of two coils are $L_d = 2.0$ mm and $L_p = 2.1$ mm. The other important geometrical quantities are the radius $R$ and the thickness $t$ of the film which changes from sample to sample.

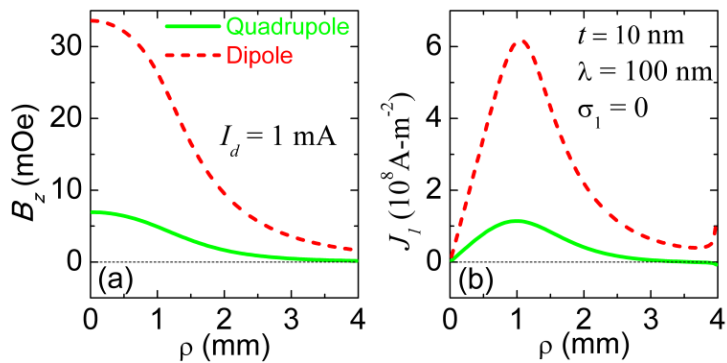

Figure 3.5. (a). Amplitude of ac magnetic ($B_z$) and (b) induced current density ($J_1$) as a function of function of radial position (ρ) for quadrupole and dipole drive coil (see text for details).

---

* Macor is a machineable glass-ceramic developed by Corning Inc. To remove the eddy current effect induced by ac drive field, Macor was used instead of Cu or any other metal.

I would like to note that for measurements in $^3$He cryostat at base temperature 0.3 K, we have developed a similar setup. The coil dimension of the bobbins of this setup is slightly different.

The Fig. 3.5.(a) shows the numerically calculated radial distributions of amplitude of ac magnetic field ($B_z$) created by ac drive current (~1 mA). The (red dashed) line represents the magnetic field distribution of a dipole coil with same geometrical configuration as our quadrupole coil except the coil is wound in one direction. We can see that the magnetic field reduces in much faster rate for quadrupole coil than the dipolar one. Fig. 3.4.(b) shows the induced current density in a SC film with thickness, $t = 10$ nm for $\lambda = 100$ nm and $\sigma_1 = 0$. The induced current density for quadrupole coil is almost zero at the edge of the film; however induced current density for the dipole coil shows finite value with a sharp peak at the edge of the film. Therefore the edge error is much less for a quadrupole coil.

### 3.2.1.3. Calculations of mutual inductance, $M$ ($\lambda$,$\sigma_1 t$)

The mutual inductance between two coils on the opposite side of a SC films is determined by complex screening length, $\lambda_\omega = (i\mu_0\omega\sigma)^{-1/2}$ and dimension of the film. At different limiting conditions and for different geometries, several expressions were proposed by different groups for calculating the mutual inductance between two coils with a SC film placed between them [15,22,23,24,25]. Unfortunately, all the analytical expressions are for films with infinite radius but in practical all our films are of finite radius. To take into account the effect of finite radius, S. J. Turneaure et al [19] proposed a numerical method. In this section, I will introduce first the expressions for the mutual inductance for infinite radius film and then I will discuss in detail, the numerical calculation done following S. J. Turneaure et al [19] for film with finite radius. Then, I will analyze both types of calculations from experimental perspective.

### 3.2.1.3.1. Calculation of $M$ ($\lambda$,$\sigma_1 t$) for infinite radius film

The general analytical expression of mutual inductance for film with thickness ($t$) and infinite radius was provided by Clem and Coffey [24,25],

$$M_\infty = \mu_0 \pi \sum_{i,j} r_i^d r_j^p \int_0^\infty \frac{J_1(r_i^d x) \, J_1(r_j^p x) \, e^{-D_{ij}x}}{\cosh(\gamma t) + \left[\left(\gamma^2 + x^2\right)/2\gamma x\right]\sinh(\gamma t)} dx \qquad (3.2)$$

Here $J_1$ is the cylindrical Bessel function of first kind, $D_{ij}$ is the distance between the $i$th loop of drive coil and $j$th loop of pickup coil and $\gamma \equiv x^2 + \lambda_\omega^{-2}$ where $\lambda_\omega$ is the complex screening length given by $\lambda_\omega = (i\mu_0\omega\sigma)^{-1/2} = (\lambda^{-2} + i\ \delta^{-2})^{-1/2}$, where $\delta = (\mu_0\omega\sigma_1)^{-1/2}$. At normal state above $T_c$ the conductivity of thin film is very small ($\sigma \rightarrow 0$), therefore the above equation is reduced to bare mutual inductance expression between two coils [25],

$$M_\infty(\sigma \rightarrow 0) = \mu_0\pi \sum_{i,j} r_i^d r_j^p \int_0^\infty J_1(r_i x)\ J_1(r_j x)\ e^{-D_{ij}x} dx \tag{3.3}$$

In the thin film limit i.e. for $(|\lambda_\omega|\ ,\ t) << (D_{ij}, r_i, r_j)$ which is the usual case for most of the measurements, the eqn. (2.3) is reduced to [25],

$$M_\infty = \mu_0\pi \sum_{i,j} r_i^d r_j^p \int_0^\infty \frac{J_1(r_i^d x)\ J_1(r_j^p x)\ e^{-D_{ij}x}}{1 + \dfrac{1}{2x\lambda_\omega}\sinh(t\ /\ \lambda_\omega)} dx \tag{3.4}$$

This expression can be used for calculating the mutual inductance between two coils with an infinite radius SC thin film placed between them.

### 3.2.1.3.2. Calculation of M ($\lambda,\sigma_1,t$) for finite radius film

To minimize the error in measured $\lambda$ due to finite radius of films, a numerical method was proposed by J. Turneaure et al [19] to calculate the mutual inductance by solving coupled Maxwell and London electromagnetic equations numerically. Following their calculations, the mutual inductance between two coils is given by,

$$M = \sum_j^{n_p} M_j = \frac{1}{I_d}\sum_j^{n_p} \oint \vec{A}_j\ .\ d\vec{l} \tag{3.5}$$

Here $\vec{A}_j$ is the vector sum of vector potentials due to the drive current, $I_d$ and induced screening current at the film by excitation magnetic field created by $I_d$ in drive coil. To find the vector potential, the electromagnetic equations can be solved in cylindrical coordinate system by taking advantage of the azimuthal symmetry of the coaxial coil system and the film. Therefore, the vector potential will have only $\phi$ component in cylindrical coordinate system and the total mutual inductance is given by,

$$M = \frac{2\pi}{I_d}\sum_j^{n_p} r_j^p A_j^p \tag{3.6}$$

Here $r_j^p$ is the radius of $j$th loop of pickup coil and $A_j^p$ is the sum of vector potentials due to the drive current and screening current at the film. Since we know the current in the drive coil, the calculation of vector potential for drive coil is straight forward and given by [26],

$$A_j^{d \to p} = \frac{\mu_0 I_d}{4\pi} \sum_i \frac{4r_i^d}{m_{ij}} \left[ \frac{(2-k_{ij}^2)K(k_{ij}) - 2E(k_{ij})}{k_{ij}^2} \right] \tag{3.7(a)}$$

$$k_{ij}^2 = \frac{4r_i^d r_j^p}{(r_i^d + r_j^p)^2 + D_{ij}} \tag{3.7(b)}$$

$$m_{ij} = \sqrt{(r_i^d + r_j^p)^2 + D_{ij}} \tag{3.7(c)}$$

Here $r_j^d$ is the radius of the $i$th drive loop and $D_{ij}$ is the distance between $i$th drive loop and $j$th pickup loop. $E(k)$ and $K(k)$ are the complete elliptic integrals. To find the vector potential for the induced screening current, we need to know first the induced screening current density in the films. The induced current density can be found by solving the following self consistence equation,

$$\vec{A}_{film}(\vec{r}) = \vec{A}_{drive}(\vec{r}) + \frac{\mu_0}{4\pi} \int d^3\vec{r}' \frac{\vec{J}_{film}(\vec{r}')}{|\vec{r}-\vec{r}'|} \tag{3.8(a)}$$

$$\vec{A}_{film}(\vec{r}) = -\frac{\vec{J}_{film}(\vec{r})}{\mu_0 \lambda_\omega^2} \tag{3.8(b)}$$

The integral over the film is done in cylindrical coordinates, so that these vectors have only φ component. Therefore integral depends only on ρ and $z$. Considering that the current density decays exponentially from both sides of the film with characteristic length scale, $\lambda_\omega$, S. J. Turneaure et al [19] showed that the integration over z can be approximated by a multiplicative factor representing the effective thickness, $t_{eff}(d,\omega,\sigma) = \lambda_\omega \sinh(t/\lambda_\omega)$. Thus the integral equation is reduced to,

$$A_{drive}(\rho) = -\mu_0 \lambda_\omega^2 J_{film}(\rho) - \frac{\mu_0}{4\pi} t_{eff} \int d\rho' J_{film}(\rho') \times \frac{(\rho+\rho')}{\rho} \left[ (2-m^2)K(m) - 2E(m) \right] \tag{3.9(a)}$$

$$m^2 = \frac{4\rho\rho'}{(\rho+\rho')^2} \tag{3.9(b)}$$

The above integral equation can be solved numerically by partitioning the film into concentric annular ring of width $\Delta R = R/N$ where $R$ is the radius of the film and $N$ is the number of concentric ring (for converging solution, $N \geq 100$) [19]. Thus the integral equation is reduced to the summation over $N$ is given by,

$$A_{drive}^{k}(\rho_k) = -\mu_0 \lambda_\omega^2 J_{film}^{k}(\rho_k) - \frac{\mu_0 t_{eff}}{4\pi} \sum_{l}^{N} \Delta R \times J_{film}^{l}(\rho_l) \times \frac{(\rho_k + \rho_l)}{\rho_k} \Big[ (2-m^2)K(m) - 2E(m) \Big] \quad (3.10(a))$$

$$m^2 = \frac{4\rho_k \rho_l}{(\rho_k + \rho_l)^2} \quad (3.10(b))$$

In the above equation the continuous variable $\rho$ and $\rho'$ are replaced by $\rho_k$ and $\rho_l$. In the numerical calculation when $\rho_k = \rho_l$, $m = 1$ which leads *elliptical integral* to infinity. Thus the numerical calculation of self term contribution to the vector potential is failed. Gilchrist and Brandt [27] have shown that when $\Delta R \ll \rho_l$, the self term for the ring can be calculated from self mutual inductance of the ring using following relation,

$$A_{ll}(\rho_l) = \frac{M_{ll}I_l}{2\pi\rho_l} = \frac{\mu_0}{2\pi} \Big( \Delta R \times t_{eff} \times J_{film}^{l}(\rho_l) \Big) \left( \ln \left[ \frac{16\pi\rho_l}{\Delta R} \right] - 2 \right) \quad (3.11))$$

Therefore the above equation can be written in matrix form given by,

$$a_{kl}c_l = b_k \quad (3.12(a))$$

$$a_{kl} = \left[ \left( \frac{\lambda_\omega^2}{t_{eff}} + \frac{\Delta R}{2\pi} \left( \ln \left[ \frac{16\pi\rho_l}{\Delta R} \right] - 2 \right) \right) \delta_{kl} + \frac{\Delta R(\rho_k + \rho_l)}{4\pi\rho_k} \Big[ (2-m^2)K(m) - 2E(m) \Big] (1-\delta_{kl}) \right] \quad (3.12(b))$$

$$b_k = \frac{A_{drive}^{k}(\rho_k)}{\mu_0} \quad (3.12(c))$$

$$c_k = -d_{eff} J_{film}^{k} \quad (3.12(d))$$

The above matrix equation can be solved numerically which gives the induced current density as a function of $\rho$. Then the vector potential at the $j$th loop of pickup coil due to induced screening current at the film can be calculate using the relation,

$$A_j^{film \to p} = \frac{\mu_0}{4\pi} \sum_k^N \frac{4 I_{film}^k \rho_k}{m_{kj}} \left[ \frac{(2 - k_{kj}^2) K(k_{kj}) - 2 E(k_{kj})}{k_{kj}^2} \right] \qquad (3.13(a))$$

$$I_{film}^k = (d_{eff} J_{film}^k) \times \Delta R \qquad (3.13(b))$$

$$k_{kj}^2 = \frac{4 \rho_k r_j}{(\rho_k + r_j)^2 + D_{kj}} \qquad (3.13(c))$$

$$m_{ij} = \sqrt{(\rho_k + r_j)^2 + D_{kj}} \qquad (3.13(d))$$

Therefore the mutual inductance between two coils is given by,

$$M = \frac{2\pi}{I_d} \sum_j^{n_p} r_j^p A_j^p \qquad (3.14(a))$$

$$A_j^p = A_j^{d \to p} + A_j^{film \to p} \qquad (3.14(b))$$

The above numerical procedure developed by S. J. Turneaure et al [19] provides very accurate mutual inductance between two coils with a finite radius SC film placed between them.

### 3.2.1.4. Experimental considerations

The complex screening length, $\lambda_\omega = (i\mu_0\omega\sigma)^{-1/2}$ of the film is extracted from measured complex mutual inductance, $M_{Exp}$ by comparing with the theoretically calculated mutual inductance, $M_{Theo}$ as function of $\lambda_\omega$. For comparison, a look up table of $M_{Theo}$ for many different set of $Re(\lambda_\omega^{-2})$ and $Im(\lambda_\omega^{-2})$ is created in the form of 2D matrix (typically 100 by 100 or so). Then to get $\lambda_\omega^{-2}$ from measured $M_{Exp}$, the $M_{Theo}$ is interpolated using bilinear interpolation correspond to the measured $M_{Exp}$.

The computation time for preparing the 2D matrix of $M_{Theo}$ increases exponentially with the matrix size. Although the numerical method provides better accuracy, the computation time for calculation $M_{Theo}$ (~ 50 hours in my Del Laptop having 1.86 GHz Pentium dual core processor) is much high. Calculation of mutual inductance using analytical expression (only available for infinite diameter films) saves the computation time. However, all our films are of finite radius.

To use the analytical expression, there are some proposals in literatures by different groups [15,19,22] about the finite diameter correction. However, we have followed the numerical method for better accuracy. In the numerical calculation, to minimize the computation time, we have followed the procedure mentioned below.

The response of SC films in kHz frequency is purely inductive except very close to $T_c$. Therefore, in the above calculation the resistive part is zero and the $M_{Theo}$ is calculated only as function of $\lambda$ taking $\sigma_1 = 0$, which saves the computation time. Now close to $T_c$ we need to consider the finite value of $\sigma_1$, therefore the $M_{Theo}$ is calculated as function of $\lambda$ and $\sigma_1$ which gives a 2D matrix of $M_{Theo}$. Then using the conversion program, $M_{Exp}$ is converted to $\lambda_\omega^{-2}$ by comparing $M_{Exp}$ with the $M_{Theo}$.

***Error correction and conversion of $M_{Exp}$ to complex conductivity:***

Before the conversion, we need to take into account the parasitic coupling between primary and secondary coil, which also contributes to the experimentally measured $M_{Exp}$ and introduce a small phase shift in the pickup voltage. The phase of $M_{Exp}$ is adjusted to about $0^o$ based on the assumption that for thin films the coupling between two coils is purely inductive above $T_c$. To remove the parasitic coupling, a base line "zero position" mutual inductance ($\lambda \rightarrow 0$) is subtracted from $M_{Exp}$ and finally subtracted mutual inductance is normalized to $M_{Exp}$ measured above $T_c$. The background mutual inductance is measured using a Lead disc of same diameter as the SC film and same thickness as the substrate thickness. We refer to this base line mutual inductance as the zero position, since it is the mutual inductance when $\lambda=0$, i.e. $\lambda \ll t$. The subtraction procedure removes the parasitic couplings between the primary and secondary circuits and the normalization procedure removes uncertainty associated with the nonideal aspects of coil windings, coil geometry, error in substrate thickness etc.

After the normalization we get the experimental normalized mutual inductance, $m_{Exp.}(T) = \frac{M_{Exp.}(T) - M_{Lead}(4.2K)}{M_{Exp.}(T > T_c)}$. In calculation, we prepare a 2D square matrix (typically 100 by 100) of normalized theoretical mutual inductance, $m_{Theo.} = \frac{M_{Theo.}(\lambda, \sigma_1) - M_{Theo.}(0,0)}{M_{Theo.}(\lambda \rightarrow \infty, 0)}$ for different set of $\lambda$ and $\sigma_1$. Then $\lambda_\omega^{-2}$ is obtained by comparing $m_{Exp}(T)$ with the elements of 1000 by 1000 square matrix obtained by interpolating the matrix $m_{Theo}$ correspond to the matrix of $\lambda_\omega^{-2}$ of same dimension.

The same procedure can be followed for finite diameter correction of analytical expression and extraction of $\lambda_\omega^{-2}$ from measured $M_{Exp}$. Only difference is in preparation of look up table.

## 3.2.2. Summary

The two coil mutual inductance setup is capable of measuring complex conductivity of SC films over a wide range of temperature without any model dependent assumptions. This setup can be used for measurement of complex conductivity of all kinds of SC thin films with thickness, t ≤ $\lambda/2$. In this technique, the major source of error comes from thickness measurement of SC films like in any other thickness related measured quantities.

For calculation of mutual inductance, $M_{Theo}$ as a function $\lambda_\omega^{-2}$ and conversion of measured $M_{Exp}$ to $\lambda_\omega^{-2}$ were done using Matlab codes which are given in **appendix 3A** and **appendix 3B** respectively.

# 3.3. High frequency electrodynamics response

High frequency electrodynamics response provides crucial information on the quasi particle excitations and Cooper pairs of superconductors [28], thus the study of high frequency electrodynamics response of a superconductor is desirable to elucidate the nature of superconductivity. Historically high frequency electrodynamics response was studied mostly through measurements of surface impedance using microwave radiations. Most of the popular microwave measurement techniques used for this purpose rely on the resonance method such as cavity resonator [29,30], strip line resonator [31,32,33] etc. Although these methods provide very good sensitivity on the surface impedance as function of temperature and magnetic field, they are limited to some discrete resonance frequency and unable to provide the frequency dependent response.

To get the frequency dependent electrodynamics response, it is necessary to have resonance less broadband measurement technique which is being done using broadband microwave spectroscopy through broadband measurement of complex reflection coefficient $S_{11}$ developed by J. C. Booth et al[34] in 1996. We have implemented this powerful technique in our lab to study the high frequency electrodynamics response of SC thin films. Our setup can measure the complex conductivity over wide frequency range from 50MHz to 20 GHz at base temperature 0.5K.

### 3.3.1. Broadband microwave spectrometer

The reflection coefficient of microwave radiations in a coaxial transmission line terminated by load impedance, $Z_L$ is given by,

$$S_{11} = \frac{Z_L - Z_0}{Z_L + Z_0},\qquad(3.15)$$

where $Z_0$ is the characteristic impedance of the transmission line, usually 50 $\Omega$ and real. Therefore the reflection coefficient is a dimensionless complex quantity which measures the impedance mismatch between transmission line and terminator load. The load impedance $Z_L$ depends on the dimension of the terminator. In this technique, the transmission line is terminated by a Corbino shaped SC thin film (see Fig. 3.6 (c)). Considering only TEM mode of microwave radiations in the transmission line and the sample thickness is much smaller than the screen depth of the sample, the $Z_L$ can be expressed in term of complex conductivity, $\sigma_L(\omega) = \sigma_1 - i\sigma_2$ of the film given by,

$$Z_L = \frac{\ln(b/a)}{2\pi t \sigma_L},\qquad(3.16)$$

where $a$ and $b$ are inner and outer radius and $t$ is the thickness of the films. Therefore the complex conductivity, $\sigma_L$ can be measured as a function of frequency through broadband measurement of reflection coefficient, $S_{11}$.

### 3.3.1.1. Experimental setup

The schematic diagram of our measurement setup is shown in Fig. 3.6. The complex reflection coefficient, $S_{11}$ was measured using a "Rhode and Schwarz ZVB 20" [35] vector network analyzer (VNA) in the frequency range from 10 MHz to 20 GHz. The details of our coaxial microwave probe are shown in Fig. 3.7. The coaxial probe is connected to microwave cable assembly of VNA through a long 0.181" SS coaxial cable.

The Corbino shaped sample (see Fig. 3.7) was prepared by depositing few hundred nm thick Ag contact pad using thermal evaporation of Ag through a washer shaped shadow mask. The inner and outer diameter of the Corbino shaped sample, were about 1.5 mm and 4 mm equivalent to our coax probe. To make the direct electrical contact with the sample, a female SMA connector was modified by removing its thread partially and carefully attached to the SS cable with the help of the top part of the copper sample housing. For stable electrical contact at low temperature, a tiny KITA spring probe made of beryllium-copper [36], was inserted inside the center conductor of the female connector. The Corbino shaped sample was then electrically connected to the modified SMA connector with the help pedestal pressed by a massive spring from bottom with the help of the bottom part of copper housing. With this spring assembly, very good electrical contact was maintained over wide temperature range from room temperature to 0.5 K.

In addition to the microwave measurements, this technique allows simultaneous measurement of two probe dc resistance through a bias port available at the back of the VNA. The dc resistance measurement was carried out using a Keithley 2400 source meter [37]. The simultaneous dc and microwave measurements provide additional advantage to have consistency

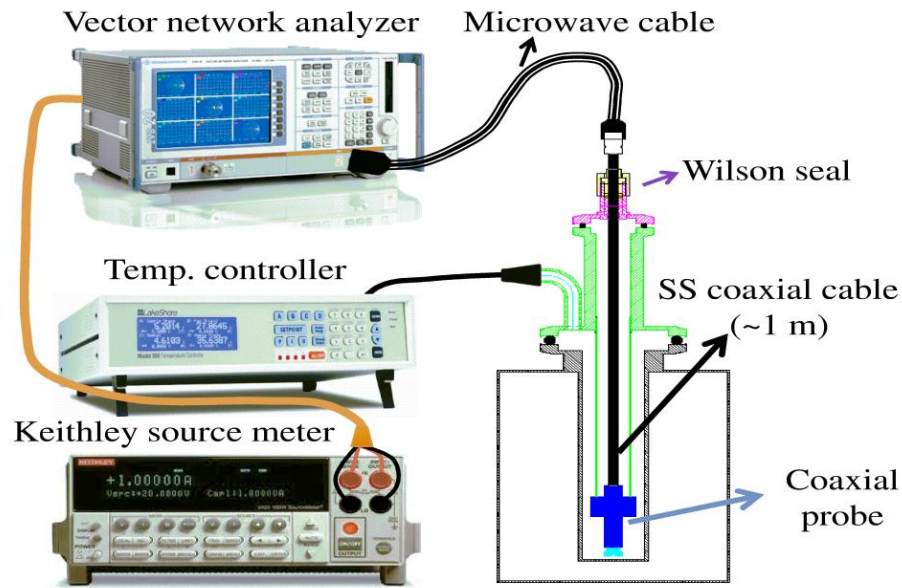

Figure 3.6. Schematic diagram of our broadband microwave measurement setup. For details about the coaxial probe see the Fig. 3.7.

check by comparing high frequency data with the more conventional dc behavior.

The temperature of the sample housing of our coaxial probe was controlled using LakeShore 340 temperature controller [38]. In our measurement setup, the sample housing is in He gas environment which provides very good thermal equilibrium. To obtain the high reproducibility, the thermal equilibrium is essential in case of any high frequency measurements. Therefore all our measurements were carried out after keeping the coaxial probe at below 10K for about 3 hours ensuring that the thermal equilibrium has been achieved.

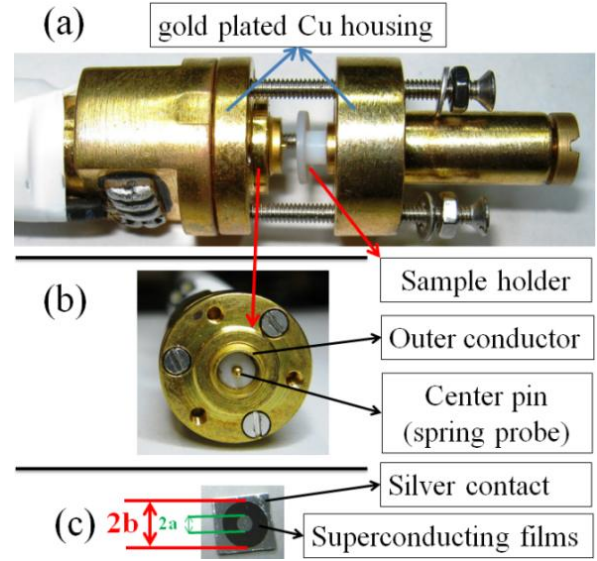

Figure 3.7. (a) Probe head of our broadband microwave spectrometer. (b) Coaxial microwave probe. (c) Corbino shaped sample with silver as contact pad.

Broadband microwave measurement is a two probe measurement technique. Therefore the measured complex reflection coefficient, $S_{11}^m$ not only depends on the sample properties but also undesired effects such as attenuation and phase shift in the long transmission line, partial reflections from the connectors, etc which add a large systematic error in measured $S_{11}^m$. The contribution from sample was extracted from measured $S_{11}^m$ through extensive calibration of our microwave probe which is described in the following section.

### 3.3.1.2. Calibration

The frequency dependent measured complex reflection coefficient $S_{11}^m$ is related to actual true reflection coefficient $S_{11}^a$ of the sample through the equation of standard error model of microwave network [39],

$$S_{11}^m = E_D + \frac{E_R S_{11}^a}{1 - E_S S_{11}^a},$$

(3.17)

Here $E_D$, $E_R$ and $E_S$ are three complex error coefficients originated from imperfect detector setup, connectors, long transmission line etc. Here $E_D$ is called directivity error arising from imperfect directional coupler at the detector port, reflections at bends, joints of the transmission line etc. The error term $E_R$ is referred as the reflection tracking error which comes from attenuation and phase shifts in transmission line. The source mismatch error, $E_S$ is the result of re-reflection of portions of microwave signal back into the VNA caused by the impedance mismatch between the transmission line and detectors of the VNA. In ideal case $E_D = E_S = 0$ and $E_R = 1$. However due to long lossy transmission line and imperfect connections, all three error coefficient are strongly frequency dependent and deviate from the ideal case. Therefore they are to be determined at each frequency of interest and which needs proper calibrations. Since threads of our modified commercial SMA connector were partially removed, we were able to use the commercially available calibration kit manufactured by Rohde-Schwarz [40], to calibrate our microwave probe at room temperature. However, after the spring probe was inserted into the center conductor of the modified connector, the microwave probe needed to be recalibrated again and the commercial calibration kit could not be connected to the probe. Also the commercial calibration kit can't be used at low temperature, since their microwave properties are not calibrated at low temperature. To overcome the calibration problem, a novel non conventional calibration procedure was suggested by different groups [34,41,42,43]. Instead of commercial calibration kit, the microwave probe was calibrated using three known references: Open, Load and Short. In principle any three reference samples can be used if the complex impedances are precisely known.

*(i) Open standard:*

After systematic study using Teflon disc, H. Kitano et al [43] suggested that Teflon disc with thickness 0.5mm or more can be considered as open with $S_{11} = 1$. To be in the safer limit, we have used a Teflon disc with thickness about 1.5 mm as an open standard.

*(ii) Load standard:*

Since a long time, NiCr films were widely used as standard resistors for its low temperature coefficient of resistance [44]. Resistance of a NiCr thin film remains almost constant over wide temperature variation from room temperature to below boiling point of $^4$He, 4.2 K. Therefore NiCr film appears as the ideal candidate for temperature dependent calibration of microwave coaxial probe. To use NiCr film as load standard, the Corbino shaped Ag contact was deposited on NiCr film in similar way as the sample to keep the same electrical environment. Thickness of NiCr film was tuned by controlling the deposition time in such way that the resistance between inner and outer contact pads was about ~20 Ω. The theoretical complex reflection coefficient $S_{11}$ was calculated from the measured dc resistance between the inner and outer Ag contact pads.

### (iii) Short standard:

Bulk copper or aluminum was used by different groups as short standard. There was also suggestion of thick gold film (~300nm) on quartz as short standard. Since we were mainly interested in temperature range from 0.3 K to a maximum of 30 K and error coefficient is almost temperature independent in this range, therefore a think SC NbN film ($t$~250nm) with $T_c$~16.5k was used at temperature ~ 2.5 K, as a short standard. To maintain the same electrical environment as the sample, the electrical contact pad was made using same procedure as the sample.

### (iv) Sample under study as standard:

Since the measurement of complex impedance, $Z_L$, using broadband microwave spectrometer is essentially a relative measurement; the sensitivity strongly depends on the calibration procedure. The impedance of a thicker sample with high $T_c$, is very small and the above calibration procedure fails because of the resolution limit of the spectrometer. In such a situation, to study the dynamical properties of superconductors close to $T_c$, H. Kitano et al [43] has proposed a new calibration procedure using the same sample which is under study.

The complex conductivity $\sigma(\omega,T)$ of sample measured well above $T_c$ is frequency independent and reduces to $\sigma_{dc}$ when the normal state Drude conductivity of the sample is in the Hagen-Rubens limit. Therefore the measured reflection coefficient well above $T_c$ can be used as load standard instead the NiCr standard. The measured complex conductivity at temperature well below $T_c$ can be used as short, if the dissipation at finite frequency is very less. The advantage of this calibration is that Short and Load are measured during the same run, therefore minimize the error. This calibration procedure is valid only close to $T_c$ and helps to explore the critical properties of superconductors close to $T_c$.

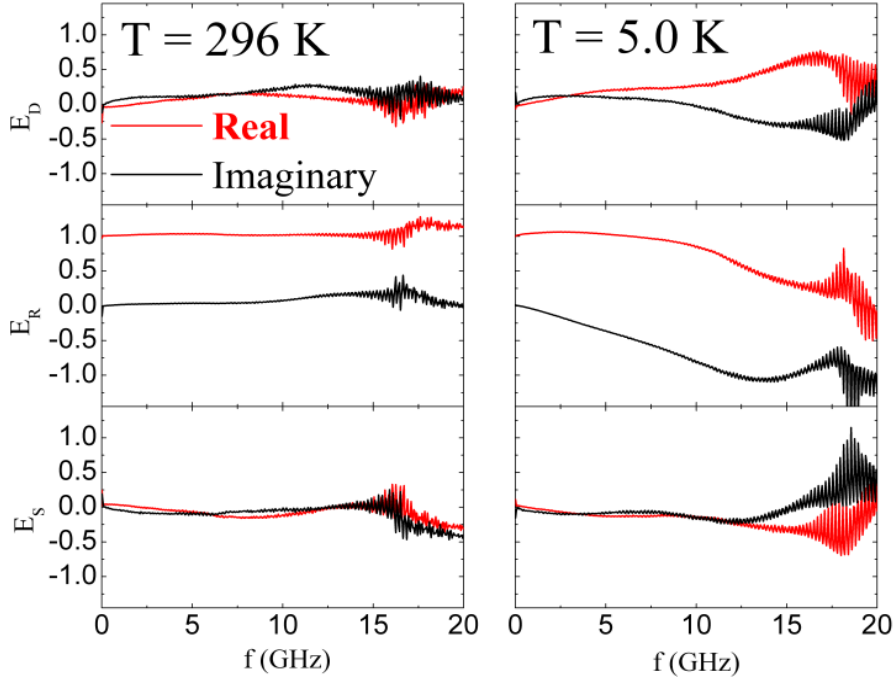

For our very low disordered sample with $T_c \geq 10$ K, the above mentioned calibration procedure was used to measure the complex conductivity. The normal state conductivity measured at temperature, $T \geq 1.5$ $T_c$ was used as load and measured complex conductivity at temperature, $T = 2.5$ K was used as short.

### 3.3.1.3. Error correction

Considering the measured and actual complex reflection coefficients of the above standard are $M_i$ ($i$=1,2,3) and $A_i$ ($i$=1,2,3), the relations for three error coefficient are given by (see ref. 43),

$$E_D(\omega) = \frac{M_1(M_2 - M_3)A_2A_3 + M_2(M_3 - M_1)A_3A_1 + M_3(M_1 - M_2)A_1A_2}{(M_1 - M_2)A_1A_2 + (M_2 - M_3)A_2A_3 + (M_3 - M_1)A_3A_1}, \quad (3.18(a))$$

$$E_R(\omega) = \frac{(M_1 - M_2)(M_2 - M_3)(M_3 - M_1)(A_1 - A_2)(A_2 - A_3)(A_3 - A_1)}{[(M_1 - M_2)A_1A_2 + (M_2 - M_3)A_2A_3 + (M_3 - M_1)A_3A_1]^2}, \quad (3.18(b))$$

$$E_R(\omega) = \frac{M_1(A_2 - A_3) + M_2(A_3 - A_1) - M_3(A_1 - A_2)}{(M_1 - M_2)A_1A_2 + (M_2 - M_3)A_2A_3 + (M_3 - M_1)A_3A_1}, \qquad (3.18(c))$$

The typical values of three error coefficients are shown in figure 3.8. The frequency dependence of $E_D$, $E_R$ and $E_S$ are very weak up to 16 GHz and all are almost zero except only the real part of $E_R$ which is unity. Therefore the errors at room temperature are clearly very small. However at low temperature, the $E_R$ shows strong frequency dependence due to the change in attenuation and phase shift in the long coaxial cable and frequency dependence of $E_D$ and $E_S$ also become prominent at low temperature. By applying these three error coefficients, the corrected reflection coefficients were obtained using eqn. 3.18. Fig. 3.9 shows the typical result for a NbN film before and after calibration. We can clearly see that the spurious frequency dependence mostly disappears in corrected $S_{11}$ (black dashed line) using the three references (i) Teflon as open (ii) Thick NbN films at 2.5K as short and (iii) NiCr as load. At high frequency, some spurious frequency dependent behavior is still observed which is due to the resolution limit of our spectrometer. To remove the spurious frequency dependent component, we have used a second calibration procedure. The measured $S_{11}$ for the sample at temperature $T \sim 1.5\ T_c$ was used as the reference data for load and the measured $S_{11}$ at temperature $T \sim 2.5$ K was used as the reference data for short. We can see that the spurious frequency dependence is completely gone in corrected $S_{11}$ (blue solid line). The corrected $S_{11}$ was then used to compute the complex conductivity of our SC thin films.

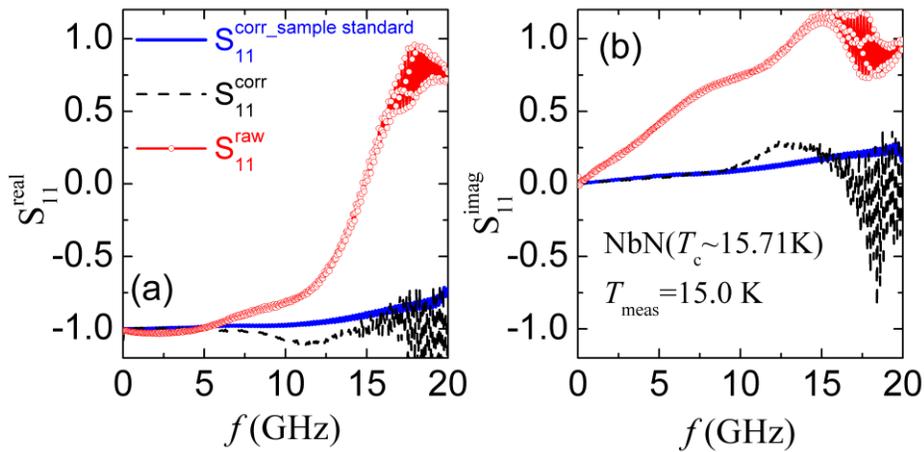

Figure 3.9. The real and imaginary part of $S_{11}$ before and after correction. The red scattered plots represent the measured raw $S_{11}$. The black dashed lines are the corrected $S_{11}$ using three references: (i) Teflon as open (ii) Thick NbN films at 2.5 K as short and (iii) NiCr as load. The solid blue lines represent the corrected $S_{11}$ using low temperature ($T$=2.5 K) and high temperature ($T \geq 1.5\ T_c$) data of sample as short and load reference respectively (see text for details).

### 3.3.2. Summary

Our broadband microwave spectrometer can measure frequency dependent electrodynamics response through measurement of complex conductivity in the frequency range from 50 MHz to 20 GHz provided all calibrations were done properly. The broadband technique provides major advantage over resonance based techniques, for studying dynamical behavior of thin films of various exotic superconductors. This technique suffers from the reproducibility and sensitivity problems. For good reproducibility and sensitivity the shorter coaxial cable and good thermal equilibrium are desirable.

## 3.4. Cryogenic setups

All our measurements were carried out in our lab involving various kinds of cryogenic setups for achieving very low temperatures. Brief descriptions of our cryogenic setups extensively used for our experiments are provided below.

**(a) Continuous flow cryostat:** It is a very simple cryostat and maintenance is also very easy. This cryostat manufactured by Oxford Instruments, [45] provides a base temperature of 2.1 K. Some of the transport measurements, low frequency electrodynamics response measurements using two coil mutual inductance technique and high frequency electrodynamics response measurements using broadband microwave spectrometer were done in this cryostat.

**(b) $^4$He cryostat with 12 T magnet**: This cryostat provides excellent temperature stability over very wide range of temperatures from minimum 1.7 K to room temperature. The best thing about this cryostat is that sample temperature can be raised to room temperature in presence of an applied magnetic field upto 12 T generated by a SC magnet kept in liquid $^4$He bath of the cryostat. Some of the magnetic field dependence study of transport properties and electrodynamics properties were done in this cryostat.

**(c) $^3$He cryostat with 5.5 T magnet**: This cryostat can achieve minimum base temperature 0.3 K and provides excellent temperature stability ($\leq$1mK) for temperatures $T \leq$ 6K. It also has provision for applying magnetic field upto 5.5T using a split Helmholtz SC magnet. Some of the transport, magneto transport and low frequency electrodynamics response measurements were carried out in this cryostat.

This was originally a bottom loading $^3$He cryostat manufactured by Oxford Instruments and later on it was modified to top loading one for quick sample transfer without taking out the VTI. The additional wired advantage of this change is that that the sample is directly dipped in liquid $^3$He or gas depending on the sample temperature, which provides a very good thermal equilibrium.

For temperature measurement and control, commercially available Cernox sensors and LakeShore temperature controllers were used.

# Appendix 3A.    Matlab code for calculating $M_{\text{Theo}}(\lambda_\omega, t)$

```
% Mintu Mondal%
% Program for calculating mutual inductance for finite diameter film

c = clock;
Start_time=fix(c)

clear
%Film dimension
d1 = input (' Give film thickness in nm = ');
Sd = input (' Give substrate thickness in mm = ');
strd1 = num2str(d1);
strSd = num2str(Sd);

m_filename=['M_' strd1 'nm_' strSd 'mm_cont.txt'];
l_filename=['lamda_' strd1 'nm_' strSd 'mm_cont.txt'];

tic()

r1 = 4;              % Radius of the film in mm
N = 100;             % Number of annular ring the film will be divided
%End of film dimension

%Drive current details...
Id1 = 0.5;           % Drive current in mA
Id = Id1/1000;
Ido=-Id;
freq=60000;          %measured frequency 60 kHz
% End of current details

%Coil configuration
%primary coil       = 14+(-14) turns..............
%secondary coil     = 30*4=120 turns ..............
Ld = 2.0;            % Length of the drive coil in mm
Lp = 2.1;            % Length of the pickup coil in mm
Nd = 14;             % number of turn in half layer of quadruple coil
Np = 30;             % number of turn per layer of pickup coil

a1  = 1+ .03;        % Radius of drive coil in mm
ap1 = 1+.03;         % Radius of pickup coil in mm

Hd1  = .4+Sd;        % Distance of drive coil from the film in mm
Hdo1 = Hd1+Ld/2;     % Distance of the opposite turn drive coil in mm
Hp1  = .4;           % Distance of pickup coil from the film in mm

dHd=.0714/1000;      % Distance between two loops of drive coil in m
dHp=.066666/1000;    % Distance between two loops of pickup coil in m
da=.06/1000;         % wire diameter in m
%End of coil configuration

% conversion to SI unit
a=a1/1000;
```

```
app=ap1/1000;
r=r1/1000 ;
d2=d1/10^9;

Hd=Hd1/1000;
Hdo=Hdo1/1000;
Hp=Hp1/1000;
%End of coil configuration

%Inputs
lamda1_in1 = input (' Give the lower limit of penetration depth in nm = ');
lamda2_in1 = input (' Give the upper limit of penetration depth in nm = ');

lamda1_in=(lamda1_in1*10^(-9))^(-2);
lamda2_in=(lamda2_in1*10^(-9))^(-2);

Sigma1_1_in_1 = input (' Give the lower limit of sigma_1 in SI unit= ');
Sigma1_2_in_1 = input (' Give the upper limit of sigma_1 in SI unit= ');

D_p_lam = input (' Give the number of data points to be calculated for lamda = ');
D_p_Sig1 = input (' Give the number of data points to be calculated for sigma_1= ');
%End of input

Sigma1_1_in = (2*pi*freq)*(4*pi*10^(-7))*Sigma1_1_in_1; % complex screening length in nm
Sigma1_2_in = (2*pi*freq)*(4*pi*10^(-7))*Sigma1_2_in_1; % complex screening length in nm

del_lam_2 =  (lamda1_in-lamda2_in)/(D_p_lam-1);

% For calculation of mutual inductance for purely inductive coupling where
% sigma1=0
if D_p_Sig1==1
del_sigma =0;
else
del_sigma =(Sigma1_2_in-Sigma1_1_in)/(D_p_Sig1-1);
end
%End if

Sigma1_1=complex(0,Sigma1_1_in);

for l_v1=1:D_p_lam
for l_v2=1:D_p_Sig1

    lamda_com_lam2(l_v1)= lamda1_in - (l_v1-1)*del_lam_2;

    lamda2_com_sig1(l_v2)= Sigma1_1 + complex(0,((l_v2-1)*del_sigma));

%-----------------------------------------------------
lamda_com_2(1,l_v2)=(lamda_com_lam2(l_v1)+(lamda2_com_sig1(l_v2)));

lamda_com(1,l_v2)=(lamda_com_2(1,l_v2))^(-1/2);

lamda =lamda_com(1,l_v2);
d=lamda*sinh(d2/lamda);
```

```
z2=Hd+Hp;
zo2=Hdo+Hp;

% Calculation of current density
p=zeros(N,1);
q=zeros(N,1);
matrixJbyId=zeros(N,1);
matrixA=zeros(N,N);
matrixB=zeros(N,1);
matrixJbyId=zeros(N,1);

for i_v1=1:Nd

    z=Hd+(i_v1-0.5)*dHd;
    zo=Hdo+(i_v1-0.5)*dHd;
    dq=r/N;

for i_v=1:N

    p(i_v)=i_v*dq-dq/2;
    k1=(a+ p(i_v))^2 + z^2;
    k1o=(a+ p(i_v))^2 + zo^2;

    k2=sqrt(k1);
    k2o=sqrt(k1o);

    m = (4*a*p(i_v)) /k1;
    mo = (4*a*p(i_v)) /k1o;

    [K,E]=ellipke(m);
    matrixB1= (a*((2-m)*K-2*E))/(pi*k2*m);

    [K,E]=ellipke(mo);
    matrixB2= (a*((2-mo)*K-2*E))/(pi*k2o*mo);

    matrixB(i_v)= matrixB1-matrixB2;

    for j_v=1:N
        q(j_v)=j_v*dq-dq/2;

        k1=(q(j_v)+ p(i_v))^2;
        m = (4*q(j_v)*p(i_v)) /k1;

        [K,E]=ellipke(m);

        if j_v==i_v
            dr=dq/2;
            matrixA(i_v,j_v)= (lamda^2)/(d*r)+ (dq*(log(8*pi*q(j_v)/dr)-2))/(2*pi*r) ;
        else
            matrixA(i_v,j_v)= ((1-m/2)*K-E)*(p(i_v)+q(j_v))/(2*pi*N*p(i_v));
        end
    end
end
```

```
end
matrixC = matrixA\matrixB;
matrixJbyIdn = matrixC./(d*r);
% End of calculation of Matrix element

for n=1:N
matrixJbyId(n)=matrixJbyIdn(n)+matrixJbyId(n);
end
end
matrixJ = (Id*matrixJbyId);
%End of calculation of current density

%Mutual inductance calculation start
M=0;
Mfp=0;
for l=1:4
    ap=app+(l-1)*da;

Mp=0;
Mf = 0;
for j_v=1:Np
    z1=z2+(j_v-.5)*dHp;
    zo1=zo2+(j_v-.5)*dHp;

Md=0;

%Calculation of mutual inductance for drive coil and single pickup loop
for i_vd=1:Nd

    z=z1+(i_vd-0.5)*dHd;
    zo=zo1+(i_vd-0.5)*dHd;

    k1=(a+ ap)^2 + z^2;
    k1o=(a+ ap)^2 + zo^2;

    k2=sqrt(k1);
    k2o=sqrt(k1o);

    m = (4*a*ap) /k1;
    mo = (4*a*ap) /k1o;

    [K,E]=ellipke(m);

    Adrive = (10^(-7)*4*Id*a*((2-m)*K-2*E))/(k2*m);

    [K,E]=ellipke(mo);

    Adriveo = (10^(-7)*4*Ido*a*((2-mo)*K-2*E))/(k2o*mo);

    Atotal=Adrive+Adriveo;

    M1 =2*pi*ap*Atotal/Id;
    Md=Md+M1;
```

```
end

%Calculation of mutual inductance for film and single pickup loop
Afilmp=0;
zf = Hp + (j_v-.5)*dHp;
for j1=1:N
    k1=(ap+q(j1))^2 + zf^2;
    k2=sqrt(k1);

    m = (4*q(j1)*ap) /k1;
    [K,E]=ellipke(m);

    Afilmpn = (10^(-7)*4*(dq*d*matrixJ(j1))*q(j1)*((2-m)*K-2*E))/(k2*m);
    Afilmp  = Afilmp+Afilmpn;
end
Mf1 =2*pi*ap*Afilmp/Id;
Mf=Mf+Mf1;
Mp=Mp+Md;
end
Mfp=Mfp+Mf;
M=Mp+M;
end
Mfp=(10^9)*Mfp;
Mc=(10^9)*M;
M_lamda_com(1,l_v2)=Mc-Mfp;
end
dlmwrite (l_filename,lamda_com_2,'-append','delimiter','\t','newline','pc');
dlmwrite (m_filename,M_lamda_com,'-append','delimiter','\t','newline','pc');
l_v1
end
toc()
c = clock;
Stop_time=fix(c)
%End of program................
%Mintu Mondal%
```

## Appendix 3B:    Matlab code for conversion of $M_{\text{Exp.}}(T)$ to $\lambda_\omega = (i\mu_0\omega\sigma)^{-1/2}$

```matlab
%Mintu Mondal%
%Conversion program for complex conductivity
clear
c = clock;
Start_time=fix(c)
tic()
% values of M for normalization.............
M_inf=complex(97.46889,0);      %Experimental value
M_0=complex(-0.04458,0);

M_theo_inf=complex(97.1126,0);  %Theoretical value
M_theo_0=complex(-0.4543,0);
%--------------------------------

%File read
File_M_Exp= 'R_MKBT_01_19Feb10.txt';
M_exp =dlmread(File_M_Exp);

Lamda2_in =dlmread('lamda_6nm_0.5mm_cont.txt');
M_theo_in =dlmread('M_6nm_0.5mm_cont.txt');

Final_file=['Lamd_w_2_' File_M_Exp];
%--------------------------------

% Normalization and interpolation............
M_exp_com=complex(M_exp(:,2),M_exp(:,3));
m_exp=(M_exp_com-M_0)/M_inf;
[n,m]=size(m_exp);

M_theo=(M_theo_in-M_theo_0)/M_theo_inf;

Lamda2 =imresize(Lamda2_in,10,'bilinear');
m_theo =imresize(M_theo,10,'bilinear');

%End of normalization and interpolation...

m_theo_final=zeros(n,1);
Lamda2_final=zeros(n,1);

for i=1:n
   m=m_theo-m_exp(i);
   [r,c]=find(m==min(min(m)));

   m_theo_final(i)=m_theo(r,c);
   Lamda2_final(i)=Lamda2(r,c);
end

Final(:,1)=M_exp(:,1);
Final(:,2)=m_exp;
```

```
Final(:,3)=m_theo_final;
Final(:,4)=real (Lamda2_final);
Final(:,5)=imag(Lamda2_final);

dlmwrite (Final_file,Final,'delimiter','\t','newline','pc');
%End of conversion

%Plotting......
subplot(2,1,1)
plot(Final(:,1),Final(:,4))
grid on
subplot(2,1,2)
plot(Final(:,1),Final(:,5))
grid on

c = clock;
Stop_time=fix(c)
toc()
%End.....
%Mintu Mondal%
```

# CHAPTER 4

## Berezinskii-Kosterlitz-Thouless (BKT) transition in ultrathin NbN films

## 4.1. Introduction

In 2D or quasi 2D superconductors, the phase disordering superconducting transition has been predicted to be of the Berezinskii-Kosterlitz-Thouless (BKT) universality class [1,2], where the proliferation of free vortices induced by transverse phase fluctuations, destroy the superconductivity at temperature, $T=T_{BKT}$, before $\Delta$ goes to zero as predicted within BCS theory (see section 1.6.2.3).

BKT transition in 2D superconductors can be studied through two different schemes. When approaching the transition temperature $T_{BKT}$ from below, the superfluid density, $n_s$ ($\propto \lambda^{-2}$) is expected to go to zero discontinuously at the transition with a "universal" relation between $n_s$ and $T_{BKT}$ given by (section 1.6.2.3),

$$J(T_{BKT}^{-}) \equiv \frac{2}{\pi} T_{BKT} \,, \qquad (4.1(a))$$

$$J(T) = \frac{\hbar^2 n_s t}{4m} = \frac{\hbar^2 t}{4\mu_0 e^2 \lambda^2} \,, \qquad (4.1(b))$$

Here $t$ is the thickness of the films and $\lambda$ is the magnetic penetration depth. Approaching the transition from above, we can identify the BKT transition from superconducting fluctuations which leave its signature in the temperature dependence of various quantities such as resistivity, magnetization etc [1,2,3]. In the second scheme, information on the BKT transition is encoded in the correlation length $\xi(T)$ which diverges exponentially at $T_{BKT}$, in contrast to power-law dependence expected within Ginzburg-Landau (GL) theory[4]. Both approaches should in principle give identical information.

Although superfluid He films follow the BKT relation quite precisely [5], the BKT transition in 2D superconductors has remained controversial [6]. For instance, the jump in $n_s$ is often observed at a temperature lower than $T_{BKT}$ and at a $J$ ($\propto n_s$) larger than the expected value within BKT theory [7,8]. There are other complicacies in superconductors, such as inhomogeneity which tends to smear the sharp signatures of BKT transition compared to the clean case and difference in vortex-core energy, $\mu$, from the predicted value within the 2D XY model originally investigated by Kosterlitz and Thouless [9,10]. This can give rise to different manifestation of vortex physics, even without the change of the order of transition [11]. Recently, the relevance of $\mu$ for the BKT transition has attracted a renewed interest in different contexts, ranging from the case of layered high temperature superconductors [12,13,14] to the SC interfaces in artificial hetero structures [15,16,17,18] and liquid gated interface superconductivity[19].

In the following sections, by a systematic study of $\lambda^{-2}(T)$ and $\rho(T)$, we will elucidate the nature of BKT transition in ultrathin NbN films. We will show that when film thickness becomes comparable to coherence length, $\xi_0 \sim 5$nm, the superconducting transition is governed by transverse phase (vortex) fluctuations belonging to BKT universality class while low vortex core energy is taken into account. The apparent discrepancies is result from the effect of quasi particle excitations which modifies the vortex core energy, $\mu$ from the value expected within 2D XY model and intrinsic disorder in the system.

## 4.2. Experimental details

### 4.2.1. Sample preparation

To study the BKT physics in NbN thin films, four 8 mm diameter epitaxial NbN thin films with thickness, $t \sim 3$, 6, 12 and 18 nm were grown on (100) oriented MgO substrates by reactive dc magnetron sputtering at optimum deposition condition. The thickness, *t*, of NbN films was controlled by controlling the deposition time keeping all other deposition conditions fixed. The detailed procedure of sample preparation is described in section 3.1.

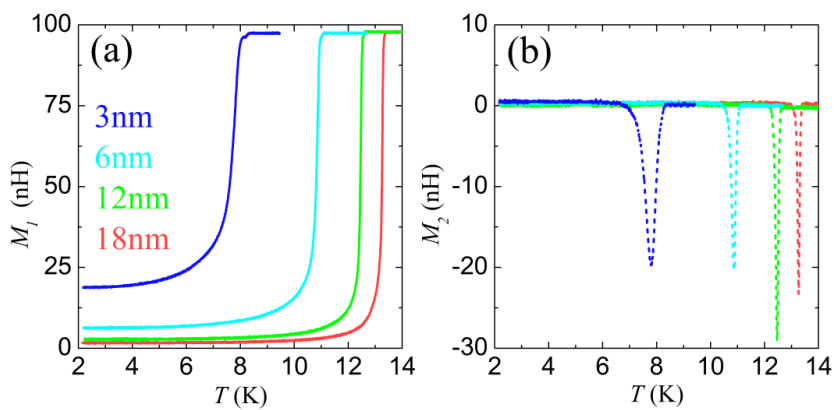

Figure 4.1. shows the mutual inductance, $M=M_1+iM_2$ between two coils for four epitaxial NbN thin films with thickness, $t \sim 3$, 6, 12 and 18 nm.

### 4.2.2. Magnetic penetration depth measurements

The magnetic penetration depth, $\lambda$, was measured using two coil mutual inductance technique. Detailed procedure of measurements of mutual inductance, $M_{Exp}$, and conversion of $M_{Exp}$ to complex screening length, $\lambda_\omega$, is described in section 3.2.

Figure 4.1.(a) and (b) show the experimentally measured mutual inductance, $M_{Exp} = M_1 + iM_2$ as a function of temperature, between two coils with the superconducting film of 8 mm diameter placed in between them. Here the real part of $M_{Exp}$ i.e. $M_1$ is the inductive coupling and imaginary part of $M_{Exp}$ i.e. $M_2$ is the resistive coupling between two coils. The $M_2$ shows single reasonably narrow peak for all samples, therefore confirms that all samples under study are of

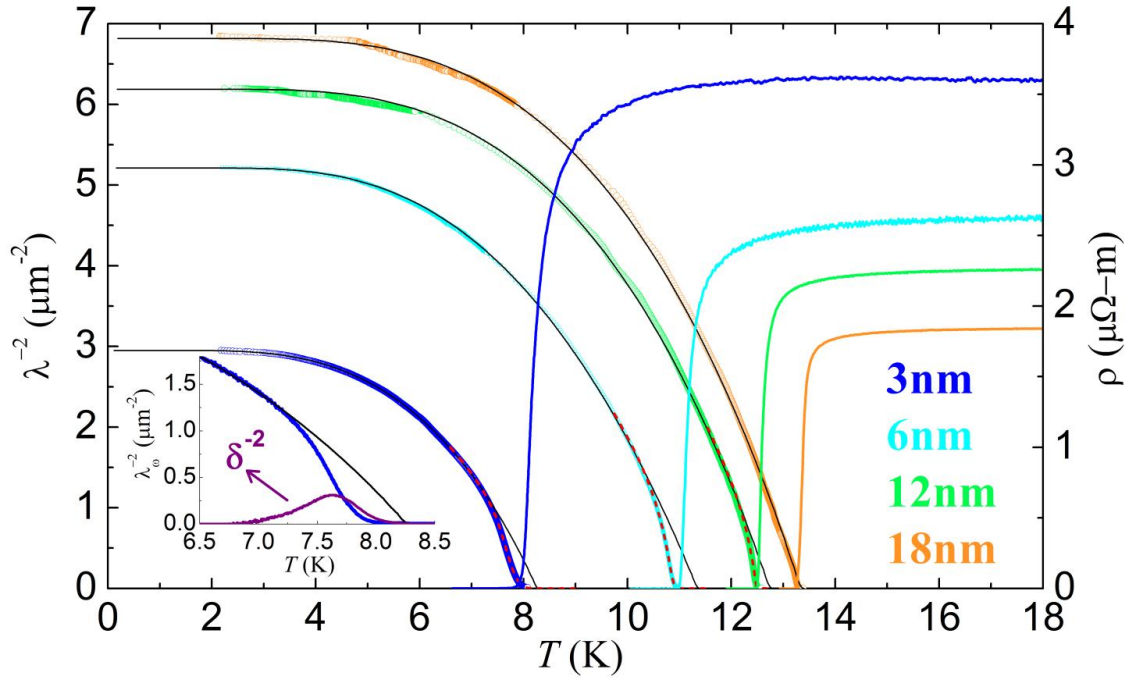

Figure 4.2. shows the temperature dependence of $\lambda^{-2}(T)$ and $\rho(T)$. The (black) solid lines and (red) dashed lines correspond to the BCS and BKT fits of the $\lambda^{-2}(T) - T$ data, respectively. The *inset* shows the expanded view of real and imaginary part of $\lambda_\omega^{-2}(T)$ close to $T_{BKT}$ for 3 nm film (for details see text).

single phase and good quality.

The complex screening length, $\lambda_\omega = (\lambda^{-2} + i\delta^{-2})^{-1/2}$ where $\lambda = (\mu_0\omega\sigma_2)^{-1/2}$ and $\delta = (\mu_0\omega\sigma_1)^{-1/2}$, was extracted from experimentally measured mutual inductance, $M_{Exp.}$ (see section 3.2). The Fig. 4.2. shows the obtained $\lambda^{-2}(T)$ ($\propto n_s(T)$) as a function of temperature. The *inset* shows the real and imaginary part of $\lambda_\omega^{-2}(T)$ for the 3nm thick film.

### 4.2.3. Transport measurements

The normal state Hall carrier density ($n_H$) was deduced from Hall coefficient ($R_H$) measured on standard Hall bar geometry which was prepared by cutting a rectangular piece from the same circular sample used for magnetic penetration depth measurements. $R_H$ was calculated from the measured Hall voltage by sweeping the magnetic field from +12 to −12 T and then subtracting the resistive contribution.

The resistivity, $\rho$, of the films at different temperatures was measured using standard four-probe technique on 1mm width strip prepared using Ar ion beam milling. In this experiment, for each thickness, we have measured the $\lambda^{-2}(T)$, $R_H$ and $\rho$ on the same film.

The Fig.4.3.(a) shows the measured normal state resistivity, $\rho$, as a function of temperatures. The 18 nm thick film shows small negative temperature coefficient caused by electron-electron Coulomb interactions [20], however as the thickness reduces the resistivity shows strong negative temperature dependence due to increase in Coulomb interaction with reducing thickness [21].

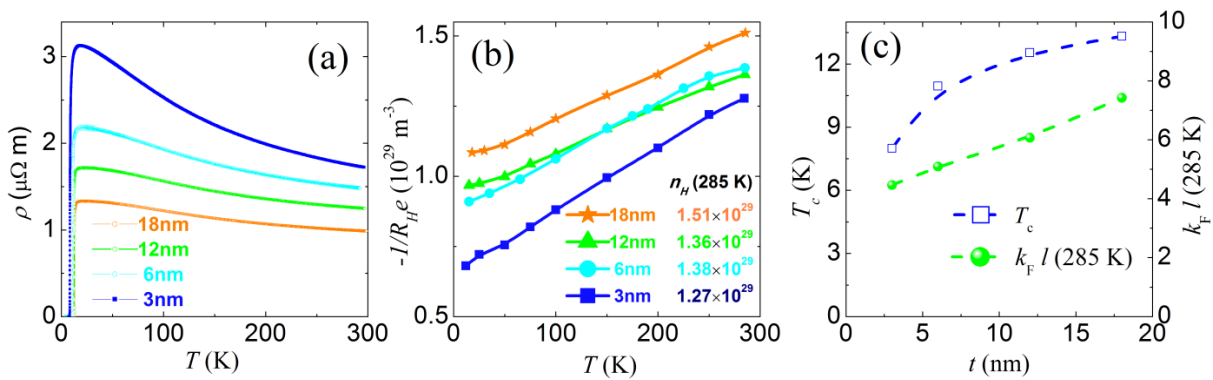

Figure 4.3.(a) $\rho$ *vs* $T$ for the same set of films. (b) Experimentally measured *-1/$R_H e$* as a function of temperature. (c) Shows the variation of superconducting transition temperature, $T_c$, and obtained $k_F l(285)$ from measured $n_H$ and $\rho$ at 285 K with film thickness. Here $T_c$ is defined as the temperature where resistance goes below our measurable limit.

The Fig.4.3.(b) shows $-1/R_H e$ as a function of temperature. The $-1/R_H e$ is monotonically decreasing with decreasing temperature due to increase in electron-electron interactions [20]. In absence of interactions the Hall carrier density can be estimated as $n_H = -1/R_H e$. Since the electron-electron interaction is expected to be small at high temperature [22], therefore we have determined the carrier density from Hall coefficient $R_H$ measured at room temperature which is the maximum achievable temperature in our setup. The room temperature carrier densities shown in Fig. 4.3.(b), are scattered in a small range of values, indicating that they have similar compositional disorder [23].

The Ioffe-Regel parameter, $k_F l$ which is the measure of disorder in our films, is extracted from experimentally measured ρ and $n_H$ at temperature, $T \sim 285$ K considering free electron model (see section 2.1.1.). Figure 4.3.(c) shows the variation of $T_c$ and $k_F l$ with thickness. The $T_c$ changes from 13.37 K for the 18nm to 7.99 K for the 3nm thick film. The value of $k_F l$ reduces from $k_F l \sim 8$ for 18 nm to $k_F l \sim 5$ for 3nm thick film; therefore the effective disorder increases with decreasing film thickness. However films deposited under identical deposition conditions, have similar compositional disorder as described in section 2.1[23]. The effective disorder in our films increases mainly due to decrease in phase space volume with decreasing film thickness.

## 4.3. Discussions

### 4.3.1. Superfluid density

#### 4.3.1.1. Observation of BKT transition in temperature variation of $\lambda^{-2}(T)$

Since in our films the electronic mean free path, $l$ << coherence length, $\xi_0$, we fit the temperature variation of $\lambda^{-2}(T)$ with the dirty limit BCS expression (see section 1.4.2.1) [24],

$$\frac{\lambda^{-2}(T)}{\lambda^{-2}(0)} = \frac{\Delta(T)}{\Delta(0)} \tanh\left(\frac{\Delta(T)}{2k_B T}\right), \qquad (4.2)$$

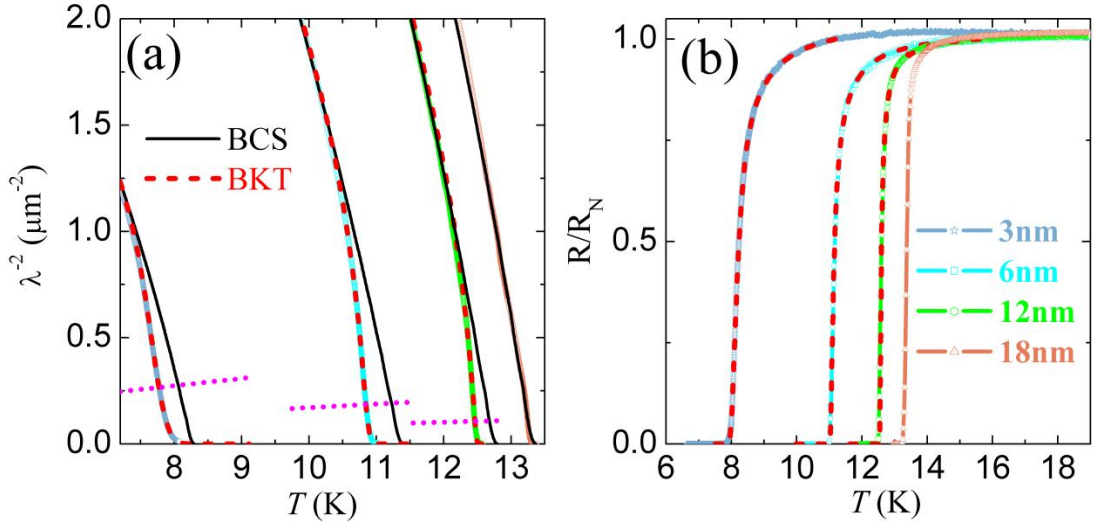

Figure 4.4.(a) Temperature dependence of $\lambda^{-2}(T)$ close to $T_{BKT}$. The (black) solid lines and (red) dashed lines correspond to the BCS and BKT fits of the $\lambda^{-2}(T)-T$ data respectively. The intersection of BCS curve with (magenta) dotted lines where the BKT transition is expected within 2D XY model. (b) The temperature variation of $R/R_N$. The (red) dashed lines correspond to theoretical fits to the normalized resistivity data as described in text.

using $\Delta(0)$ as a fitting parameter. We observe that for thinner films $\lambda^{-2}(T)$ starts to deviate downwards from expected BCS behavior close to $T_c$ (see Fig. 4.2.). In conventional 3D superconductors, $\lambda^{-2}(T)$ smoothly goes to zero at $T = T_{BCS}$ where $\Delta$ goes to zero predicted within BCS theory. However in 2D superconductors within BKT theory, the superconducting transition is controlled by proliferation of vortices and expected to show a sharp jump in $\lambda^{-2}(T)$ given by the eqn. (4.1). Therefore, the sharp downturn in $\lambda^{-2}(T)$ close to $T_c$ which becomes more and more prominent as the film thickness decreases, signifying the destruction of superconductivity by phase disordering belongs to universality class BKT transition. To understand the nature of BKT transition, $\lambda^{-2}(T)$ and universal BKT line predicted within 2D XY model given by eqn. (4.1) are plotted in Fig. 4.4.(a). We observe that the jump in $n_s$ apparently starts around a temperature lower than the expected $T_{BKT}$ and therefore at a larger $J$ than expected from eqn. (4.1). In addition, we can see in Fig. 4.4.(a) that the transition is slightly rounded near $T_{BKT}$ and the sharp jump of BKT transition is replaced by rapid down turn around the intersection with the universal line $2T/\pi$. Since the real superconductors always have some amount of intrinsic inhomogeneity which smears the sharp signature of BKT transition, sharp jump becomes rounded.

### 4.3.1.2. Role of vortex core energy

The apparent discrepancy mentioned above, can be reconciled by taking into account that the superfluid stiffness, $J(T)$ is not only affected by the presence of quasi particle excitations but also by the presence of thermally activated vortex-antivortex pairs in the system. When vortex core energy, $\mu$ is large, the latter effect is negligible for $T < T_{BKT}$ and $T_{BKT}$ can be estimated from the intersection of the line $2T/\pi$ and $J(T)$ according to eqn. (4.1). However in real superconductors, the presence of quasiparticle excitations reduces the vortex core energy from the value expected within 2D XY model. Therefore $J(T)$ gets renormalized due to thermally excited vortex-antivortex pairs even below $T_{BKT}$.

Within 2D XY model, the vortex core energy, $\mu$ is determined by the short distance cutoff of the energy of vortex line at short length scale as,

$$E = \pi J \left[ \log\left( \frac{L}{\xi_0} \right) + \alpha \right],$$

(4.3)

where $L$ is the system size, $\xi_0$ is the coherence length and the vortex core energy, $\mu \equiv \pi J \alpha$. We obtain $\alpha \cong \pi^2/2$, when the 2D XY lattice model is mapped to continuum Coulomb gas problem [25]. Therefore in 2D XY model, the vortex core energy, $\mu$ in energy scale of $J$, is given by,

$$\frac{\mu_{XY}}{J} \cong \frac{\pi^2}{2} \cong 4.9 \,,$$

(4.4)

However in real superconductors the $\mu$ is determined by the loss of condensation energy in vortex core. Therefore we can estimate $\mu$ from the loss of condensation energy within a vortex core of the size of the order of coherence length $\xi_0$ [26] as,

$$\mu = \pi \xi_0^2 \varepsilon_{cond} \,,$$

(4.5)

where $\varepsilon_{cond}$ is the condensation energy density for the superconductor. In a clean conventional superconductor, $\mu$ can be expressed in terms of $J$ by means of the BCS relations for $\varepsilon_{cond}$ and $\xi_0$. Now $\varepsilon_{cond} = N(0)\Delta^2/2$, where $N(0)$ is the electronic DOS at the Fermi level and $\Delta$ is the BCS gap and $\xi_0 = \xi_{BCS} = \hbar v_F/\pi\Delta$, where $v_F$ is the Fermi velocity. Assuming $n_s \sim n$ at $T = 0$, where $n = 2N(0)v_F^2 m/3$, the value of $\mu_{BCS}$ can be determined as

$$\mu_{BCS} = \frac{\pi\hbar^2 n_s}{4m}\frac{3}{\pi^2} = \pi J \frac{3}{\pi^2} \approx 0.95 J \, , \qquad (4.6)$$

Therefore the value of $\mu$ is quite smaller than $4.9J$ expected within 2D XY model. The small $\mu/J$ explains the deviations from the BCS behavior before the renormalized superfluid stiffness reaches the predicted universal value $2T_{BKT}/\pi$ given by eqn. (4.1).

### 4.3.1.3. Analysis of $\lambda^{-2}(T)$ using general model of BKT transition

To take into account the effect of low vortex core energy, $\mu$, and inhomogeneity, we have numerically solved the renormalization group (RG) equations of the original BKT formalism (see section 1.6.2.3) [9],

$$\frac{dK^{-1}}{dl} = 4\pi^3 y^2 \quad \text{and} \quad \frac{dy}{dl} = (2 - \pi K) y \qquad (4.7)$$

where reduced renormalized stiffness $K = J/T$ and vortex fugacity $y = e^{-\mu/T}$. The above RG equations were solved using only one free parameter: $\mu/J(T)$, where $J(T)$ is obtained from the BCS fit to the experimental data at low temperatures [black solid lines in Fig. 4.2. and Fig. 4.4(a)], where vortex excitations are suppressed.

Table 1 Magnetic penetration depth ($\lambda(T\to0)$), $T_{BTK}$, $T_{BCS}$ along with the best fit parameters obtained from BKT fits of the $\lambda^{-2}(T)$ and $R(T)$ data for NbN thin films of different thickness. The $T_{BCS}$ corresponds to the mean field transition temperature obtained by extrapolation of the BCS fit of $\lambda^{-2}(T)$ at $T<T_{BKT}$.

| $d$ | $\lambda(0)$ | $T_{BKT}$ | $T_{BCS}$ | From best fit of $\lambda^{-2}(T)$ | | | From best fit of $\rho(T)$ | |
|---|---|---|---|---|---|---|---|---|
| (nm) | (nm) | (K) | (K) | $\mu/J$ | $\delta/J$ | $b_{theo}$ | $A$ | $b$ |
| 3 | 582 | 7.77 | 8.3 | 1.19±.06 | 0.020±0.002 | 0.108 | 1.35±0.14 | 0.108±0.006 |
| 6 | 438 | 10.85 | 11.4 | 0.61±.05 | 0.005±0.0007 | 0.048 | 1.30±0.13 | 0.067±0.008 |
| 12 | 403 | 12.46 | 12.8 | 0.46±.05 | .0015±0.0003 | 0.027 | 1.21±0.12 | 0.039±0.006 |
| 18 | 383 | ---- | 13.4 | ---- | ---- | ---- | ---- | ---- |

To take into account the effect of inhomogeneity in $J$, we assume distribution of local superfluid stiffness, $J^i(T)$ values around the BCS value and perform an average of $\lambda^{-2}(T)$ associated to each local $J^i(T)$[13,26]. For simplicity we take the occurrence probability $w_i$ of each $J^i(T)$ is a Gaussian distribution with width $\delta$ as,

$$P(J) = \frac{1}{\sqrt{2\pi}\delta} \exp\left[\frac{-(J - J_{BCS})^2}{2\delta^2}\right] \qquad (4.8)$$

Hence, $\sum_i w_i = 1$. Then we rescale proportionally the local mean field SC transition temperature $T_{BCS}^i$ and calculate the resultant $J$ from the RG equations [13,15]. This procedure leads to excellent fits to our experimental data shown in Fig. 4.2. and Fig. 4.4.(a), in the whole temperature range. The best fit values are listed in Table 1. We can see that obtained values of vortex core energy are of the order of estimated $\mu_{BCS}$ from eqn. (4.6).

## 4.3.2. Resistivity

To further establish our findings, we have analyzed our resistivity data above $T_{BKT}$ by considering BKT fluctuations and GL fluctuations using same set of parameters obtained from the analysis of $\lambda^{-2}(T)$ below $T_{BKT}$. In 2D, the contribution of SC fluctuations to conductivity is encoded in the temperature dependence of SC correlation length, $\delta\sigma \propto \xi^2(T)$. The functional form of $\xi(T)$ depends on the character of the SC fluctuations. The correlation length, $\xi(T)$ is power law dependence for Gaussian fluctuations and exponential for BKT fluctuations. Due to proximity effect between the $T_{BKT}$ and $T_{BCS}$, it is expected that most of the fluctuation regime will be accounted for by GL fluctuations while KT fluctuations will be relevant only between $T_{BKT}$ and $T_{BCS}$ (see section 1.6.2.3.2). We interpolate between these two regimes using the Halperin-Nelson [2] interpolation formula for the correlation length,

$$\frac{\xi}{\xi_0} = \frac{2}{A}\sinh\left(\frac{b}{\sqrt{t_r}}\right), \qquad (4.9)$$

where $t_r = (T - T_{BKT})/T$ and $A$ is a constant of order unity. Here $b$ is the most relevant parameter which determines the shape of resistivity above $T_{BKT}$, given by [26],

$$b \approx \frac{4}{\pi^2} \frac{\mu}{J} \sqrt{t_c} \ , \qquad (4.10)$$

where $t_c = (T_{BCS}\text{-}T_{BKT})/T_{BKT}$. Thus the normalized resistance corresponding to the SC correlation length is given by [26],

$$\frac{R}{R_N} = \frac{1}{1+(\Delta\sigma/\sigma_0)} \equiv \frac{1}{1+\left(\xi/\xi_0\right)^2} \ , \qquad (4.11)$$

where $R_N$ is the normal state resistance taken as $R_N = R(T=1.5T_{BKT})$. To take into account the sample inhomogeneity, we map the spatial inhomogeneity of the sample in a random resistor network problem by associating each local superfluid stiffness, $J^i$ to a normalized resistance $r_i = R_i/R_N$ obtained from eqn. (4.11) by using local values of $T^i_{BCS}$ and $T^i_{BKT}$ computed in superfluid density data analysis. Then the overall normalized resistance, $r = R/R_N$ can be calculated using effective medium theory (EMT) [27], where $r$ is the solution of the self consistent equation,

$$\sum_i \frac{w_i(r-r_i)}{(r+r_i)} = 0 \ , \qquad (4.12)$$

where $w_i$ is the occurrence probability of each resistor $r_i$, which is the same as the occurrence probability of each $J^i$ determined from the superfluid density data analysis. We apply the above procedure to analyze our resistivity data using probability distribution width $\delta$ determined from the analysis of superfluid density data and taking $A$ and $b$ as free parameters. The resulting fits are in excellent agreement with the experimental data shown in Fig. 4.4.(b). Considering that the interpolation formula is an approximation, the values of $b$ are in very good agreement with theoretically estimated values from equ. (4.10), $b_{theo}$ obtained from analysis of superfluid density data listed in table 1.

### 4.3.3. Error analysis

Now we try to estimate the error in our different parameters obtained by fitting our experimental data. The Fig. 4.5. (a) and (b) show the experimental $\lambda^{-2}(T)$ of 3 nm films, with the best fit curves represented by black solid line for best fit parameters $\mu/J \cong 1.2$ and $\delta \cong 0.02$. In Fig 4.4.(a) the green dashed lines and red dashed lines represent the curves for deviation ±5% and ±10% in the best fit value of $\mu/J$. In Fig 4.4.(b), the red dashed line represents the curve for deviation ±10% in $\delta$. The figure 4.4.(c) shows the experimentally measured normalized resistance with the best fit theoretical curve for fitting parameter $b \cong 0.106$, $\delta \cong 0.02$ and $A = 1.35$. The red dashed line represents the plot for ±10% deviation in the best fit value of fitting parameter $A$. Therefore, from the above analysis, we conclude that the errors in our estimated values are not more than 10% in all our obtained parameters.

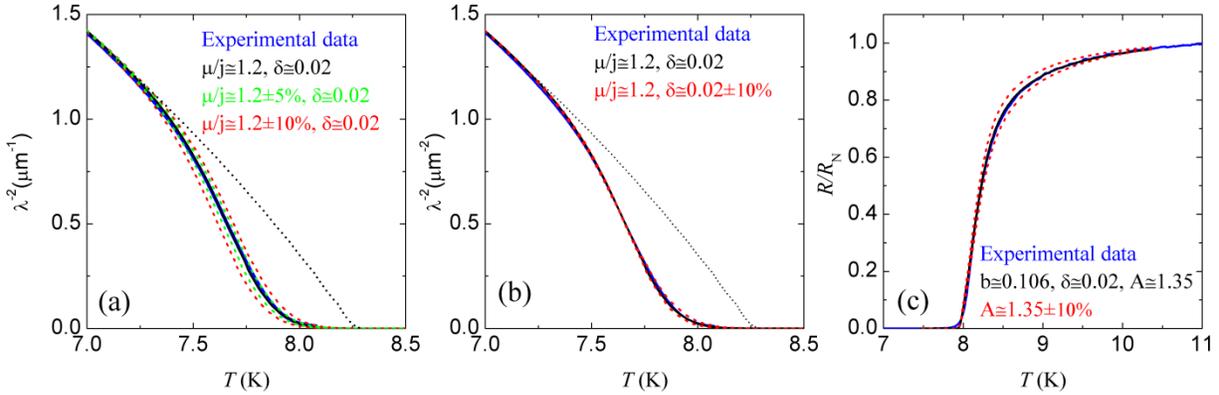

Figure 4.5. (a) and (b) show the $\lambda^{-2}(T)$ as function of temperature for 3nm films. Blue solid line is the experimentally measured data and black dotted line is the BCS fit. Black solid line is the theoretical fit with best fit parameters (see table). (a) The green dashed lines represent the fit for ±5% deviation and the red dashed lines represent the fit for ±10% deviation from best fit value of $\mu/J$. (b) The red dashed lines represent the fit for ±10% deviation from best fit value of $\delta$. (c) shows the normalized resistance as a function $T$. The blue solid line is the experimentally measured data and black solid line is the theoretical best fit to the data. The red dashed lines represent the fit for ±10% deviation from best fit value of $A$.

### 4.3.4. Effect of disorder on vortex core energy

Since the effect of disorder can alter the relation between $\varepsilon_{cond}$, $\Delta$ and $J$, the value of $\mu$ can be affected significantly due to presence of disorder in the system. The Fig. 4.6.(a) shows the $\mu/J$ as a function $k_F l$ which is the measure of disorder in our samples. We can see that the value of $\mu/J$ monotonically increases with decreasing $k_F l$ i.e. increasing disorder. The increase in value of $\mu/J$

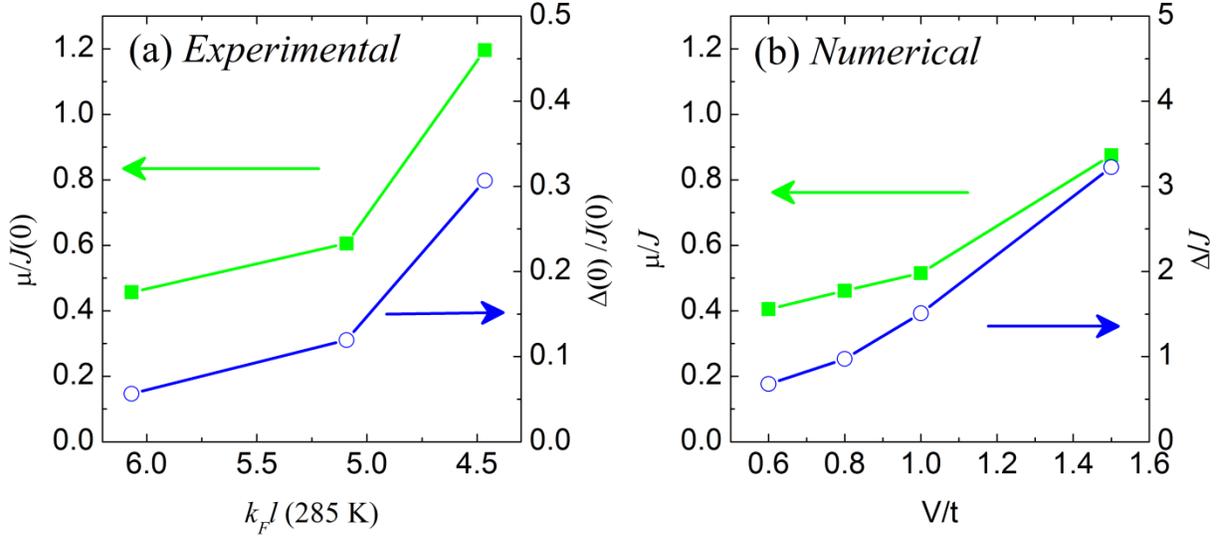

Figure 4.6 (a) Experimental values of μ/*J* and Δ/*J* in our NbN films, plotted as a function of $k_F l$.(b) Numerical results for the disorder dependence of μ/*J* and Δ/*J* as a function of disorder for the attractive Hubbard model.

with increasing disorder can be understood by considering the effect of disorder on different energy scales associated with the superconducting ground state.

To properly account for the effect of disorder, μ and *J* were computed explicitly within attractive two dimensional Hubbard model with onsite local disorder [28],

$$H = -t \sum_{(ij)\sigma} c_{i\sigma}^{\dagger} c_{i\sigma} + h.c. - U \sum_i n_{i\uparrow} n_{i\downarrow} + \sum_{i\sigma} V_i n_{i\sigma}, \qquad (4.13)$$

The above Hamiltonian has been solved in the mean field using Bogoliubov-de Gennes equatins [29], on a N=$N_x \times N_y$ system. The local potential $V_i$ is randomly distributed between $0 \leq V_i \leq V_0$. The superfluid stiffness *J* is computed from the change in ground state energy in presence of vector potential [28], while the vortex core energy, μ is obtained using eqn. (4.5) by determining $\varepsilon_{cond}$ and ξ in the presence of disorder at doping level, $n_d$=0.87 and coupling U/t=1. The resulting value of μ/*J* at *T*= 0 is plotted in Fig. 4.6.(b). It is of the order of BCS estimate and it shows a steady increase as disorder increases, in agreement with the experimental results, shown in Fig. 4.5.(a) where $k_F l$ is taken as the measure of disorder.

This behavior can be understood as a consequence of the increasing separation between the energy scales associated, respectively, to the $\Delta$ which controls $\varepsilon_{cond}$ and $J$, as it is shown by the ratio $\Delta/J$ that we report in the two panels of Fig. 4.6 for comparison. Notice that, the values of $\Delta/J$ are much larger than experimental ones because the calculation was done at larger coupling strength as compared to our NbN samples due to constraint in numerical analysis. Nonetheless, our approach already captures the experimental trend of $\mu/J$ as function of disorder and its correlation with behavior of $\Delta/J$.

## 4.4. Summary

In summary, we have shown that the phase transition in 2D or quasi-2D superconductors can be reconciled with the standard BKT model when the small vortex core energy is taken into consideration. Due to low vortex-core energy, the number of thermally activated vortex-antivotex pairs is significant enough to renormalize the bare superfluid density even at temperature well below $T_{BKT}$. Therefore the downturn in superfluid density starts at higher superfluid density value and lower temperature than expected within 2D XY model, even though actual BKT transition occurs where the universal BKT line intersects with the experimentally measured $\lambda^{-2}(T)$. We also observed steady increase of the ratio $\mu/J$ with decreasing film thickness. This effect can be understood considering the increasing separation between the energy scales associated with $\Delta$ and $J$ due to increase in effective disorder with decreasing thickness. Our work finally provides a complete paradigm description of the BKT transition in real superconductors.

It should be noted that the recently discovered interface superconductivity in complex hetero structures [16,17,18] and liquid gated interface [19] expected to show superconducting transition belongs to the BKT universality class. However, the nature of BKT transition in these materials were studied mostly by transport measurements, which can give misleading results if the correct value of vortex core energy is not taken into account. Therefore to indentify the true nature of BKT transition, the study of superfluid density is preferable to have better estimate of the correct value of $\mu$ as compared to energy scale given by superfluid stiffness, $J$.

It would also be interesting to do similar study in other 2D superconductors such as TiN [30], InO$_x$ [7] including layered high temperature superconductors [31], to understand the nature of BKT transition in those materials and effect of disorder on vortex core energy.

# CHAPTER 5

# Effect of phase fluctuations in strongly disordered 3D NbN films

## 5.1. Introduction

The remarkable zero resistance property of superconductors arises from the macroscopic quantum phase coherent state characterized by complex order parameter, $\psi = |\Delta|e^{i\theta}$, where $|\Delta|$ is the measure of binding energy of the Cooper pairs which manifests as a gap in the electronic excitation spectrum, and $\theta$ is the phase of the macroscopic condensate. In clean conventional superconductors which are well described by Bardeen-Cooper-Schrieffer (BCS) theory [1], the superconductivity is destroyed at a characteristic temperature, $T_c$, where $|\Delta|$ goes to zero.

Based on BCS theory, Anderson theorem states that the superconducting transition temperature ($T_c$) remains unchanged in moderately disorder limit [2]. However it turns out that with increasing level of disorder, superconductivity is progressively suppressed and can result in superconductor to normal transition at a critical level of disorder. In addition, strongly disordered superconductors show many exotic phenomena resulting from competition between different kinds of interaction [3,4,5,6,7]. Some of the phenomena, namely, giant magneto resistance peak in strongly disordered s-wave superconductors [3], persistence of flux quantization beyond the destruction of superconductivity [4] and finite superfluid phase stiffness above $T_c$ [5,6] show evidences of superconducting correlation persisting well after the global superconducting state is destroyed. Furthermore, recent scanning tunneling spectroscopy (STS) measurements on strongly disordered $s$-wave superconductors such as TiN, InO$_x$ and NbN [8,9,10,11], reveal the appearance novel pseudogap (PG) state which persist at a temperature well above $T_c$ contrary to BCS prediction [1]. The observation of PG characterized by a soft gap in electronic density of states (DOS) without zero resistance, similar to the case of high temperature superconductors, raises obvious question of whether the strong disorder can destroy the superconductivity without destroying Cooper pairs, leading to the system having finite Cooper pair density but no global superconductivity.

In strongly disordered superconductors, superconductivity can be suppressed through two distinct mechanisms. The first mechanism is where electron-electron interaction increases due to loss of effective screening with increasing disorder [12,13]. The increase in repulsive Coulomb interaction partially reduces the attractive pairing interaction mediated by phonon, therefore suppresses the superconductivity.

The second and more intriguing mechanism is where the superfluid density, $n_s$ ($\propto \lambda^{-2}$) is suppressed with increasing disorder scattering. Superconductors characterized by low phase stiffness, $J$, which is proportional to $n_s$ are susceptible to phase fluctuations [14]. Therefore in strongly disordered superconductors, phase fluctuations can destroy the SC state by phase disordering even when the system may retains the Cooper pairs [14] which manifests as the gap in electronic DOS.

The effect of phase fluctuations in a superconductor can be assessed by superfluid phase stiffness, $J$, which is the measure of energy cost of twisting the phase, given by (see section 1.5),

$$J = \frac{\hbar^2 a n_s}{4m}; \quad n_s = \frac{m}{\mu_0 e^2 \lambda^2}, \tag{5.1}$$

where $m$ is electronic mass, $n_s$ is the superfluid density and $\lambda$ is the magnetic penetration depth. The length scale, $a$ is the characteristic length scale for phase fluctuations which is of the order of dirty limit coherence length, $\xi_0$. In clean bulk conventional superconductors, $J >> T_c$, and therefore phase fluctuations play negligible role, consistent with BCS theory. However, when $n_s$ is reduced through strong disorder scattering, $J/k_B$ decreases and eventually becomes smaller than the mean field $T_c$ (defined by BCS theory) at some critical value of disorder. In such a situation, superconductivity can get destroyed through phase disordering, giving rise to novel electronic states with finite density of Cooper pairs but no global superconductivity [14,15].

In 3D disordered superconductors, there are two types of phase fluctuations about the BCS ground state, which can destroy superconductivity: (i) the classical (thermal) phase fluctuations (CPF) and (ii) the quantum phase fluctuations (QPF) associated with number phase uncertainty. QPF results from the fact that, there will be Coulomb energy cost associated with number fluctuations when phase coherence is established between neighboring regions.

Therefore if the electronic screening is poor, such as in a strongly disordered system it becomes energetically favorable to relax the phase in order to reduce number fluctuations.

In this chapter, I will explore the role of phase fluctuations in superconducting properties of disordered NbN thin films by a thorough study of finite frequency electrodynamics response [16]. Thickness of our films used for this study is about 50 nm while coherence length ξ~ 5nm, therefore all our films are in 3D limit. Our study provides compelling evidences of role played by phase fluctuations in suppressing of superconductivity and appearance of novel PG state in strongly disordered superconductors [16].

## 5.2. Low frequency electrodynamics response

To study the effect of phase fluctuations, λ(T) was measured using low frequency two coil mutual inductance technique (see section 3.2) for a set of 3D NbN films of thickness, $t \geq 50$ nm, with progressively increasing disorder with $T_c$ varying from 16 K to 2.27 K.

### 5.2.1. Experimental results

The Fig. 5.1 shows the raw mutual inductance, $M_{Exp}=M_1+iM_2$ between two coils as a function of temperature for a set of NbN films with varying $T_c$ from 15 K to 2.27 K. The real part of mutual inductance, $M_1$ corresponds to inductive coupling and the imaginary part of the mutual inductance, $M_2$ corresponds to resistive coupling between two coils. In the Fig. 5.1, the mutual inductance measured by two coil setup for [4]He cryostat is shown by solid lines and by a similar two coil setup for [3]He cryostat is shown by dashed lines. The difference in the absolute value of $M$ between two set of measurements in the normal state is due to difference in coil dimensions. This difference is taken into account while calculating the penetration depth where we use the actual coil geometry.

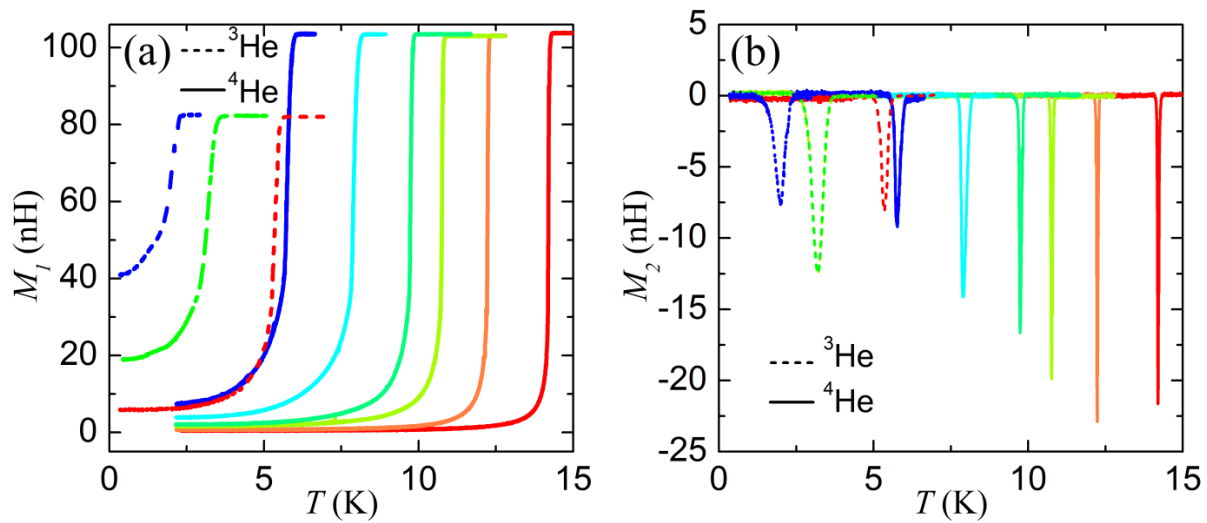

**Figure 5.1. Experimentally measured raw mutual inductance, $M_{Exp}=M_1+iM_2$, as a function of temperature for a set of NbN thin films with varying $T_c$ from 2.27 K to 15 K. (for details see text).**

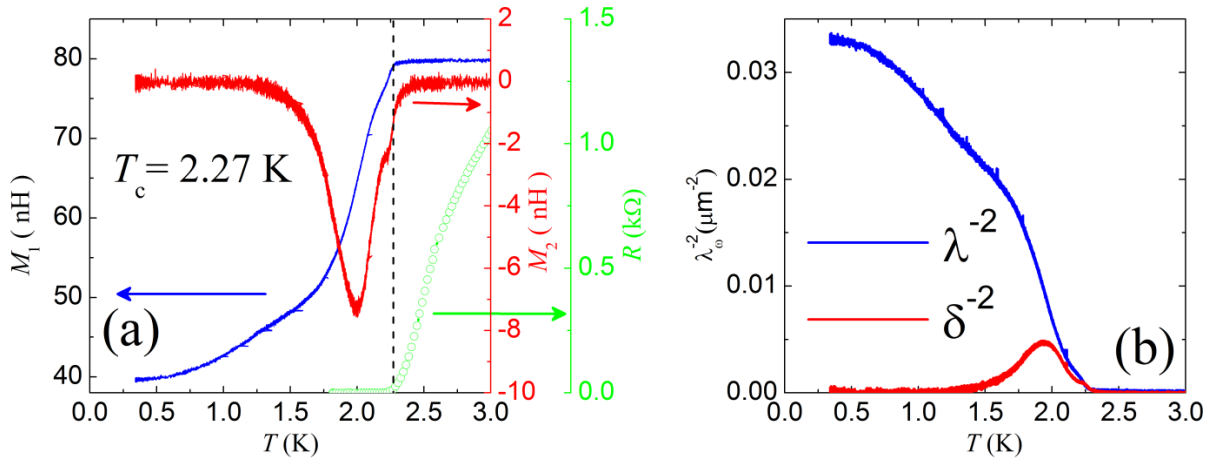

**Figure 5.2.** (a) Mutual inductance, $M_{Exp}=M_1+iM_2$, as a function of temperature for a strongly disordered NbN thin film with $T_c$ ~2.27 K. The circular scattered plot shows the resistance of the cut peace as a function of temperature. (b) The extracted complex screening length as function of temperature for the same film.

All films show reasonably narrow single peak in $M_2$ confirming good quality and single phase of our films. Figure 5.2.(a) shows the mutual inductance for the strongly disordered NbN thin film with $T_c$ ~ 2.27 K, as a function of temperature. The four probe resistivity is measured on a rectangular piece of width 1 mm, which was cut from the same circular film used for mutual inductance measurement. The measured resistance is shown by (green) open circle scattered plot in Fig. 5.2.(a). We can see that the resistance appears at a temperature, $T=T_c$ where the diamagnetic response of the film goes to zero showing the loss of superconductivity at $T=T_c$. The Fig. 5.2.(b) shows the extracted $\lambda^{-2}$ and $\delta^{-2}$ from measured mutual inductance, $M_{Exp}$.

## 5.2.2. Discussion

Fig. 5.3.(a) and (b) show the temperature variation of $\lambda^{-2}(T)$ as a function $T$ for set of NbN films with varying $T_c$. For the films with low disorder, the temperature variation of $\lambda^{-2}(T)$ follows the dirty-limit BCS behavior (black solid line) [17],

$$\frac{\lambda^{-2}(T)}{\lambda^{-2}(0)} = \frac{\Delta(T)}{\Delta(0)} \tanh\left(\frac{\Delta(T)}{2k_B T}\right), \tag{5.2}$$

However as with increasing disorder, the $\lambda^{-2}(T)$ progressively deviates from the expected BCS temperature variation and shows a gradual evolution towards a linear-$T$ variation which saturates at low temperatures for samples with $T_c \leq 6$ K. This trend is clearly visible in strongly disordered sample with $T_c \sim 2.27$ K (See Fig. 5.2.(b)). In this disorder regime i.e. for samples having $T_c \leq 6$ K, we have also noticed that the scanning tunneling spectroscopy (STS) study shows the appearance of PG gap in the tunneling DOS. (see section 2.4.3).

### 5.2.2.1. Superfluid phase stiffness and phase fluctuations

To find the correlation between the deviation of $\lambda^{-2}(T)$ from BCS temperature dependence and the observed PG in tunneling DOS we compare two energy scales: superfluid stiffness, $J$ ($\propto n_s \propto \lambda^{-2}$) and SC energy gap, $\Delta$. Using relation (5.1) we have estimated the values of $J$ (Fig. 5.4) from experimentally measured $\xi_0$ (ref.23) and $\lambda^{-2}(T \rightarrow 0)$. For NbN films, we observed $\Delta(0) \approx 2.05 k_B T_c$ from tunneling measurements performed at low temperatures ($T < 0.2 T_c$) on planar tunnel junctions fabricated on a number of low disordered samples with $T_c \geq 6$ K [18] (see section 2.4.1.4). Since the tunneling spectra for strongly disordered samples with $T_c < 6$ K, do not follow the BCS behavior, $\Delta$ can't be determined accurately from the tunneling measurements. Therefore we have obtained $\Delta(0)$ for strongly disordered sample by extrapolating $\Delta(0) \approx 2.05 k_B T_c$. We would like to note that the obtained $\Delta(0)$ from the extrapolation may have some error but it will always be less than the actual value of $\Delta$ [10,11] which is very important to draw our conclusions as will be discussed latter. The Fig. 5.4 shows the measured $\lambda^{-2}(0)$, $J$ and $\Delta$ as function of $T_c$.

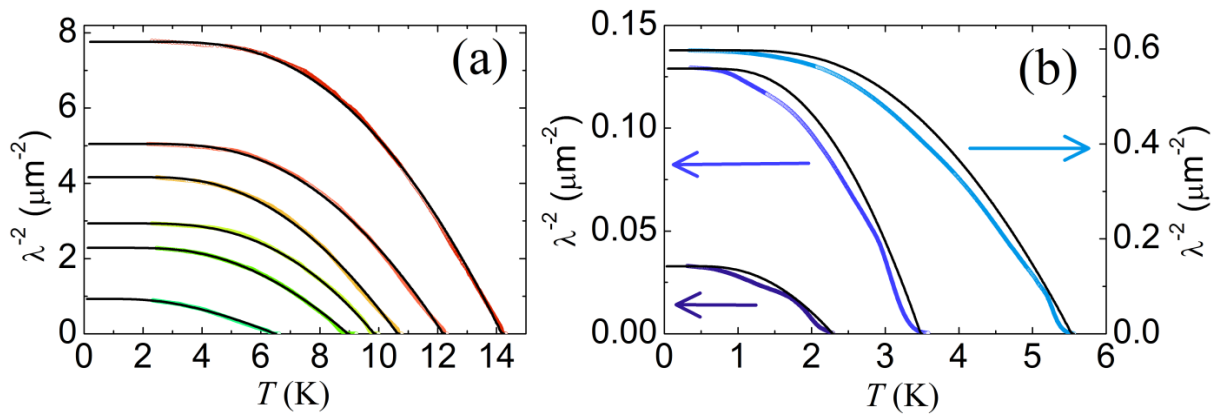

**Figure 5.3.** (a)-(b) $\lambda^{-2}(T)$ vs $T$ for a set of disordered NbN films; the solid black lines are the expected temperature variations from dirty limit BCS theory.

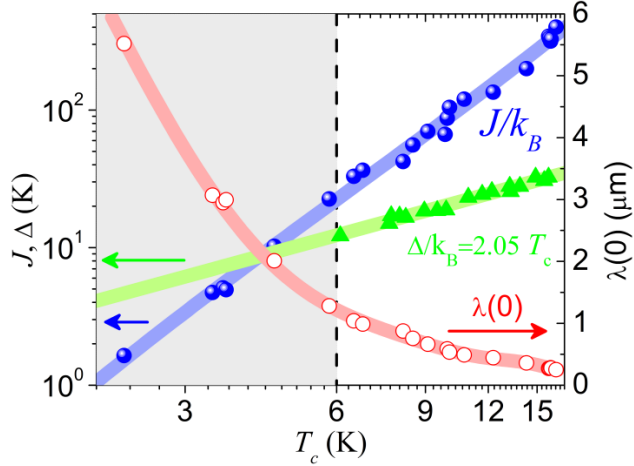

As expected, in the low disorder regime, $J$ is very large and thus the effect of phase fluctuations is negligible. However, as the disorder increases, $J$ rapidly reduces and becomes comparable to $\Delta$ for sample with $T_c \leq 6$ K. Therefore phase fluctuations are expected to play a significant role in superconductivity and we expect deviation in $\lambda^{-2}(T)$ from BCS temperature dependence [10,11].

### 5.2.2.2. Effect of phase fluctuations on superfluid density

To understand the nature of phase fluctuations, we now concentrate on the value of $\lambda$ at $T \rightarrow 0$. In absence of phase fluctuations, the disorder scattering reduces $\lambda^{-2}(0)$ according to the BCS relation [17] (see section 1.4.2.1),

$$\lambda^{-2}(0)_{BCS} = \frac{\pi \mu_0 \Delta(0)}{\hbar \rho_0}, \tag{5.3}$$

where $\rho_0$ is the resistivity just above $T_c$. The $\Delta(0)$ was determined using relation $\Delta(0) \approx 2.05 k_B T_c$ obtained from tunneling measurements on low disordered samples [10]. Fig. 5.5(a) shows the $\lambda^{-2}(0) \approx \lambda^{-2}(0)_{BCS}$ within experimental error for samples with $T_c > 6$K. However, as we approach the critical disorder $\lambda^{-2}(0)$ becomes gradually smaller than $\lambda^{-2}(0)_{BCS}$, reaching a value which is 50% of $\lambda^{-2}(0)_{BCS}$ for the sample with $T_c \sim 2.27$K. As the estimated $\Delta(0)$ using relation $\Delta(0) \approx 2.05 k_B T_c$, will always be less than the actual $\Delta(0)$ of strongly disordered samples [10,11], error in estimated suppression in superfluid density would be in negative direction.

Since the suppression of $\lambda(0)^{-2}$ from its BCS value and observed linear-$T$ dependence of $\lambda^{-2}(T)$ are characteristic features associated with QPF and CPF [19] respectively, we now try to quantitatively analyze our data from the perspective of phase fluctuations. The importance of quantum and classical phase fluctuations is determined by two energy scales [14]: The Coulomb energy $E_c$, and the superfluid stiffness, $J$. For a 3D superconductor $E_c$ can be estimated from the following relation [20],

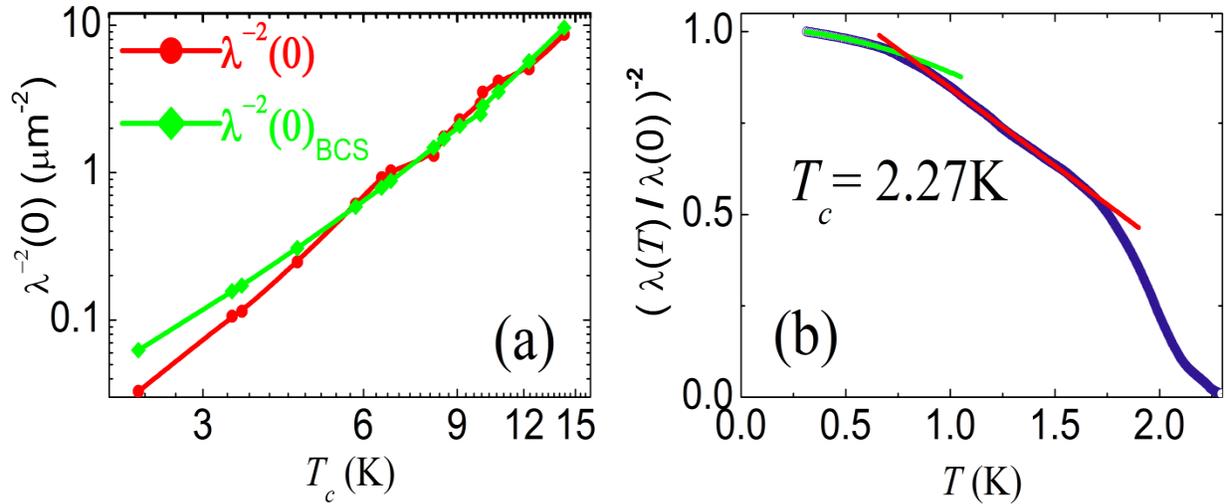

**Figure 5.5.** (a) $\lambda^{-2}(0)$ and $\lambda^{-2}(0)_{BCS}$ as a function of $T_c$. (b) Temperature variation of $(\lambda(T)/\lambda(0))^{-2}$ for film with $T_c = 2.27$ K; the solid lines (green) are fits to the $T^2$ dependence of $\lambda^{-2}(T)/\lambda^{-2}(0)$ at low temperature ($T \leq 0.65$K) and the $T$ dependence (red) at higher temperature.

$$E_c = \frac{16\pi e^2}{\varepsilon_\infty a} \ , \tag{5.4}$$

where $\varepsilon_\infty$ is the background dielectric constant from the lattice and $a$ is the characteristic length scale for phase fluctuations. The suppression of $\lambda^{-2}(0)$ from QPF over its bare value can be estimated using the self consistent harmonic approximation [21] which predicts (in 3-D) (see section 1.6.2.2),

$$\frac{n_s\left(T=0\right)}{n_{s0}\left(T=0\right)} = e^{\frac{-\Delta\theta^2(T=0)}{6}} \ , \tag{5.5(a)}$$

$$\Delta\theta^2\left(T=0\right) = \frac{1}{2}\sqrt{\frac{E_c}{J}} \ , \tag{5.5(b)}$$

Here $n_{s0}$ is the bare value of the superfluid density in absence of phase fluctuations. At the same time, the temperature scale above which phase fluctuations become classical is given by the Josephson plasma frequency,

$$k_B T_{cross} \approx \hbar\omega_p = \sqrt{4\pi e^2 n_s / m * \varepsilon_\infty} = \sqrt{E_c J} \tag{5.6}$$

We now calculate these values for the sample with $T_c \sim 2.27$ K and compare with the data. For NbN, we estimate $\varepsilon_\infty \approx 30$ from the plasma frequency (12600 cm$^{-1}$) measured at low temperatures [22]. Taking the characteristic phase fluctuation length-scale $a \approx \xi \sim 8$nm [ref.23], we obtain, $E_c \approx 0.3$eV and $J \approx 0.14$meV at $T=0$. This corresponds to $n_s(T=0)/n_{s0}(T=0) \approx 0.02$. While this numerical value is likely to have some inaccuracy due to the exponential amplification of any error in our estimate of $E_c$ or $J$, the important point to note is that this suppression is much larger than our experimental estimate, $\lambda^{-2}(0)/\lambda_{BCS}^{-2}(0) \approx 0.5$. On the other hand the crossover temperature from QPF to CPF is estimated to be , $T_{cross} \approx 75$ K which implies that CPF cannot be responsible for the observed linear temperature dependence of $\lambda^{-2}(T)$ in this sample.

### *Effect of phase fluctuations in presence of dissipation:*

These two apparent contradictions can be resolved by considering the role of dissipation. In *d*-wave superconductors, the presence of low energy dissipation has been theoretically predicted [24] and experimentally observed from high frequency conductivity [25,26] measurements. In recent microwave experiment [5] on amorphous InO$_x$ films reveals that low energy dissipation can also be present in strongly disordered *s*-wave superconductors. While the origin of this dissipation is not clear at present, the presence of dissipation has several effects on phase fluctuations: (i) QPF are less effective in suppressing $n_s$; (ii) QPF contribute to a $T^2$ temperature dependence of $n_s$ of the form $n_s/n_{s0} = 1 - BT^2$ at low temperature where B is directly proportional to the dissipation and (iii) the crossover to the usual linear temperature dependence of $n_s$ due to CPF, $n_s/n_{s0} = 1 - (T/6J)$, occurs above a characteristic temperature that is much smaller than predicted temperature. In the sample with $T_c \sim 2.27$ K, the $T^2$ variation of $\lambda(T)^{-2}/\lambda(0)^{-2}$ can be clearly resolved below 650 mK. In the same sample, the slope of the linear-*T* region is 3 times larger than the slope estimated from the value of *J* calculated for $T = 0$. This discrepancy is however minor considering the approximations involved. In addition, at finite temperatures $n_{s0}$ gets renormalized due to QE. With decrease in disorder, QE eventually dominates over the phase fluctuations, thereby recovering the usual BCS temperature dependence at low disorder. Since CPF eventually lead to the destruction of the SC state at temperature less than the mean field transition temperature, the increased role of phase fluctuations could naturally explain the observation of a PG state in strongly disordered NbN films. We would also like to note that in all disordered samples $\lambda^{-2}(T)$ shows a downturn close to $T_c$, reminiscent of the BKT transition, in ultrathin SC films [27]. However, our samples are in the 3D limit where a BKT transition is not expected. In latter section, in the scaling analysis of frequency dependent fluctuation conductivity we will show that transverse phase (vortex) fluctuations fail to explain the extended fluctuation region above $T_c$ in strongly disordered samples (see section 5.3.3.2.2). Since there are many effects which play significant role close to $T_c$, at present we do not know the exact origin of this behavior.

## 5.2.3. Summary of low frequency electrodynamics response

In summary, we have observed a progressive increase in phase fluctuations in strongly disordered NbN thin films. The above observations lead us to conclude that the SC state in strongly disordered superconductors is destroyed by phase fluctuations. In scanning tunneling spectroscopy measurements [10,11], it was observed that at strong disorder the superconductor spontaneously segregates into domains separated by regions where the SC order parameter is suppressed (see Fig. 5.6. (b)) [28]. Therefore, one can visualize the SC state in strongly disorder films, as a network of Josephson junctions with a large distribution in coupling strength (see Fig. 5.6.(a)), where the SC transition is determined by phase disordering. One would expect that the phase fluctuations between these domains result in destruction of the global SC state whereas Cooper pairs continue to survive in localized islands.

In this scenario, $T_c$ corresponds to the temperature at which the weakest couplings are broken. Therefore, just above $T_c$ the sample consist of large phase coherent domains (consisting of several smaller domains) fluctuating with respect to each other. As the temperature is increased further, the large domains will progressively fragment giving rise to smaller domains till they completely disappear at $T = T^*$ corresponds to the pseudogap temperature observed in STS study. In such a scenario $J$ will depend on the length scale at which it is probed. When probed on a length scale much larger than the phase coherent domains, $J \rightarrow 0$. On the other hand, when probed at length scale of the order of the domain size, $J$ would be finite, however $J(\omega)$ would vanish at a temperature where the phase coherent domain becomes much smaller

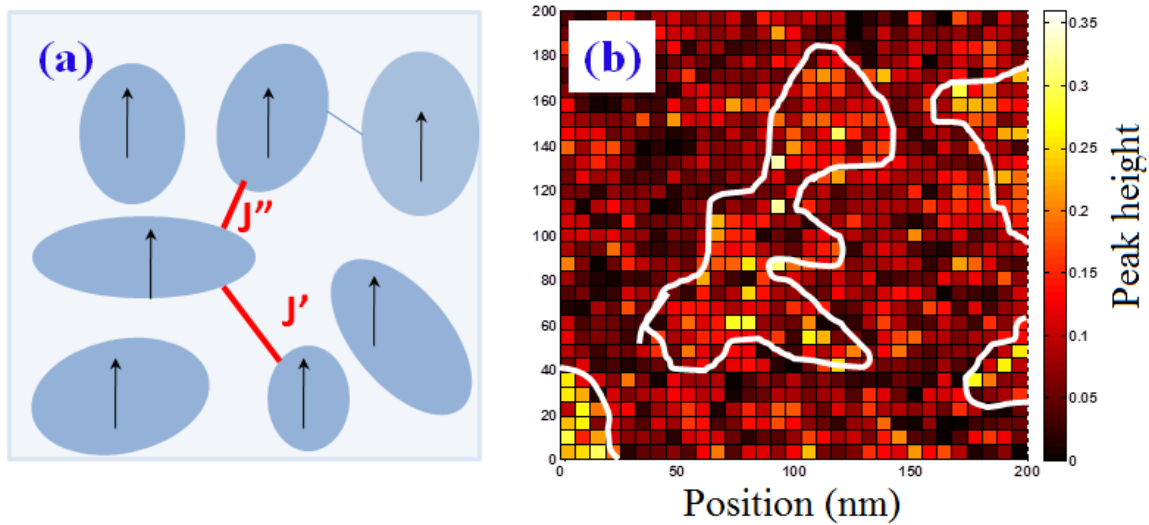

Figure 5.6. (a) Cartoon diagram of our system similar to Josephson junctions network but with large distribution in puddles size and coupling strength, $J$. (b) The spatial variation of coherence peak height measured using STM as the measure of local order parameters at 500 mK, is shown for sample with $T_c$ ~2.9 K over a 200 nm × 200 nm area. (figure is adapted from Ref. 28)

than probing length scale. Therefore in a very strongly disordered system the superfluid stiffness, $J$ is expected to be strongly frequency dependent above $T_c$ and it will be zero over a large length scale but will remain finite in shorter length scale in the PG regime.

To confirm our phase fluctuations scenario, we have done high frequency measurements of electrodynamics response using broadband microwave spectrometer. Detailed investigation is described below.

## 5.3. High frequency electrodynamics response

The study of high frequency electrodynamics response of disordered NbN films were carried out through measurements of ac complex conductivity, $\sigma(\omega) = \sigma_1(\omega) - i\sigma_2(\omega)$ using microwave radiation with the help of broadband microwave Corbino spectrometer [see section 3.3] in the frequency range 0.4 to 20 GHz.

### 5.3.1. Probing length of microwave radiations

The advantage of this technique is that it is sensitive to the length scale set by the probing microwave frequency, given by the relation,

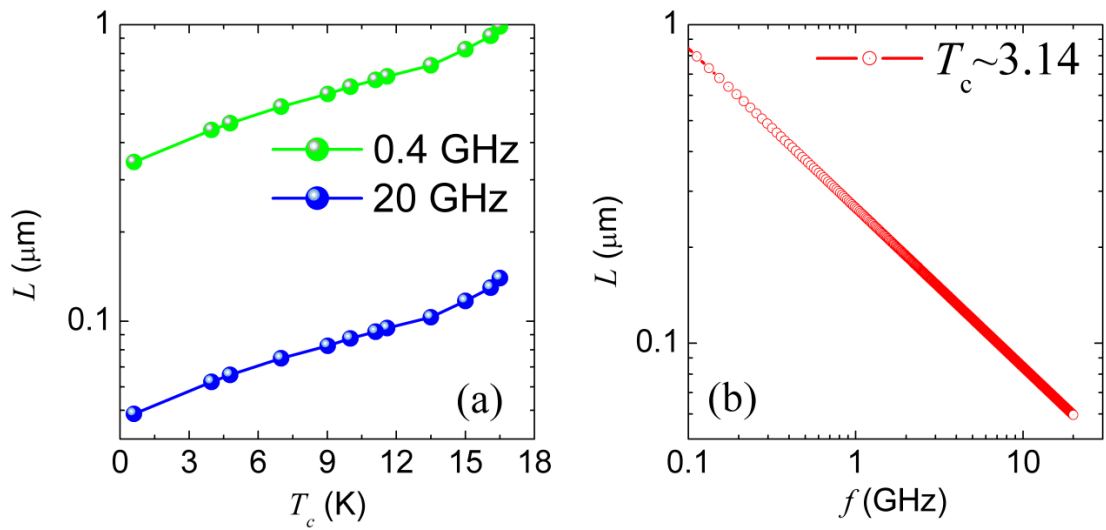

Figure 5.7. (a) shows the probing length, $L$ as a function of SC transition temperature, $T_c$ of disordered NbN thin films at frequency 0.4 GHz and 20 GHz. (b) The probing length, $L$ as a function of frequency, $f$ for the sample with $T_c \sim 3.14$ K.

$$L = \left(\frac{D_{diff}}{2\pi\omega}\right)^{1/2} \tag{5.7}$$

Here, $D_{diff}$ is the electronic diffusion constant given by, $D_{diff} \approx v_F l/d$, where $v_F$ is the Fermi velocity, $l$ is the electronic mean free path and $d$ is the dimension of the films. Since all our films are in 3D limit respect to electronic mean free path, $l \sim 4$ Å, we have calculated probing length by taking $d = 3$ and experimentally measured $l$ and $v_F$ [11]. Fig. 5.7.(a) shows the obtained probing length, $L$ as a function of $T_c$ for a set of samples. Fig. 5.7(b) shows the probing length, $L$ as a function of frequency for a disordered film with $T_c \sim 3.14$ K. With increasing frequency, $L$ decreases from about 420 nm at 0.4 GHz to ~58nm at 20 GHz. Thus using our measurement technique, we can probe the electrodynamics response in local length scale set by $D_{diff}$ and probing frequency, $f$ through measurements of complex conductivity.

## 5.3.2. Experimental results

The complex conductivity, $\sigma(\omega)$ was measured of a set of epitaxial NbN thin films of thickness, $t \geq 50$ nm, with different levels of disorder having $T_c$ varying in the range $T_c \approx 15.7$ to 3.14 K.

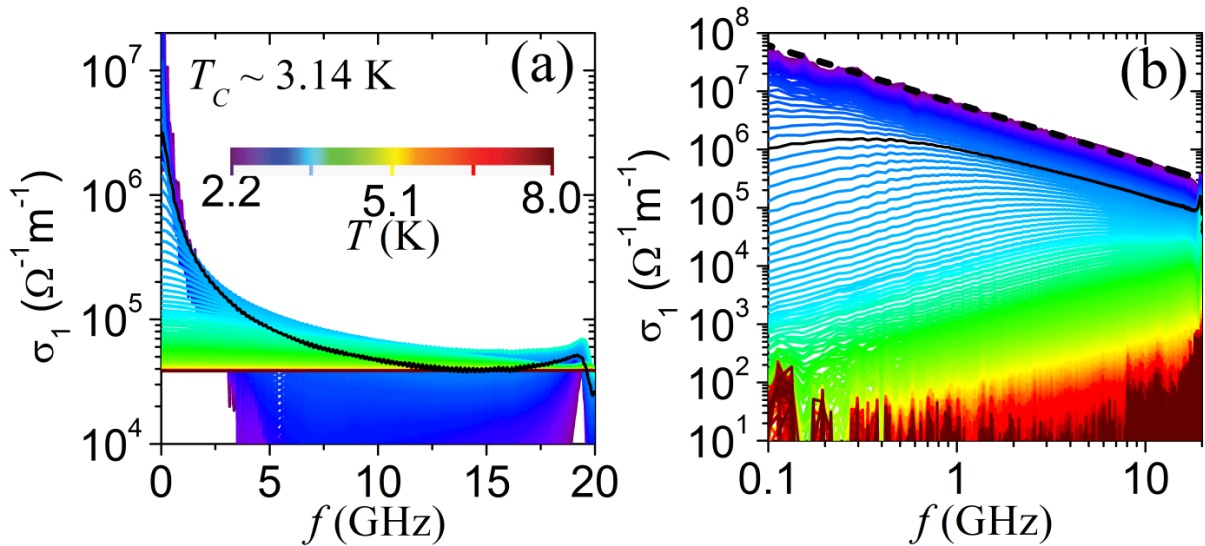

Figure 5.8. Frequency dependence of real and imaginary part of conductivity for a disordered NbN film with $T_c$=3.14 K. The solid (black) lines are the conductivity at $T_c$. In panel (b) dashed black line is 1/ω fit to σ$_2$ below $T_c$. The residual features in conductivity about 19GHz is due to the calibration error of our spectrometer.

Fig. 5.8(a) and (b) show the representative data for $\sigma_1(\omega)$ and $\sigma_2(\omega)$ as functions of frequencies at different temperatures for the sample with $T_c \sim 3.14$ K. At low temperatures $\sigma_1(\omega)$ shows a sharp peak at $\omega \rightarrow 0$ whereas $\sigma_2(\omega)$ varies as $1/\omega$ (dashed line), consistent with the expected behavior in the SC state. Well above $T_c$, $\sigma_1(\omega)$ is flat and featureless and $\sigma_2(\omega)$ is within the noise level of our measurement, consistent with the behavior in a normal metal. In the SC state where phase coherence is established at all length and time scales, the superfluid density ($n_s$) and $J$ can be determined from $\sigma_2(\omega)$ using the relation,

$$\sigma_2\left(\omega\right) = \frac{n_s e^2}{m\omega} \text{ and } J = \frac{\hbar^2 n_s a}{4m},$$ (5.8)

where $e$ and $m$ are the electronic charge and mass respectively, and $a$ is the characteristic length scale associated with phase fluctuations which is of the order of the dirty limit coherence length, $\xi_0$. Fig. 5.9(a)-(d) shows $\sigma_1(\omega)$-$T$, $\sigma_2(\omega)$-$T$ and $J$-$T$ at different frequencies for four samples with different $T_c$. One should notice that, all samples show a dissipative peak in $\sigma_1(\omega)$ close to $T_c$ and the peak becomes more and more prominent as we increase the disorder in our samples. In low disorder samples for all frequencies, $\sigma_2(\omega)$ dropped close to zero at $T_c$. On the other hand samples with higher disorder show an extended fluctuation region where $\sigma_2(\omega)$ remains finite up to a temperature well above $T_c$. We convert $\sigma_2(\omega)$ into $J$ (from eqn. 5.8) using the experimental values of $\xi_0$ (see section 2.4.1.3) [23]. For $T < T_c$, $J$ is frequency independent, showing that the phase is rigid at all length and time scales. However, for the samples with higher disorder (Fig. 5.9(c) and 5.9(d)), $J$ becomes strongly frequency dependent above $T_c$. While at 0.4 GHz $J$ falls below our experimental threshold very close to $T_c$, with increase in frequency, it acquires a long tail and remains finite well above $T_c$.

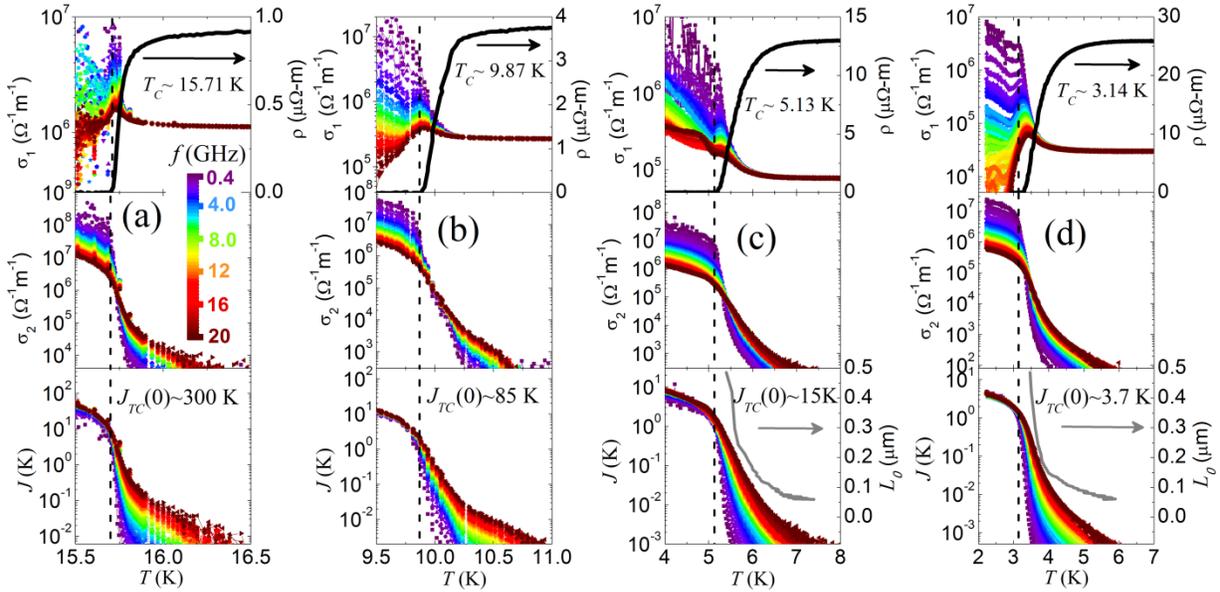

**Figure 5.9.** Temperature dependence $\sigma_1$ (upper panel), $\sigma_2$ (middle panel) and $J$ (lower panel) at different frequencies for four samples with (a) $T_c \sim 15.7$ K (b) $T_c \sim 9.87$ K (c) $T_c \sim 5.13$ K and (d) $T_c \sim 3.14$ K. The corresponding $J_{TC}$ is the zero temperature superfluid stiffness obtained from measurements using low frequency two coil mutual inductance technique. The color scale representing different frequencies is displayed in panel (a). The solid (black) lines in the top panels show the temperature variation of resistivity. Vertical dashed lines correspond to $T_c$. The solid (gray) lines in the bottom panels of (c) and (d) show the variation of $L_0$ above $T_c$.

Above $T_c$ the sample consists of large phase-coherent domains (consisting of several smaller domains) fluctuating with respect to each other. As the temperature is increased further, the large domains progressively fragment eventually reaching the limiting size observed in STS measurements at a temperature close to $T^*$. In such a scenario $J$ will depend on the length scale at which it is probed. When probed on a length scale much larger than the phase coherent domains, $J \rightarrow 0$. On the other hand, when probed at length scale of the order of the domain size $J$ would be finite. Therefore, for each temperature, the frequency-dependent stiffness $J(\omega)$ would vanish at a frequency such that the corresponding $L(\omega)$ becomes much larger than the size of the phase-coherent domains at the same temperature, $L_0(T)$. Using this criterion we can estimate the temperature dependence of $L_0(T) = L(\omega)$, corresponding to the frequency for which $J(\omega)$ goes below our measurement resolution (Fig 5.9 (c) and 5.9(d)). The limiting value of $L_0$ at $T \approx T^*$ is between 50-60 nm which is in the same order of magnitude as the domains observed in STS measurements [28] on NbN films with similar $T_c$ (see Fig. 5.6).

## 5.3.3. Discussion

### 5.3.3.1. Connection with pseudogap state observed in STS study

The STS measurements on disordered NbN thin films [10,11] with $T_c \lesssim 6K$ show a pronounced PG state above $T_c$. To understand the relation between these observations and the PG state observed in STS measurements, we define a temperature, $T_m$, above which $J \leq 5 \times 10^{-4} J(T=T_c)$. The variation of $T_m$ with frequency shows a trend which saturates at high frequencies and can be fitted well with empirical relation of the form, $(T_m(f) - T_c) = A(1 - e^{-f/f_0})$ (Fig. 5.10 (a) and (b)). Using the best fit values of $A$ and $f_0$, we determine the limiting value, $T_m^* = T_m(f \to \infty)$. In Fig. 5.10 (c), we plot $T_m^*$ and $T_c$ for several samples obtained from microwave measurement along with the variation of $T^*$ and $T_c$ obtained from STS measurements, as a function of $k_F l$. Within the error limits of determining these temperatures, $T^* \approx T_m^*$, showing that the onset of the PG state in STS measurements and onset of the finite $J$ at 20 GHz take place at the same temperature. Furthermore, only the samples in the disorder range where a PG state appears, show a difference between $T_c$ and $T_m^*$. We therefore attribute the frequency dependence of $J$ to a fundamental property related to the PG state.

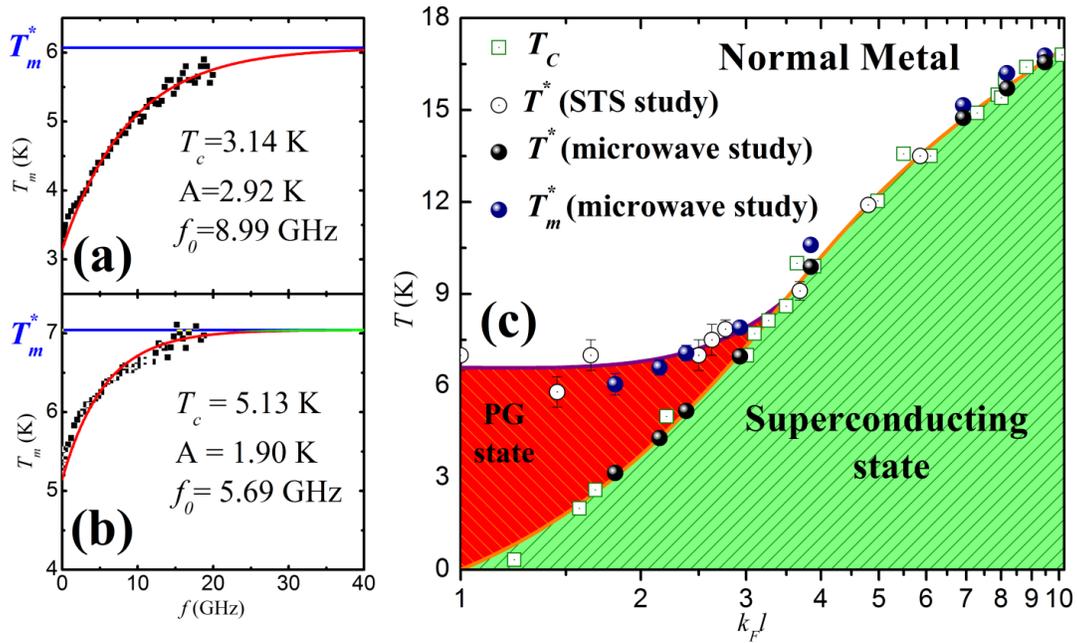

**Figure 5.10.**(a)-(b) Variation of $T_m$ as a function of frequency for the sample with $T_c \sim 3.14$ K and $T_c \sim 5.13$ K respectively. The red line shows a fit of the data with the empirical relation$(T_m(f) - T_c) = A(1 - e^{-f/f_0})$; the best fit values of $A$ and $f_0$ are shown in the panels. The blue lines show $T_m^* = T_m(f \to \infty)$ (c) Phase diagram showing $T_c$ and $T^*$ obtained from STS measurements [Ref.11] as a function of $k_F l$ along with $T_c$ and $T_m^*$ obtained from microwave measurements. Samples with $T_c < 6$ K show a PG state above $T_c$. The PG temperature, $T^*$, obtained from STS and is almost identical to the temperature, $T_m^*$, at which the superfluid stiffness goes below our measurable limit at 20 GHz.

## 5.3.3.2. Fluctuation conductivity above $T_c$

Having established the relation between the PG state and the finite high frequency phase stiffness, we now concentrate on the fluctuation region above $T_c$. A superconductor above $T_c$ shows excess conductivity due to presence of unstable SC pairs induced by fluctuations. The fluctuation component of conductivity can be obtained from the experimentally measured conductivity by subtracting the normal state conductivity, $\sigma_N$, as,

$$\sigma_{fl}(\omega,T) = \sigma(\omega,T) - \sigma_N , \tag{5.9}$$

Here the normal state conductivity, $\sigma_N$, is the experimentally measured conductivity at temperature $T{\sim}1.5\ T_c$ for low disordered films and for strongly disordered superconductors $T \sim 1.5\ T^*$ where $T^*$ is the PG gap temperature.

### 5.3.3.2.1. Zero frequency fluctuation conductivity

The fluctuation conductivity in dc electric filed due to unstable SC pairs in a dirty superconductor predicted by Aslamazov and Larkin (AL) [29],

$$\sigma_{dc,fl}^{2D\,AL} = \frac{1}{16}\frac{e^2}{\hbar t}\varepsilon^{-1} \tag{5.10}$$

$$\sigma_{dc,fl}^{3D\,AL} = \frac{1}{32}\frac{e^2}{\hbar \xi_0}\varepsilon^{-1/2} \tag{5.11}$$

where $\varepsilon = \ln(T/T_c)$, $t$ is the thickness of the sample and $\xi_0$ is the BCS coherence length. This fluctuation conductivity is the result of direct acceleration of unstable SC pairs. Since all our films are in dirty limit, the Maki-Thomson contribution is negligible.

Fig. 5.11 (a)-(d) show the dc fluctuation conductivity ($\sigma_{fl}^{dc}$) for four samples on which we have done microwave measurements, obtained from measured two probe resistivity, $\rho$ vs $T$ in the same run. In the case of less disordered sample, the $\sigma_{fl}^{dc}(T)$ follow the 2D AL prediction very well [Fig. 5.11 (a) and (b)] instead of the 3D AL prediction as the correlation length above $T_c$ becomes very large and effectively the sample behaves as 2D. However when we increase the disorder, the temperature dependence of $\sigma_{fl}^{dc}(T)$ starts to deviate from AL predictions and in very strong disorder sample, the $\sigma_{fl}^{dc}(T)$ decreases in much slower rate with temperature than the expected temperature variation from AL prediction. This anomalous behavior of $\sigma_{fl}^{dc}(T)$ with respect to temperature, gives us indication that in a strongly disordered superconductor, the amplitude fluctuations alone can't explain the fluctuation region above $T_c$.

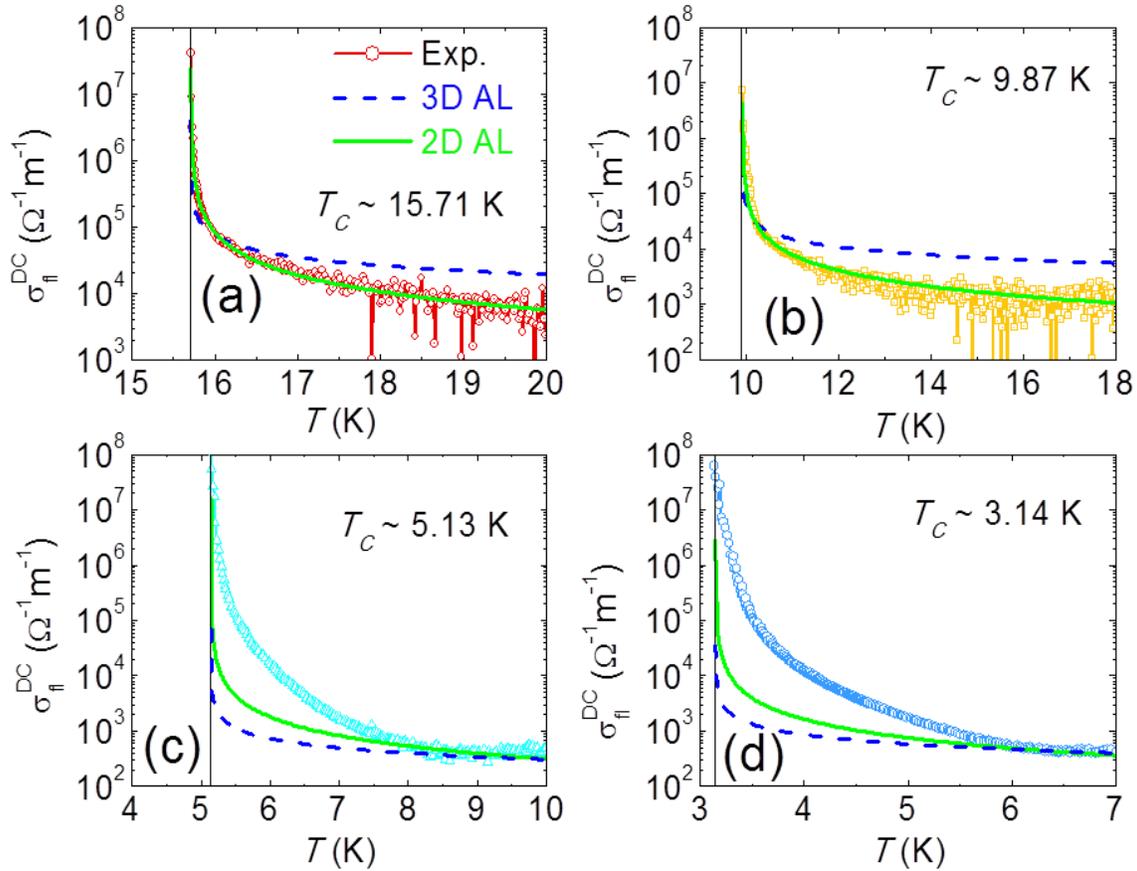

Figure 5.11. Temperature dependence of DC fluctuation conductivity {$\sigma_{fl}^{dc} = \sigma^{dc}(T) - \sigma_N$} for four samples with (a) $T_c \sim 15.7$ K (b) $T_c \sim 9.87$ K (c) $T_c \sim 5.13$ K and (d) $T_c \sim 3.14$ K. The scattered color plots are experimental data. The solid green lines are theoretical fits according to 2D AL prediction and dashed blue lines are for 3D Al prediction of fluctuation conductivity.

## 5.3.3.2.2. Finite frequency fluctuation conductivity

For further understanding about the fluctuation region, we now concentrate on the frequency dependence of fluctuation conductivity. To study the critical fluctuation region, D. Fisher, M. Fisher and D. Huse [30] have proposed a dynamical scaling theory where the fluctuation conductivity, $\sigma_{fl}(\omega)$ is predicted to scale as,

$$\sigma_{fl}(\omega)/\sigma_{fl}(0) = S(\omega/\omega_0) \qquad (5.12)$$

Here $\omega_0$ is the characteristic fluctuation frequency, $\sigma_{fl}(0)$ is the zero frequency fluctuation

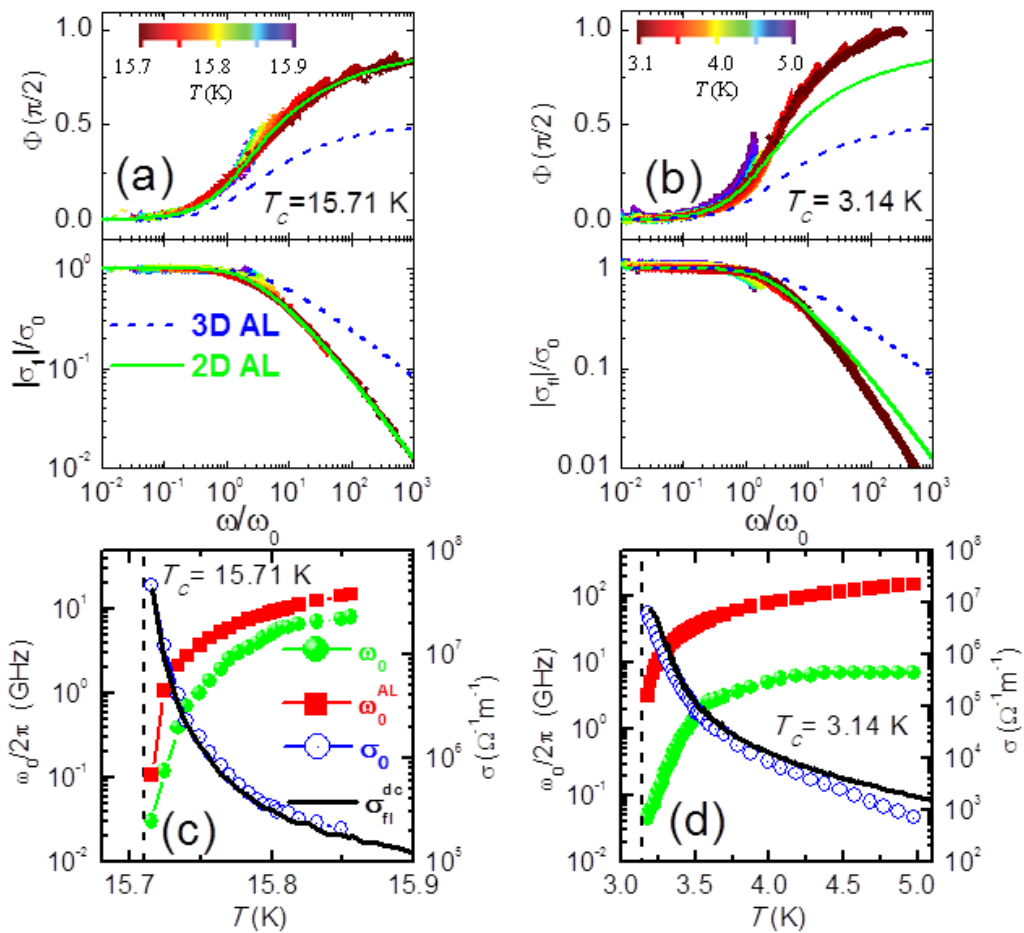

**Figure 5.12** (a)-(b) Rescaled phase (upper panel) and amplitude (lower panel) of $\sigma_{fl}(\omega)$ using the dynamic scaling analysis on two films with $T_c \sim 15.7$ and $3.14$ K respectively. The solid lines show the predictions from 2D and 3D AL theory. The color coded temperature scale for the scaled curves is shown in each panel. (c)-(d) Show the variation of $\omega_0$, $\omega_0^{AL}$, $\sigma_0$ and $\sigma_{dc}$ as function of temperature. The dashed vertical lines correspond to $T_c$.

conductivity at that temperature and $S$ is the universal scaling function. We experimentally obtain $\sigma_{fl}(\omega)$ from measured $\sigma(\omega)$ by subtracting the normal state dc conductivity, $\sigma_N$ using eqn. (5.9). Since the phase angle $\phi(\omega) = \tan^{-1}(\sigma_{fl}^2 / \sigma_{fl}^1)$ is the same as the phase angle of $S$, $\phi(\omega)$ expected to collapse into single curve by scaling $\omega$ differently at each temperature. For the amplitude, the data would similarly scale when normalized by $\sigma_{fl}(0)$. Such a scaling works for all the samples in the frequency range 0.4 GHz to 12 GHz, as shown in Fig. 5.12(a) and (b) for the samples with $T_c \sim 15.7$ K and 3.14 K respectively. Fig. 5.12(c) and (d) show the variation of $\omega_0$ and $\sigma_{fl}(0)$ obtained for the best scaling of the data. In both plots $\omega_0 \rightarrow 0$ as $T_c$ is approached from above showing the critical slowing of fluctuations as the SC transition is approached. We observe a perfect consistency between the temperature variation of $\sigma_{fl}(0)$ and dc fluctuation conductivity, $\sigma_{fl}^{dc}$.

The scaling function $S$ is constrained by the physics of the low and high frequency limits: For $\omega \rightarrow 0$, $S \rightarrow 1$ corresponding to the normal state conductivity, and for $\omega \rightarrow \infty$, $S \rightarrow c(\omega/\omega_0)^{[(d-2)/z-1]}$ where c is a complex constant, $d$ is the dimension and $z$ is the dynamical exponent, with $z = 2$ for models based on relaxational dynamics. This is the case for example of ordinary Ginzburg-Landau (GL) amplitude and phase fluctuations which are the possible candidates for the observed fluctuation conductivity, at least in moderate disorder limit[31]. Indeed, a direct comparison with the data for the sample with $T_c \sim 15.71$ K shows that $S$ matches very well with the Ashlamazov-Larkin (AL) prediction in $d=2$ dimensions (see section 1.6.1), whereas the Maki-Thomson (MT) corrections are suppressed due to disorder, in agreement with earlier measurements on low-disorder NbN films [31]. Since the correlation length above $T_c$ is larger than the film thickness $\sim 50$ nm, the sample behaves as effectively 2D system. On the other hand, the corresponding curve for the sample with $T_c \sim 3.14$ K does not match with any of these models. Whereas for both samples $\omega_0 \rightarrow 0$ as $T \rightarrow T_c$, showing the critical slowing of fluctuations as the SC transition is approached, a clue as to the origin of this deviation is obtained from a comparison of the temperature variation of $\omega_0$ (Fig. 5.12(c) and 5.12(d)) with the prediction from 2D AL theory, i.e.

$$\omega_0^{AL} = \frac{16 k_B T_c}{\pi \hbar} \ln\left(\frac{T}{T_c}\right) \tag{5.13}$$

Whereas for the film with $T_c \sim 15.71$ K the best scaling values of $\omega_0$ is in agreement with $\omega_0^{AL}$ within a factor of the order of unity, in the films with $T_c \sim 3.14$ K the $\omega_0$ is more than one order of magnitude smaller than $\omega_0^{AL}$. The low characteristic fluctuation in the disordered sample signals a breakdown of the amplitude fluctuation scenario, which cannot be accounted for by any simple adjustment of parameters in these models. Therefore in our sample the superconducting transition is governed by longitudinal phase fluctuations between superconducting domains and gives rise to novel PG state.

### *Fluctuation conductivity and Berezinskii-Kosterlitz-Thouless (BKT) vortex fluctuations:*

In the above discussion, we have proposed that superconducting transition is driven by mainly longitudinal phase fluctuations between domains. Nevertheless transverse (vortical) phase fluctuations of BKT type also can destroy the superconductivity giving rise to similar type of novel PG state above $T_c$. However the transverse phase fluctuations are expected to be relevant only in a small range of temperatures above $T_c$, as we demonstrated in a recent analysis of the BKT transition in ultra thin NbN films in chapter 4. Here we show in more details why not only the ordinary GL theory but also the standard BKT one also fails in explaining the fluctuation regime at strong disorder.

Fig. 5.13(a) shows the superfluid density at various frequencies of most disordered sample with $T_c \sim 3.14$ K. The BKT jump in superfluid stiffness, $J$, is expected to occur in the zero-frequency limit given by [27],

$$J = \frac{\hbar^2 n_s t}{4m} = \frac{2T}{\pi} \tag{5.14}$$

Since the BKT transition occurs in 2D system where the thickness, $t$, plays the role of characteristic length scale of fluctuations, with respect to eqn. (5.1) we have replaced $a$ with $t$. When the superfluid stiffness is probed at finite frequency as in our case the universal jump in eqn. (5.14) is expected to be smeared out[6,32,33]. In particular, $J$ measured at different frequencies start to deviate from each other at the temperature, $T_V$, where (bound) vortex-antivortex pairs become thermally excited. As we discussed in the case of thin films [33], the $T_V$ is usually smaller than the real BKT critical temperature due to a small value of the vortex-core energy. This can be seen in Fig. 5.13(b) where we report for comparison also the data for a 3 nm thick NbN sample measured using two coil mutual inductance technique. Here $T_V$ can be easily identified by the temperature where experimental data deviates from the BCS fit valid at lower temperatures. Notice that the downturn of $J$ at $T_V$ observed in this thin film (panel b) is much more pronounced than the smooth temperature dependence observed in our thick sample (panel a) even at the lowest accessible frequency. In the case of the 3nm sample one can also estimate the mean-field BCS critical temperature $T_{BCS}$ by extrapolating the BCS fit. The BKT fluctuations extend only up to $T_{BCS}$ (see section 1.6.2.3.2), as it is seen in Fig. 5.13(b) where the BKT regime is less than 1 K wide. Therefore in our much thicker samples this range of BKT fluctuations is expected to be even smaller, consistent with the fact that $T_c$ approaches rapidly $T_{BCS}$ when the film thickness increases.

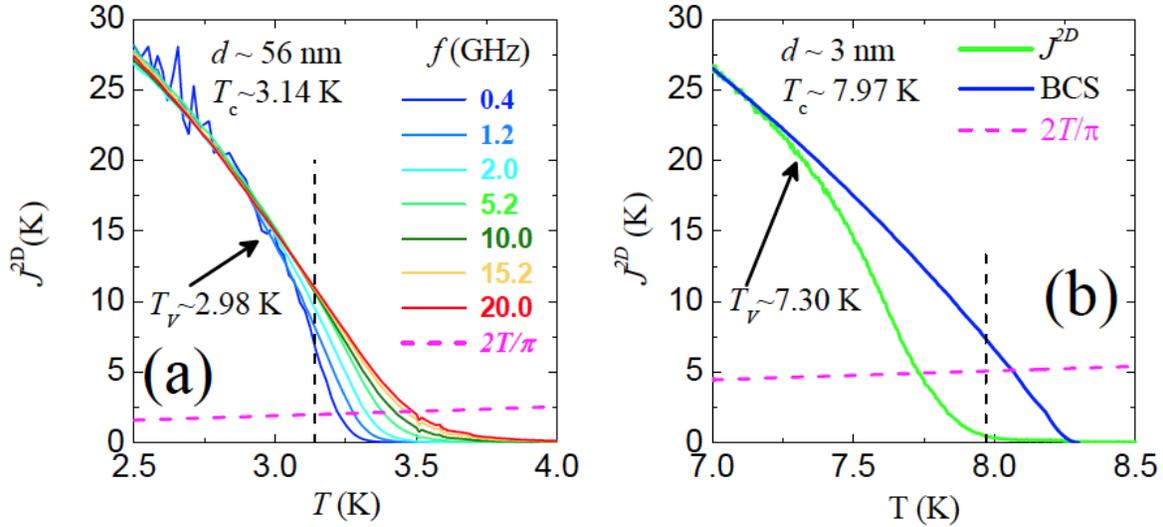

Figure 5.13. (a) Superfluid stiffness of the $T_c$=3.14 K sample as given in Eq. (5.14) at various frequencies. The BKT transition is expected to occur when the zero-frequency stiffness intersect the universal line $2T/\pi$, see Eq. (9). The effect of vortex-antivortex pairs is expected to occur at a lower temperature $T_V$, where the frequency dependence of the data starts to develop. (b) Superfluid stiffness of a thin 3nm NbN sample taken from Ref. [27]. In this case the temperature $T_V$ can be easily identified by the rapid downturn of the data with respect to the BCS fit valid far from the transition. In both panels the dashed vertical line marks the dc $T_c$.

To further support this conclusion, we now concentrate on scaling analysis of fluctuation conductivity of most disordered samples. Within BKT theory the fluctuation conductivity can be written as [32]:

$$\sigma = \frac{4e^2}{\hbar^2 d} \frac{J_0(T)}{\Omega(T)} \frac{1}{1 - i\omega/\Omega(T)} = \frac{4e^2}{\hbar^2 d} \frac{J_0(T)}{\Omega(T)} S(\omega/\Omega(T)), \quad S(x) = \frac{1}{1-ix} \qquad (5.15)$$

where

$$\Omega(T) = \frac{4\pi^2 J_0(T)}{k_B T} \frac{D_V}{\xi^2(T)} \qquad (5.16)$$

Here $J_0(T)$ is the "bare" superfluid density in absence of vortex fluctuations, $D_V \sim \hbar/m \sim 10^{-4} \ m^2/s$ is the vortex diffusion constant [33] and $\xi(T)$ is the BKT correlation length. A direct comparison between the data and $S$ (eqn. 15) shows that the scaled amplitude and phase do not follow the prediction from BTK theory (see Fig 5.14(a-b)). We also note that the value of the vortex diffusion constant is of the same order of the electron diffusion constant. In this respect, the estimate of $L_0$ shown in Fig. 5.9 is independent on the possible specific nature (longitudinal vs transverse) of the phase fluctuations in the pseudogap state. This demonstrates that the enhanced fluctuation regime observed in the pseduogap regime cannot be accounted for by the standard BKT approach.

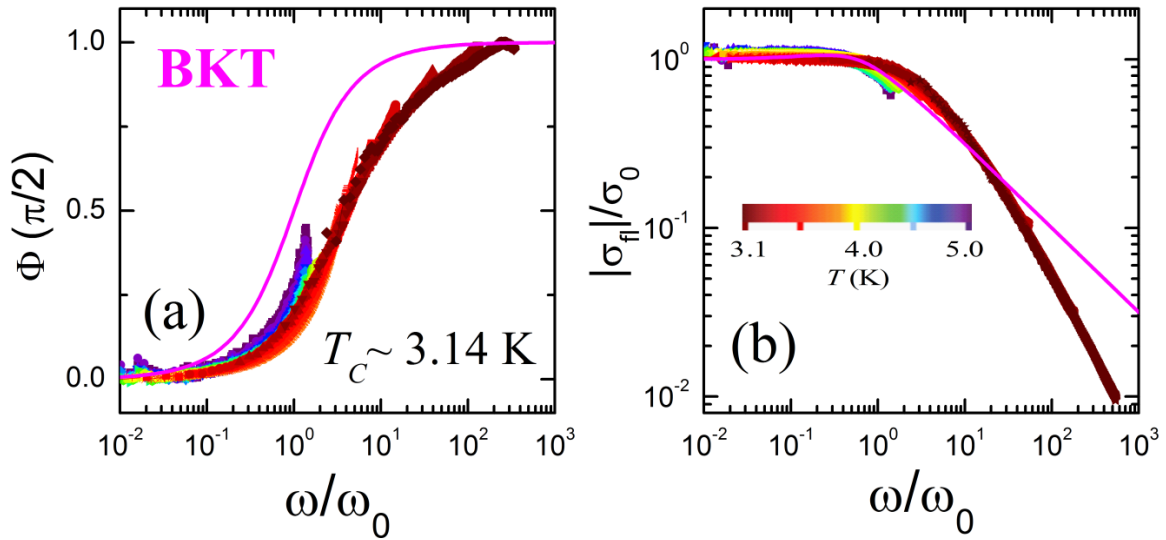

Figure 5.14. (a)-(b) Rescaled phase (a) and amplitude (b) of $\sigma_{fl}(\omega)$ using the dynamic scaling analysis on the film with $T_c \sim 3.14$ K. The solid line shows the prediction from BKT theory. The color coded temperature scale for the scaled curves is shown in panel (b).

### 5.3.2. Summary of high frequency electrodynamics response

In a strongly disordered NbN thin films which display PG state, $J$ becomes dependent on the temporal and spatial length scale in the temperature range, $T_c < T < T^*$. The remarkable agreement between $T^*$ determined from STS and $T^*_m$ from microwave measurements is consistent with the notion that the SC transition in these systems is driven by phase disordering due to longitudinal phase fluctuations.

## 5.4. Conclusions

We have shown that a 3D conventional s-wave superconductor, NbN thin film progressively becomes susceptible to phase fluctuations with increasing disorder. In the strong disorder limit, the superconductivity is destroyed by phase disordering due to longitudinal phase fluctuations although $|\Delta|$ remains finite well above $T_c$ contrary to BCS prediction, which gives rise to a novel pseudogap state with finite Cooper pair density above $T_c$ but no global superconductivity.

# CHAPTER 6

## Conclusions, open questions and future directions

In this thesis we have explored the role of phase fluctuations induced by reduced dimensionality and homogeneous disorder on superconducting properties of conventional s-wave superconductor, NbN thin films through measurements of finite frequency electrodynamics response.

We have shown that the superconducting transition in 2D superconductors, which is predicted to be governed by transverse phase (vortex) fluctuations belonging to Berezinskii-Kosterlitz-Thouless (BKT) universality class, greatly influenced by the vortex core energy and inhomogeneity in the system [1]. Therefore, to elucidate the nature of BKT transition in real 2D superconductors, correct value of vortex core energy should be taken into account. We would like to note that many investigations on BKT transition in 2D superconductors rely on the effect of BKT fluctuations on transport properties above $T_{BKT}$. Our study underpins the need to constrain the BKT fitting parameters using an independent measurement such as the temperature variation of superfluid density below $T_{BKT}$, while analyzing the fluctuation regime above $T_{TBK}$. In the absence of such a constraint, one risk attributing the entire fluctuation regime including GL fluctuations to the BKT fluctuations, while BKT fluctuations only important in the narrow temperature range in between $T_{BKT}$ and $T_{BCS}$. Our study also indicates that disorder in the system can modify the vortex core energy. Therefore, it would be interesting to investigate the effect of disorder on vortex core energy by changing disorder while keeping thickness fixed in future study.

In strongly disordered 3D NbN thin films, we have shown that the superconducting transition is governed by long wave length phase fluctuations giving rise to a novel psedogap state with frequency dependent superfluid stiffness above $T_c$ but no global superconductivity [2,3,4]. More detailed picture of this novel pseudogap state can be revealed by scanning tunneling spectroscopy which is being carried out by other group members [5]. We have also observed that above a critical value of disorder, the superconducting ground state is completely destroyed and giving rise to a quantum critical point at the Superconductor-Insulator transition.

The detailed investigation close to the quantum critical point using microwave measurements at very low $T$ may reveal novel physics associated with the quantum phase transition.

It would also be interesting to carry out similar investigation in other system, for example ultrathin TiN films [6] and strongly disordered InO$_x$ films [7], to see up to what extent the phase fluctuations influence the superconducting properties in these materials. Finally, I would like to note that many of these observations are similar to under doped high temperature superconductors [8]. It would therefore be interesting to explore, through similar measurements, whether the superconducting transition is driven by phase disordering

The role of disorder on the superconducting properties is a well studied problem and over the decade the supersession of superconductivity due to strong disorder or reduced dimensionality has been understood to a great extent. However there is an outstanding effect of disorder in some superconductors such as Al, In, Sb, Be, W etc [9,10] where superconducting transition temperature is enhanced with reducing dimensionality or increasing disorder. The most dramatic effect has been observed in W where $T_c$ is enhanced by several orders of magnitude from ~10 mK for clean bulk W to ~ 6 K for disordered film of W [11]. So far there is very little understanding on the physical mechanism giving rise to this enhancement [12].In my opinion, this anomalous enhancement of $T_c$ in these materials should form the future direction in the study of disordered superconductors in days to come.

# APPENDIX A

## Point contact Andreev reflection spectroscopy on a non-centrosymmetric superconductor, BiPd

*During my PhD, when I was developing our broadband microwave spectrometer, our old Vector Network Analyzer (VNA) broke down. The very old VNA could not be repaired and we had to purchase a new VNA which took almost a year to procure. In the mean time, there was an interesting development in our department. One of my friends, Bhanu Joshi working with Prof. Srinivasan Ramakrishnan and Dr. Arumugam Thamizhavel had grown very good single crystal of noncentrosymmetric superconductor (NCS), BiPd. Noncentrosymmetric superconductors having no inversion symmetry are predicted to show many exciting exotic physics (for example, mixing of singlet and triplet order parameter, topologically protected state, novel mixed state and many more). However experimental evidences are so far very little. I took this opportunity to explore this new field using point contact Andreev reflection spectroscopy and observed interesting new phenomenon which will be discussed in the following sections. I believe, our study is a significant step towards understanding of exotic physics theoretically predicted in the family of noncentrosymmetric superconductors.*

*This work [1] doesn't have direct correlation with the rest of my PhD thesis; therefore I am putting it in the appendix.*

## A1. Introduction

The discovery of superconductivity in the non-centrosymmetric superconductor (NCS) $CePt_3Si$[2], has generated widespread interest in this class of systems. In superconductors where inversion symmetry is present, the superconducting order parameter (OP) is characterized by a distinct parity corresponding to either a spin-singlet or a spin-triplet pairing. However in NCS, the lack of inversion symmetry combined with antisymmetric (Rashba-type) spin orbit coupling (ASOC) [3] can cause an admixture of the spin-singlet and spin-triplet pairing [4]. In the simplest situation of a single band contributing to superconductivity, this mixing is expected to give rise to a two component OP. In a real system, the order parameter would therefore have two or more components, depending on the complexity of the Fermi surface, giving rise to unusual temperature and field dependence of superconducting parameters [5,6,7,8,9,10,11,12].

Despite numerous theoretical predictions, experimental evidence of an unconventional superconducting state in NCS has been very few, possibly due to the small spin-orbit coupling. The vast majority of NCS, (e.g. $Re_3W$, $Mg_{10}Ir_{19}B_{16}$, $Mo_3Al_2C$, $Re_{24}Nb_5$) display predominantly conventional s-wave behavior and occasionally multiband superconductivity [13,14,15,16]. In some systems such as $CePt_3Si$ [ref. 17] and UIr [ref. 18], the study of parity broken superconductivity is complicated by strong electronic correlations and by the coexistence of magnetism. One notable exception is $Li_2Pt_3B$, in which penetration depth [19,20] and nuclear magnetic resonance [21] measurements provide evidence for the existence of nodes in the gap function. However, a direct spectroscopic evidence for the presence of unconventional order parameter has not been reported for any of these materials.

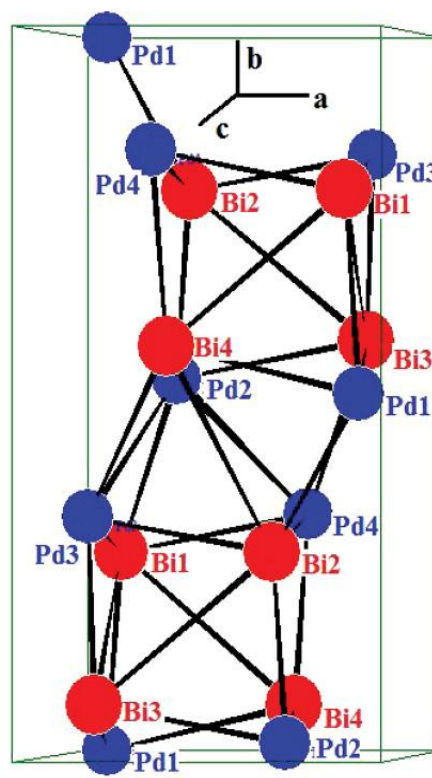

In this work, the directional Point Contact Andreev reflection (PCAR) measurements were carried out on a BiPd single crystal [ref. 22], which is a recent addition to the family of NCS superconductors. Since the spin-orbit coupling depends on the square of the atomic number of the elements involved, the presence of Bi (Z=83) is expected to result in a large spin-orbit coupling in this material. Fig. A1. shows the crystal structure of $\alpha$-BiPd which has a monoclinic crystal structure with lattice constants $a$=5.63Å, $b$=10.66Å, $c$=5.68Å, $\alpha=\gamma=90^0$, $\beta=101^0$. Recent thermodynamic and transport measurements[22] on high quality BiPd single crystals (residual resistivity, $\rho$~0.3μΩ-cm and residual resistivity ratio ~160) revealed that the specific heat jump at $T_c$ is much smaller than expected for a BCS superconductor,

Figure A1. showing the Monoclinic (P21) crystal structure of $\alpha$-BiPd single crystal. It has 16 atoms in a unit cell (8 formula unit) with four inequivalent Bi sites and four inequivalent Pd sites (Bhanu Joshi et al, (2011)) (figure is adapted from ref. 22).

suggesting the possibility of multiple superconducting order parameters in this material.

Directional point contact Andreev reflection (PCAR) spectroscopy[23], i.e. where the conductance spectra ($dI/dV$ versus $V$) are recorded by injecting current from a normal metal through a ballistic point contact along different crystallographic directions in the superconductor, is a powerful tool to investigate the gap anisotropy in superconductors[24,25]. In this work, PCAR spectra were recorded on a BiPd single crystal by injecting current either along $b$ (I∥$b$) or perpendicular to $b$ (L⊥$b$). The central observation from these studies is the presence of a pronounced zero bias conductance peak (ZBCP) in both crystallographic directions, which coexists with more conventional gap-like features. Our results strongly suggest that a spin triplet OP coexists with a spin singlet OP in this material.

## A2. Experimental details

High quality BiPd single crystal was grown by Bhanu Joshi et al (2011) [22] using modified Bridgeman technique (for details about crystal growth see ref [22]). The directional point contact measurements were carried out on a piece of single crystal cut into a rectangular parallelepiped shaped of size 1mm×1.5mm×2mm which had large well oriented faces on the (010) and (001) planes. The superconducting transition temperature, $T_c$ ~ 3.62K of the crystal was determined by measuring ac susceptibility at 60 kHz using a two coil mutual inductance technique[26]. From the resistivity and specific heat measurements on a similar crystal, we estimate the electronic mean free path[22], $l$~2.4 μm at low temperatures. The quality of the crystal was also confirmed by observing de Haas-van Alphen (dHvA) oscillation [27]. Before doing point contact measurement, the crystal surface was polished to a mirror finish. To make ballistic point contact, a mechanically cut fine tip made from 0.25mm diameter Ag wire was brought in contact with the crystal using a differential screw arrangement in a conventional sample-in-liquid [3]He cryostat. Measurements were performed by making the contact on two different crystal faces: (i) (010) corresponding approximately to I∥$b$ and (ii) (001) corresponding approximately to L⊥$b$. I-V characteristics of the junction formed between the tip and the sample were measured at different temperature down to $T$=0.32 K using conventional 4-probe technique. The dI/dV vs. V spectra was obtained by numerically differentiating the I-V curves. For all spectra reported here, the contact resistance ($R_c$) in the normal state varied in the range $R_c$~1Ω to 30Ω. The corresponding contact diameter estimated using the formula [28], $d = \left(\frac{4\rho l}{3\pi R_c}\right)^{1/2}$ ~ 100-500Å, was much smaller than $l$. Therefore, all our point contact spectra are taken in ballistic limit. To further understand the nature of superconductivity, we have measured the upper critical field ($H_{c2}$) and its anisotropy along two crystallographic axes (H∥$b$ and H⊥$b$) by measuring ac susceptibility as function of magnetic field at different temperatures.

## A3. Results and discussion

We first concentrate on the PCAR spectra at the lowest temperature. From large statistics, we observe two kinds of PCAR spectra, corresponding to I∥$b$ and I⊥$b$ respectively. Statistically, we did not find any difference between the two orthogonal directions corresponding to I⊥$b$, with the facets approximately along [001] and [100] respectively. Fig. A2. (a) and (b) show representative evolution with contact resistance for ($dI/dV$) versus $V$ spectra for I∥b and I⊥b respectively. In both directions, the striking feature is the observation of a pronounced zero bias conductance peak (ZBCP) which coexists with more conventional gap-like features in the low $R_c$ contacts ((Fig. A2. (a) and (d))). In addition, for I‖$b$, clear coherence peaks associated with superconducting gaps are observed around 0.1 meV and 0.4 meV respectively. For L⊥$b$, the corresponding structures are observed at 0.4 meV and 0.8 meV respectively. Both the ZBCP and gap features disappear at the bulk $T_c$ confirming their superconducting origin. As the contact resistance increased by gradually withdrawing the tip in both directions the ZBCP slowly vanishes and we recover spectra with only gap-like features. To quantitatively obtain the values of the superconducting energy gaps, we fit the spectra using a two-band Blonder-Thinkham-Klapwijk (BTK) model[29,30] generalized to take into account broadening effects. In this model, the normalized conductance ($G(V)/G_N$, where $G_N=G(V>>\Delta)$) is a weighted sum of the conductance of two transport channels ($G_1(V)$ and $G_2(V)$) arising from the two order parameters: $G(V)/G_N = (1-w)G_1(V)/G_{1N} + wG_2(V)/G_{2N}$. $G_1(V)/G_{1N}$ and $G_2(V)/G_{2N}$ are calculated using the generalized BTK formalism using the relative weight factor of the two gaps ($w$) superconducting energy gaps ($\Delta_1$ and $\Delta_2$), the barrier potentials ($Z_1$ and $Z_2$) and the broadening parameters ($\Gamma_1$ and $\Gamma_2$) as fitting parameters. All spectra can be fitted very well with this two-band model if we neglect the large ZBCP that arises for contact with low $R_c$. Analyzing more than 50 spectra along I‖$b$ and L⊥$b$ (Fig. A2. (c) and 1(d)), we observed that the dominant feature is a gap, $\Delta_1$~0.4±0.1 meV present along both directions. For I‖$b$, in a about 50% of the spectra we can clearly resolve a smaller gap, $\Delta_2$~0.1±0.05 meV with $w$~0.2-0.6. On the other hand in 50% of the spectra along L⊥$b$, we can clearly resolve a larger gap $\Delta_3$~0.8±0.15 meV with $w$~0.1-0.35. We did not obtain any spectra showing the three gaps simultaneously in the same spectra. The large variation in $w$ and the dispersion in gap values arise from surface roughness which limits our inability to precisely inject current along a desired direction. Fig. A2. (e) and (f) show the temperature dependence of the ZBCP for two representative low $R_c$ contacts, corresponding to I‖$b$ and L⊥$b$ respectively. The ZBCP decreases with increasing temperature and disappears at about $0.7T_c$.

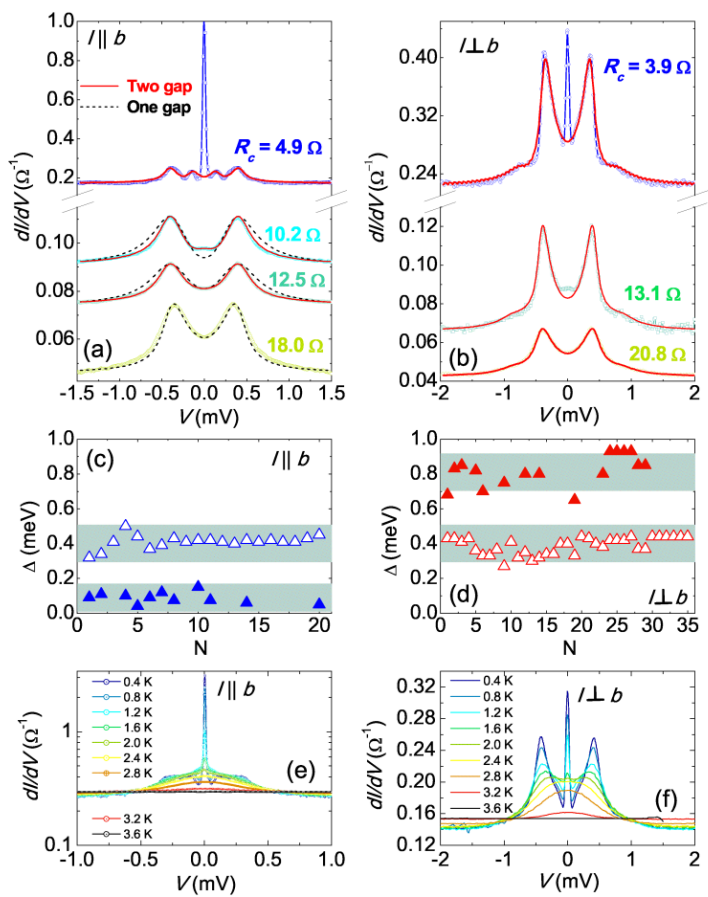

Figure A2. PCAR spectra for different $R_c$ at $T$~ 0.35 K: (a) I∥$b$ and (b) I⊥$b$. Solid lines (red) are fits to the modified two gap BTK model. The corresponding fits with a single gap model (black dash lines) are also shown for comparison for some of the spectra. (c) and (d) Scatter plot of SC energy gap obtained by fitting the modified BTK model to the experimental PCAR spectra for I∥$b$ and I⊥$b$ plotted as a function of the serial number of the spectra. The bands are guides to the eye. (e) and (f) Temperature dependence of the PCAR spectra for two low $R_c$ contacts for I∥$b$ and I⊥$b$ respectively.

To obtain the temperature variation of the superconducting energy gaps we analyze the temperature dependence of two point contacts along the two directions with large $R_c$ (Fig. A3), where the ZBCP is suppressed. Consistent with the notation used for the superconducting energy gaps we denote the barrier parameter and the broadening parameters as $Z_1$ and $\Gamma_1$ (associated with $\Delta_1$), $Z_2$ and $\Gamma_2$ (associated with $\Delta_2$) and, $Z_3$, $\Gamma_3$ (associated with $\Delta_3$). A comparison between the single and two gap fits (Fig. A3. (a) and (d)) of the spectra at the lowest temperatures shows that a single gap is clearly inadequate to fit the spectra. Fig. A3. (b) and (e) show the two-gap fits of the spectra at various temperatures for I$\|b$ and L$\perp b$ respectively. As expected, $Z_1$, $Z_2$ and $Z_3$ (Fig. A3. (g) and (h)) are temperature independent [31] for both I$\|b$ and L$\perp b$. The temperature variation of $\Gamma_1$, $\Gamma_2$ and $\Gamma_3$ on the other hand is more complicated. Formally, the broadening parameters are introduced as an inverse lifetime [32] of the excited quasi particles. From this perspective one expects this parameter to be small at low temperatures and increase close to $T_c$ due to recombination of electron and hole-like Bogoliubons. This is consistent with the temperature variation of $\Gamma_1$ in both Fig. A3. (g) and (h). However, $\Gamma_2$ and $\Gamma_3$ decrease with temperature. To understand this discrepancy we note that phenomenologically the broadening parameter take into account all non-thermal sources of broadening, such as a distribution of gap function resulting from anisotropic gap function[24] and instrumental broadening. For a strongly anisotropic gap function, with increase in temperature, intraband scattering can partially smear out the gap anisotropy thereby causing the broadening parameter to decrease. In the present case we can also not rule out the possibility that the discrepancy is an artifact arising from the fact that our fits assume $k$-independent $\Delta_1$ and $\Delta_2$, where the anisotropy of the gap functions are ignored. Figures A3. (c) and (f) show the temperature dependence of $\Delta_1$, $\Delta_2$ and $\Delta_3$ and Figure A3. (i) shows the relative weight factors corresponding to the two-gap fits. It is instructive to note that $\Delta_1$ has similar temperature variation for both I$\|b$ and L$\perp b$ and closes at $T_c$, confirming that this gap is associated with the same gap function. For I$\|b$, $w$ remains constant with temperature whereas $\Delta_2$ decreases rapidly at low temperatures and forms a tail towards $T_c$ as expected for a multiband superconductor. For L$\perp b$, $w$ decreases with increasing temperature and above 1.6 K all the spectra can be effectively fitted with a single gap, $\Delta_1$.

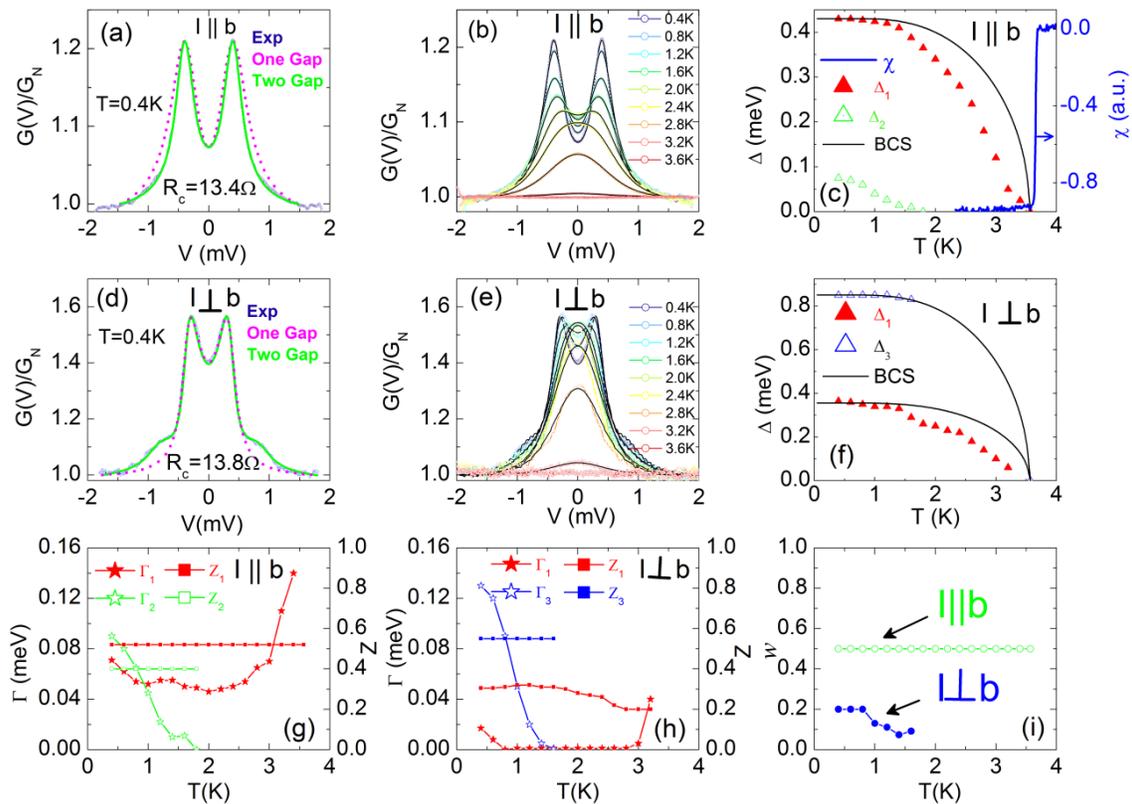

Figure A3. Comparison between one-gap and two-gap fits for PCAR spectra recorded at the lowest temperature for (a) I‖b and (d) I⊥b. Temperature variation of the PCAR spectra for (b) I‖b and (e) I⊥b. (c) and (f) show the temperature dependence of Δ extracted from the (b) and (e) respectively; the solid black lines are the expected BCS variation of Δ. (g) and (h) show the temperature variation of the broadening parameters and the barrier parameters obtained from the best fits of the spectra in (b) and (e) respectively. (i) Temperature variation of the relative weight factor for the two gaps used in the best fit of the spectra in (b) and (e). The blue solid line in panel (c) shows the real part of the ac susceptibility (χ) as a function of T showing the superconducting transition of BiPd.

### A3.1. Signature of Andreev bound states: Zero bias conductance peak

We now focus on the origin of the ZBCP in the low resistance spectra. Since ZBCP can arise from several origins it is important to analyze the observed ZBCP in BiPd critically. First, we look for extrinsic origins of the ZBCP that are not associated with genuine spectroscopic features. It has been shown that in the case where the point contact is not purely in the ballistic limit, ZBCP can arise from the current reaching the critical current [33] ($I_c$) of the point contact. However, in our case such a possibility can be trivially ruled out for two reasons. First, as we have shown before our contact is well in the ballistic limit even after considering error associated with our determination of contact diameter from $R_s$. More importantly, the conductance spectra at currents larger than $I_c$ cannot contain any spectroscopic information. In our case however, we observe clear signatures of the superconducting energy gap at bias voltages much larger than voltage range where the ZBCP appears. Other origins of ZBCP include (i) magnetic scattering[34,35] (ii) proximity induced pair tunneling (PIPT)[36] and (iii) Andreev bound state[8,37] (ABS) when the superconductor has an unconventional symmetry. The ZBCP resulting from magnetic scattering is expected to split under the application of magnetic field and PIPT should get suppressed at small fields of the order of 0.1T.

In Figure A4. (a-b) we show the evolution of the ZBCP with magnetic field ($H$) applied perpendicular to the junction (i.e. $H\|I$) for two contacts with I$\|b$ and I$\perp b$ respectively. For this orientation of $H$, the ZBCP arising from ABS is not expected to split but will gradually reduce and disappear at high fields. We observe that the ZBCP for both I$\|b$ and I$\perp b$ persists moderately high fields and does not show any splitting with magnetic field. This effectively rules out magnetic scattering and PIPT as origins of the ZBCP, but is consistent with expectation for ABS. We would also like to point out while the narrow ZBCP observed in our experiments has a superficial similarity with that originating from Josephson effect in a superconductor-superconductor tunnel junction such a possibility is extremely remote in our experimental configuration where the contact is established between a normal metal tip and a superconducting single crystal. While it is possible for some contacts to form accidental grain boundary

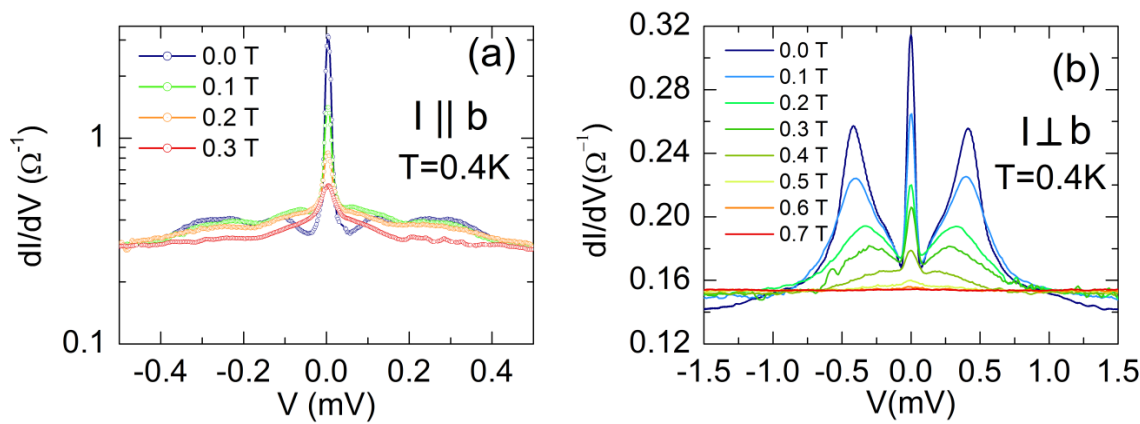

Figure A4. Magnetic field dependence of PCAR spectra showing the ZBCP for two point contacts with (a) I $\|$b and (b) I$\perp$b. The spectra are measured at 0.4K.

Josephson junction through a tiny broken piece of the crystal which comes in series during the contact formation process, it is statistically impossible for this to happen for all large area contacts that we have measured in the course of this study. We therefore conclude the ZBCP-s observed here are manifestations of ABS originating from an unconventional component of the order parameter in this material.

Further confirmation of the ABS origin of the ZBCP comes from its evolution with contact size. Since the mean size of the ABS is of the order of the dirty limit coherence length ($\xi_0$), the ZBCP originating from ABS gradually disappears as the contact diameter becomes smaller than $\xi_0$. From the upper critical field ($H_{c2}$) measured with $H||b$ and $H\perp b$ (Fig. A5. (b)), assuming the simplest situation of a triangular Abrikosov vortex lattice existing in BiPd we estimate $\xi_0$ to be of the order of 20 nm and 17 nm respectively. In figure A5. (a) we plot the height of the ZBCP (defined as the difference between the experimental zero bias conductance and the zero bias conductance obtained from the generalized two-band BTK fit) as a function of

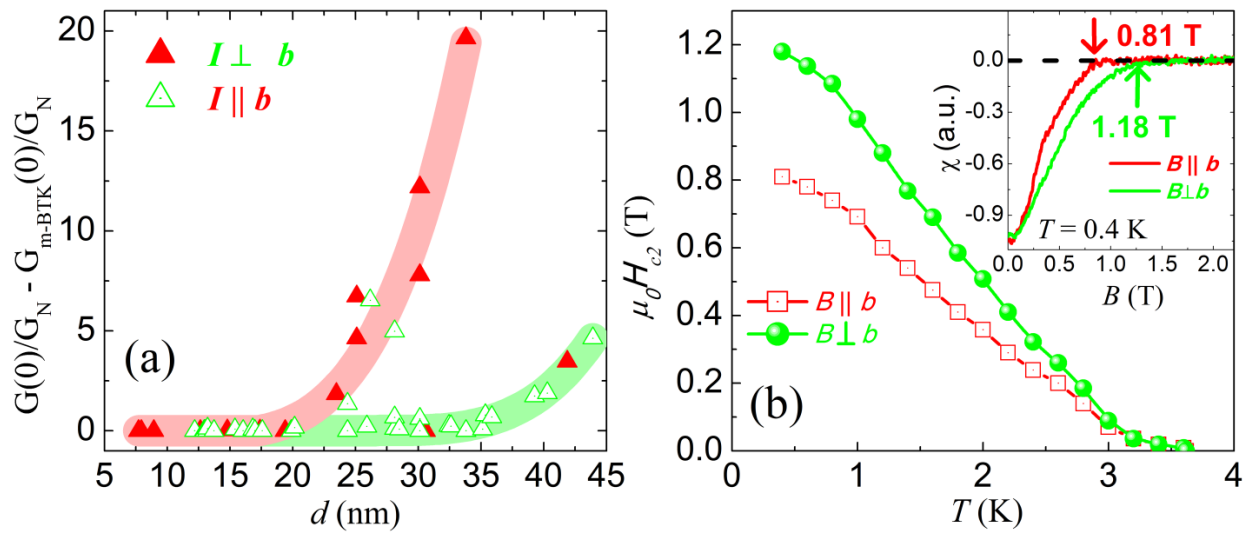

Figure A5. (a) Height of the ZBCP as function of contact diameter $d$. The solid triangle (red) corresponds $I||b$ and open triangle to $I\perp b$. The thick shaded lines are to guide eye. It shows two branches of points associated with ABS in two different directions. (b) $H_{c2}$ as function of temperature ($T$) for H$||b$ and H$\perp b$. The inset shows ac susceptibility as function of magnetic field at $T = 0.4$ K. The $H_{c2}$ (shown by arrows) has been extracted from susceptibility data taken as function magnetic field at different temperatures.

*d* calculated using the Shavrin formula [28]. The ZBCP disappears for $d \lesssim 20$ nm and $d \lesssim 32$ nm for I∥*b* and I⊥*b* respectively [38]. We believe that the slightly larger critical diameter for I⊥*b* compared to $\xi_0$ results from several approximations used in this analysis. First, the determination of *d* from $R_s$ is necessarily an approximation, which does not take into account the irregular shape of a real contact or the effect of the barrier potential that could exist between the tip and the superconductor. Secondly, the determination of $\xi_0$ assumes a triangular vortex lattice which might not necessarily be the case for a superconductor with unconventional pairing symmetry. Considering the errors involved with these approximations and the fact that the criterion for the disappearance of the ZBCP with contact size is only valid within a factor of the order of unity, the qualitative trend of the ZBCP with *d* is in excellent agreement with theoretical expectation for ABS. We therefore conclude that the ZBCP in BiPd originates from the ABS resulting from an unconventional OP for which the phase varies on the Fermi surface.

## A3.2. Superconducting order parameters

We can now put these observations in proper perspective. For a NCS, ASOC leads to a term of the form $\alpha g(k)\cdot\sigma$ in the Hamiltonian, where $\alpha$ is the spin-orbit coupling constant, $\sigma$ the Pauli matrices and vector $g(k)$, representing orbital direction, obeys the antisymmetric property such that $g(k)=-g(-k)$. In general the explicit form of $g(k)$ is determined by details of the crystal structure. The ASOC breaks the spin degeneracy which leads to two bands characterized by ± helicities for which the spin eigenstates are either parallel or antiparallel[39] to $g(k)$. This is schematically illustrated in Fig. A6. These helicity eigenstates are therefore coherent superposition of spin-up and spin-down eigenstates. The superconducting gap functions for the intraband pairs are $\Delta_+(k)$ and $\Delta_-(k)$ for the respective helicity bands. When the ASOC is large, interband pairing is suppressed and in such a case the superconducting transition temperature is maximized[7] when the quantization direction of the triplet symmetry becomes parallel to $g(k)$. Since the pairing occurs between intraband electrons only and the bands are of helicity eigenstates, the superconducting gap function will be an admixture of spin-siglet and spin-triplet symmetries: $\Delta(k)=(\Delta_s(k)I+\Delta_t(k) \hat{g}(\boldsymbol{k}) \cdot \sigma)(i\sigma_y)$, where I is the 2×2 identity matrix, $\hat{g}(\boldsymbol{k})$ is the unit vector pointing along $g(k)$ and $\Delta_s(k)$ and $\Delta_t(k)$ are the singlet and triplet amplitude of the gap function respectively. A PCAR experiment will thus see two gap functions, $\Delta_\pm(k)=\Delta_s(k)\pm\Delta_t(k)$ where each gap is defined on one of the two bands formed by the degeneracy lifting of ASOC. In general both the singlet and the triplet component of the order-parameter can be anisotropic and can even change sign over the Fermi surface. An ABS is formed as helical edge mode[12,40,41,42] for each $k$ when $|\Delta_t(k)| > |\Delta_s(k)|$. In such a situation, on one of the bands, say, $\Delta_-(k)$ can change sign giving rise to nodes in the superconducting gap function for the band with negative helicity. While at present we do not know $g(k)$ appropriate for the monoclinic structure for BiPd, based on the experimental data we propose the following scenario. Since $\Delta_1 \sim 0.4$ meV is observed for both I$\|b$ and L$\perp b$ and has similar temperature dependence in both directions, this is likely to originate from a one of the gap functions associated with $\Delta_+$. On the other hand $\Delta_2$ and $\Delta_3$ are likely to be both associated with a strongly anisotropic gap function ($\Delta_-$) for which the observed gap values are different for the two different directions of current injection. While in principle $\Delta_1$, $\Delta_2$ and $\Delta_3$ could also arise from a multiband scenario containing three different bands, this is an unlikely possibility for the following reasons. First, a simple multiband scenario consisting of multiple s-wave gap functions on different Fermi sheets cannot explain the existence of the pronounced ZBCP that we observe in our data. Secondly, we do not observe $\Delta_2$ and $\Delta_3$ simultaneously in any of our spectra despite the surface roughness that produces a significant scatter in their individual for both directions of injection current. It is therefore unlikely that $\Delta_2$ and $\Delta_3$ arise from two different gap functions on different Fermi sheets. In this context, we recall that in ABS is observed only when tunneling occurs in the basal plane[43] of the crystal of $Sr_2RuO_4$ whose OP has pure spin-triplet symmetry. On the contrary the ABS observed here in BiPd compound occurs for both I$\|b$ and L$\perp b$. We believe, the main reason behind this difference is that while pure triplet symmetry breaks the crystal symmetry, the mixed singlet-triplet OP in NCS can restore the full crystalline symmetry[19] and hence OP nodes in NCS can appear along all crystalline directions.

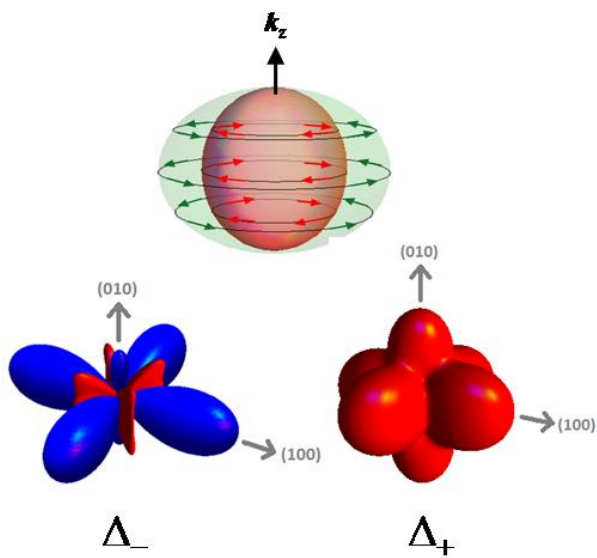

Figure A6. (a) Schematic illustration of the spin-split Fermi surface due to spin orbit coupling, forming the + and − helicity bands. (b) and (c) One possible realization of the gap functions corresponding to $\Delta_-$ and $\Delta_+$ corresponding to one possible choice of the vector $g(k)$. Red and blue correspond to positive and negative values of the gap function respectively. Note that the small monoclinic distortion of BiPd has been neglected in the symmetry of the gap functions.

## A4. Summary

We have shown evidence of mixing of spin-triplet and spin-singlet OP in the NCS BiPd. Furthermore, the presence of the pronounced ABS observed from the ZBCP suggests that the pair potential associated with the triplet OP is large enough to produce a sign change in at least one of the gap functions. Despite the absence of theoretical or experimental information on the Fermi surface that somehow limits our interpretation of experimental results, we believe that this observation is an important step towards realizing Majorana Fermionic modes which are predicted to exist in the vortex core of NCS [9]. We believe that our work will motivate further investigations on the precise nature of the order parameter symmetry in this interesting NCS superconductor.